\documentclass[11pt,twoside]{article}	% Fontsize 11 point
\usepackage{latexsym}			% load symbols like \Box
\makeatletter
\def\ps@robheadings{%
  \let\@oddfoot\@empty\let\@evenfoot\@empty
  \def\@evenhead{\thepage\hfil\slshape\leftmark}%
  \def\@oddhead{{\slshape\rightmark}\hfil\thepage}%
  \let\@mkboth\markboth
  \def\subsectionmark##1{}
  \def\sectionmark##1{\markright{##1}}}
\def\@maketitle{%
  \newpage
  \null
  \begin{center}%
  \let \footnote \thanks
    {\LARGE\bf \@title \par}%		\bf added
    \vskip 2em%				was: 1.5em
    {\large
      \lineskip .5em%
      \begin{tabular}[t]{c}%
        \@author
      \end{tabular}}%
    \vskip .3em%			\date and extra space removed
  \end{center}%
  \par}
\renewenvironment{thebibliography}[1]
     {\section*{\refname
        \markright{\refname}}%					changed
      \addcontentsline{toc}{section}{\refname}%			added
      \list{\@biblabel{\@arabic\c@enumiv}}%
           {\settowidth\labelwidth{\@biblabel{#1}}%
            \leftmargin\labelwidth
            \advance\leftmargin\labelsep
            \@openbib@code
            \usecounter{enumiv}%
            \let\p@enumiv\@empty
            \renewcommand\theenumiv{\@arabic\c@enumiv}}%
      \sloppy
      \clubpenalty4000
      \@clubpenalty \clubpenalty
      \widowpenalty4000%
      \sfcode`\.\@m}
     {\def\@noitemerr
       {\@latex@warning{Empty `thebibliography' environment}}%
      \endlist}

\makeatother
\newcommand{\hname}[1]{}
\newcommand{\href}[2]{\mbox{}{#2}}
\makeatletter
\def\@lbibitem[#1]#2{\item[\@biblabel{#1}\hfill]\hname{reference.#2}\if@filesw
      {\let\protect\noexpand
      \immediate\write\@auxout
      {\string\bibcite{#2}{\href{reference.#2}{#1}}}}\fi\ignorespaces}
\def\@bibitem#1{\item\hname{reference.#1}\if@filesw \immediate\write\@auxout
      {\string\bibcite{#1}{\href{reference.#1}{\the\value{\@listctr}}}}%
      \fi\ignorespaces}
\makeatother
\newcommand{\hlabel}[1]{\hname{#1}\label{#1}}
\newcommand{\hhref}[1]{\href{#1}{\ref{#1}}}

\newtheorem{defi}{Definition}
\newtheorem{theo}{Theorem}
\newtheorem{prop}{Proposition}
\newtheorem{lemm}{Lemma}
\newtheorem{coro}{Corollary}
\newtheorem{exam}{Example}

\newenvironment{definition}[1]{\begin{defi} \rm \hlabel{df-#1} }{\end{defi}}
\newenvironment{theorem}[1]{\begin{theo} \rm \hlabel{thm-#1} }{\end{theo}}
\newenvironment{proposition}[1]{\begin{prop} \rm \hlabel{pr-#1} }{\end{prop}}
\newenvironment{lemma}[1]{\begin{lemm} \rm \hlabel{lem-#1} }{\end{lemm}}
\newenvironment{corollary}[1]{\begin{coro} \rm \hlabel{cor-#1} }{\end{coro}}
\newenvironment{example}[1]{\begin{exam} \rm \hlabel{ex-#1} }{\end{exam}}
\newenvironment{proof}{\begin{trivlist} \item[\hspace{\labelsep}\bf Proof:]}{\hfill $\Box$\end{trivlist}}

\newenvironment{itemise}{\begin{list}{$\bullet$}{\leftmargin 18pt
                        \labelwidth\leftmargini\advance\labelwidth-\labelsep
                        \topsep 4pt \itemsep 2pt \parsep 2pt}}{\end{list}}
\newenvironment{itemise2}{\begin{list}{{\bf --}}{\leftmargin 15pt
                        \labelwidth\leftmargini\advance\labelwidth-\labelsep
                        \topsep 2pt \itemsep 1pt \parsep 1pt}}{\end{list}}
\newenvironment{itemise3}{\begin{list}{*}{\leftmargin 12pt
                        \labelwidth\leftmargini\advance\labelwidth-\labelsep
                        \topsep 0pt \itemsep 0pt \parsep 0pt}}{\end{list}}

\newcommand{\df}[1]{\href{df-#1}{Definition~\ref{df-#1}}}
\newcommand{\thm}[1]{\href{thm-#1}{Theorem~\ref{thm-#1}}}
\newcommand{\pr}[1]{\href{pr-#1}{Proposition~\ref{pr-#1}}}
\newcommand{\lem}[1]{\href{lem-#1}{Lemma~\ref{lem-#1}}}
\newcommand{\cor}[1]{\href{cor-#1}{Corollary~\ref{cor-#1}}}
\newcommand{\ex}[1]{\href{ex-#1}{Example~\ref{ex-#1}}}
\newcommand{\sect}[1]{\href{sec-#1}{Section~\ref{sec-#1}}}
\newcommand{\tab}[1]{\href{tab-#1}{Table~\ref{tab-#1}}}

\renewenvironment{abstract}{\begin{list}{}
			{\rightmargin\leftmargin
			\listparindent 1.5em
			\parsep 0pt plus 1pt}
			\small\item} {\end{list}}

\newcommand{\dl}[1]{\mbox{\rm I\hspace{-0.75mm}#1}}	% openface letter
\newcommand{\dc}[1]{\mbox{\rm {\raisebox{.4ex}{\makebox	% openface character
	[0pt][l]{\hspace{.2em}\scriptsize $\mid$}}}#1}}
\newcommand{\IT}{\mbox{\sf T\hspace{-5.5pt}T}}          % openface T (terms)
\newcommand{\plat}[1]{\raisebox{0pt}[0pt][0pt]{$#1$}}   % no vertical space
\newcommand{\goto}[1]{\stackrel{#1}{\longrightarrow}}	% transition
\newcommand{\gonotto}[1]{\hspace{4pt}\not\hspace{-4pt}	% no transition
	\stackrel{#1\ }{\longrightarrow}}
\newcommand{\bis}[1]{ \;				% bisimulation
	\raisebox{.3ex}{$\underline{\makebox[.7em]{$\leftrightarrow$}}$}
                  \,_{#1}\,}
\newcommand{\pf}{\noindent {\bf Proof:\ }}		% beginning of proof
\newcommand{\rest}{\unitlength 1mm			% restriction operator
        \begin{picture}(2.16,3.2)
        \thinlines
        \put(1.12,-0.48){\line(0,1){3.52}}
	\put(0.8,1.6){\tiny $\backslash$}
        \end{picture} }
\newcommand{\var}{{\it var}}				% variables in a term
\newcommand{\Rule}[2]{					% transition rule
	\frac{\raisebox{.7ex}{\normalsize{$#1$}}}
	{\raisebox{-1.0ex}{\normalsize{$#2$}}}}
\newcommand{\irr}{\vdash_{\rm irr}}		    % irredundantly provable

\newcommand{\IN}{\dl{N}}			    % natural numbers
\newcommand{\IP}{\dl{P}}			    % transition space
\newcommand{\IO}{\dc{O}}			    % observations
\newcommand{\pow}{{\cal P}}			    % powerset
\newcommand{\fO}{{\cal O}}			    % function into \IO

\begin{document}

\markboth{\hfill Precongruence Formats for Decorated Trace Semantics}{}

\title{Precongruence Formats for Decorated Trace Semantics\footnote{
   This paper subsumes most of \cite{Bl93} and part of \cite{vG93a}.
   An extended abstract of our work appeared as \cite{BFG00}, except that
   in the current paper we do not include the full abstraction results mentioned
   there. Those results originated from \cite{vG93a} and are planned to appear in a full
   version of \cite{vG93a}.}}
   \author{ \sf Bard Bloom\\
 \footnotesize \sl IBM T.J.\ Watson Research Center\\[-3pt]
 \footnotesize \sl Hawthorne, NY 10532, USA\\[-3pt]
 \footnotesize \tt bardb@us.ibm.com
   \and     \sf Wan Fokkink\thanks{Supported by a grant from The Nuffield
            Foundation.}\\
 \footnotesize \sl Dept.\ of Software Engineering\\[-3pt]
 \footnotesize \sl CWI\\[-3pt]
 \footnotesize \sl PO Box 94079\\[-3pt]
 \footnotesize \sl 1090 GB Amsterdam, NL\\[-3pt]
 \footnotesize \tt wan@cwi.nl\\[-3pt]
 \footnotesize \tt http://www.cwi.nl/\~{}wan
   \and     \sf Rob van Glabbeek\\
 \footnotesize \sl Computer Science Department\\[-3pt]
 \footnotesize \sl Stanford University\\[-3pt]
 \footnotesize \sl Stanford, CA 94305-9045, USA\\[-3pt]
 \footnotesize \tt rvg@cs.stanford.edu\\[-3pt]
 \footnotesize \tt http://theory.stanford.edu/\~{}rvg}

\maketitle
\thispagestyle{empty}

%%%%%%%%%%%%%%%%%%%%%%%%%%%%%%%%%%%%%%%%%%%%%%%%%%%%%%%%%%%%%%%%%%%%%%%%%%%%%%
\begin{abstract}
This paper explores the connection between semantic equivalences and
preorders for concrete sequential processes, represented by means of
labelled transition systems, and formats of transition system
specifications using Plotkin's structural approach. For several
preorders in the linear time -- branching time spectrum a format
is given, as general as possible, such that this preorder is a
precongruence  for all operators specifiable in that format.
The formats are derived using the modal characterizations of the
corresponding preorders.
\end{abstract}

\section{Introduction}

{\em Structural operational semantics} \cite{Pl81} provides process
algebras and specification languages with an interpretation. It
generates a {\em (labelled) transition system} (LTS), in which states
are the closed terms over a (single-sorted, first-order) signature,
and transitions between states may be supplied with labels. The
transitions between states are obtained from a {\em transition system
specification} (TSS), this being a signature together with
a set of proof rules called {\em transition rules}. 

In case of a TSS with only positive premises, the associated LTS simply
consists of the transitions derivable from the transition rules.  In
the presence of negative premises it is not always straightforward to
associate an LTS to a TSS. One can for instance express that a
transition holds if it does not hold.  In {\sc van Glabbeek}
\cite{vG95} the notion of derivability of transitions from a TSS is
extended to negated transitions by incorporating a notion of {\em
negation as failure} (cf.\ \cite{Cl78}): a {\em supported proof}
enables one to derive the negation of a transition by disproving a
premise in each substitution instance of each transition rule that
carries this transition as its conclusion; a {\em well-supported proof}
(inspired by the notion of a well-supported model \cite{Fa91})
features an even more powerful method to derive the negation of a
transition by demonstrating that that transition can not be derived.
A TSS is {\em complete} \cite{vG95} if for each transition
there is a well-supported proof from the TSS either for the transition
itself or for its negation.  The LTS associated
to a complete TSS consists of the transitions for which there is a
well-supported proof.
An incomplete TSS arguably does not specify a LTS in a meaningful way
at all. However, it specifies what one could call a {\em 3-valued}
LTS: an LTS in which potential transitions are either present, absent
or unknown.  This approach to the meaning of transition system
specifications can be seen as a proof-theoretic characterization of the
work of {\sc van Gelder, Ross \& Schlipf} \cite{GRS91} in logic
programming.  The notion of completeness of a TSS also coincides with
the notion of {\em being positive after reduction} of {\sc Bol \&
Groote} \cite{BolG96}.

LTSs can be distinguished from each other by means of a wide range
of semantic equivalences and preorders. These preorders are
based on the branching structure of LTSs
({\em simulation} \cite{Pa81}, {\em ready simulation} \cite{BIM95,LaSk91},
{\em bisimulation} \cite{Mi89,Pa81}, {\em nested simulations} \cite{GrV92}),
on execution sequences ({\em partial traces}, {\em completed traces},
{\em accepting traces}),
or on decorated versions of execution sequences
({\em ready pairs} \cite{BKO88,OlHo86,RoBr81}, {\em failure pairs} \cite{BKO88,BHR84,DH84},
{\em ready traces} \cite{BBK87b,Pnu85}, {\em failure traces} \cite{Phi87}).
In \cite{vG90a}, {\sc van Glabbeek} classified most equivalences and
preorders for concrete, sequential processes\footnote{A process is
{\em sequential} if it can do only one action at a time; {\em
concrete} refers to the absence of internal actions or internal
choice.} that occur in the literature, and  motivated them
by means of testing scenarios, phrased in terms of
`button pushing experiments' on generative and reactive machines.
This gave rise to {\em modal characterizations} of the preorders,
i.e.\ characterizations in terms of the observations that an
experimenter could make during a session with a process.

In general a semantic equivalence (or preorder) induced
by a TSS is not a {\em congruence} (resp.\ {\em precongruence}), i.e.\ the equivalence
class of a term $f(t_1,\ldots,t_n)$ need not be determined by the
equivalence classes of its arguments $t_1,\ldots,t_n$. Being a
(pre)congruence is an important property, for instance in order
to fit the equivalence (or preorder) into an axiomatic framework.
Syntactic formats for TSSs have been developed with respect to several
semantic equivalences and preorders, to ensure that such an equivalence or
preorder as induced by a TSS in the corresponding format is a (pre)congruence.
These formats have helped to avoid repetitive (pre)congruence
proofs, and to explore the limits of sensible TSS definitions.
A first congruence format for bisimulation equivalence was put
forward by {\sc de Simone} \cite{dS85}, which was extended to the
{\em GSOS} format by {\sc Bloom, Istrail \& Meyer} \cite{BIM95} and to the
{\em tyft/tyxt} format by {\sc Groote \& Vaandrager} \cite{GrV92}.
The tyft/tyxt format was extended with negative premises
\cite{BolG96,Gr93} to obtain the {\em ntyft/ntyxt} format
for complete TSSs. The ntyft/ntyxt format
also generalizes the GSOS format.
The congruence results of \cite{BolG96,Gr93,GrV92} were
originally only proved for TSSs satisfying a {\em well-foundedness}
criterion; in \cite{FvG96} they were shown to hold for all complete
TSSs in ntyft/ntyxt format.
To mention some formats for other equivalences and preorders,
{\sc Vaandrager} \cite{Va91a} observed that de Simone's format is a
precongruence format for the partial trace and the failure preorders,
{\sc Bloom} \cite{Bl94} introduced a more general congruence format
for partial trace equivalence,
and {\sc Fokkink} \cite{Fok00a} introduced a precongruence format
for the accepting trace preorder. Finally, {\sc Bloom} \cite{Bl93}
introduced a congruence format for readiness equivalence,
and {\sc van Glabbeek} \cite{vG93a} one for ready simulation
equivalence. This {\em ready simulation} format generalizes the GSOS format.

In this paper precongruence formats are proposed for several
semantic preorders based on decorated traces, building on results
reported in \cite{Bl93,vG93a}.
We introduce precongruence formats for the ready trace preorder,
the readiness preorder, the failure trace preorder and
the failure preorder. The precongruence formats for the last two
preorders coincide.
Following \cite{Bl93,Fok00a,Fok00b}, these three precongruence formats
distinguish between {\em frozen} and {\em liquid} arguments of function symbols.
This distinction is used in posing restrictions
on occurrences of variables in transition rules.
The ready simulation format of \cite{vG93a} is more liberal than the
format for the ready trace preorder, which is more liberal than the format
for the readiness preorder, which is more liberal than the format for
the failure trace preorder, which in turn is more liberal than de Simone's format.

The precongruence formats introduced in this paper apply to
incomplete TSSs as well. For this purpose the definitions of the
corresponding preorders are extended to 3-valued LTSs.

We also show that the tyft/tyxt format is adequate for the nested
simulation preorders.

The precongruence formats put forward in this paper
were obtained by a careful study of the
modal characterizations of the preorders in question.
The outline of the proof underlying each of our precongruence results
is as follows. First, any TSS in our ready simulation format is
transformed into an equivalent TSS---equivalent in the sense that is
proves the same transitions and negated transitions---of a special
form, in which the left-hand sides of positive premises are single
variables.  For such TSSs we show that the notions of supported
and well-supported provability coincide. Next, any such TSS is
extended with a number of transition rules with negative conclusions,
in such a  way that a (negated) transition has a supported proof from
the original TSS if and only if it has a standard proof from the extended
TSS\@. In the extended TSS, the left-hand sides of positive and negative
premises can further be reduced to single variables.
It is shown for each of the precongruence formats in this paper that
its syntactic criteria are preserved under these transformations.
Finally, the resulting transition rules are used to decompose modal
formulas; that is, a modal formula $\varphi$ for an open term $t$ is
decomposed into a choice of modal formulas $\psi(x)$ for variables $x$
such that $t$ satisfies $\varphi$ if and only if for one of those $\psi$s
the variables $x$ in $t$ satisfy $\psi(x)$.
The precongruence format for each preorder under consideration guarantees
that if a modal formula is within the modal characterization of this
preorder, then the same holds for the resulting decomposed modal formulas.
This implies the desired precongruence result.

{\sc Larsen} \cite{Lar86} and {\sc Larsen \& Xinxin} \cite{LX91} obtained
compositionality results for the {\em Hennessy-Milner logic} \cite{HeMi85}
and the {\em $\mu$-calculus} \cite{Koz82}, respectively, with respect to
{\em action transducers}, which constitute a reformulation of TSSs in
de Simone's format. The technique for decomposing modal formulas that is employed
in the current paper is firmly related to their approach, but applies to the
richer ready simulation format.

This paper is set up as follows.
\sect{preorders} gives definitions of existing semantic
preorders, both in terms of decorated traces or simulations and of
observations.
\sect{sos} presents the basics of structural operational
semantics, and extends the definitions of most of the preorders to
3-valued LTSs. \sect{formats} recalls the ntyft/ntyxt
format and formulates extra requirements
to obtain new precongruence formats for a range of preorders.
In \sect{preservation} it is shown that the syntactic
restrictions of the respective precongruence formats are preserved under
{\em irredundant} provability.
\sect{reducing} explains how well-supported
proofs are reduced to standard proofs.
In \sect{nxytt} it is shown that left-hand sides of premises
can be reduced to single variables.
\sect{crux} contains the proofs of the
precongruence results, based on material in the previous three sections,
and using the definitions of preorders in terms of observations.
In \sect{counterexamples}, counterexamples are given to show
that all syntactic restrictions are essential for the obtained
precongruence results. \sect{applications} presents some
applications of the precongruence formats to TSSs from the literature.
In \sect{partial} it is argued that our
techniques apply to other preorders as well. As an example we show
that the precongruence format for the failure trace preorder
is a congruence format for partial trace equivalence, and that
the positive variant of this format is a precongruence format for the
partial trace preorder.
In \sect{conservativity}, a conservative extension result is established for
incomplete TSSs in ready simulation format.
Finally, \sect{conclusion} discusses possible extensions of the congruence
formats derived in this paper.

\newpage

%%%%%%%%%%%%%%%%%%%%%%%%%%%%%%%%%%%%%%%%%%%%%%%%%%%%%%%%%%%%%%%%%%%%%%%%%%%%%%
\section{Preorders and equivalences on labelled transition systems}
\hlabel{sec-preorders}

\begin{definition}{LTS}
A {\em labelled transition system (LTS)} is a pair $(\IP,
\rightarrow)$ with $\IP$ a set (of {\em processes}) and
$\rightarrow\; \subseteq \IP \times A \times \IP$ for $A$ a set (of
{\em actions}).
\end{definition}
{\bf Notation:} Write $p \goto{a} q$ for $(p,a,q) \in
\rightarrow$ and $p \gonotto{a}$ for $\neg\exists q \in \IP: p
\goto{a} q$.\\[10pt]
The elements of $\IP$ represent the processes we are interested in,
and $p \goto{a} q$ means that process $p$ can evolve into process $q$
while performing the action $a$.  
We start with defining six preorders on LTSs, which are based on
execution sequences of processes. Given an LTS, we write 
$p \goto{\varsigma}q$, where $\varsigma = a_1\cdots a_n\in A^*$ for
$n\!\in\!\IN$, if there are processes $p_0,\ldots,p_n$ with
$p \!=\! p_0 \goto{a_1} p_1 \goto{a_2} \cdots \goto{a_n} p_n \!=\! q$.
For a process $p$ we define
\begin{eqnarray*}
{\sf initials}(p) & = & \{ a \in A \mid \exists p'\in\IP: p \goto{a} p'\}.
\end{eqnarray*}

\begin{definition}{decorated-traces}
Assume an LTS.
\begin{itemise}
\item
  A sequence $\varsigma \in A^*$ is a
  {\em (partial) trace} of process $p$ if
  $p \goto{\varsigma} q$ for some process $q$.
\item
  $\varsigma\in A^*$ is a {\em completed trace} of process $p$ if
  $p \goto{\varsigma} q$ for some process $q$ with
  ${\sf initials}(q)=\emptyset$.
\item
  A pair $(\varsigma,X)$ with
  $\varsigma\in A^*$ and $X \subseteq A$
  is a {\em ready pair} of process $p$ if $p\goto{\varsigma}q$ for
  some process $q$ with ${\sf initials}(q) = X$.
\item
  A pair $(\varsigma,X)$ with
  $\varsigma\in A^*$ and $X \subseteq A$
  is a {\em failure pair} of process $p$ if $p\goto{\varsigma}q$
  for some process $q$ with ${\sf initials}(q)\cap X=\emptyset$.
\item
  A sequence $X_0 a_1 X_1\cdots a_n X_n$ (with $n\in\IN$), where
  $X_i\subseteq A$ and $a_i\in A$ for $i=0,\ldots,n$, is a
  {\em ready trace} of process $p_0$ if
  $p_0 \goto{a_1} p_1 \goto{a_2} \cdots \goto{a_n} p_n$ and
  ${\sf initials}(p_i) = X_i$ for $i=0,\ldots,n$.
\item
  A sequence $X_0 a_1 X_1\cdots a_n X_n$ (with $n\in\IN$), where
  $X_i\subseteq A$ and $a_i\in A$ for $i=0,\ldots,n$, is a
  {\em failure trace} of process $p_0$ if
  $p_0 \goto{a_1} p_1 \goto{a_2} \cdots \goto{a_n} p_n$ and
  ${\sf initials}(p_i) \cap X_i = \emptyset$ for $i=0,\ldots,n$.
\end{itemise}
We write $p \sqsubseteq_{\it T} q$, $p \sqsubseteq_{\it C} q$, $p \sqsubseteq_{\it R} q$,
$p \sqsubseteq_{\it F} q$, $p \sqsubseteq_{\it RT} q$ or $p \sqsubseteq_{\it FT} q$
if the set of (partial) traces, completed traces, ready pairs, failure
pairs, ready traces or failure traces of $p$ is included in that of
$q$, respectively. Write $p \sqsubseteq_{\it CT} q$ for $p \sqsubseteq_{\it T}
q \wedge p \sqsubseteq_{\it C} q$; the preorder $\sqsubseteq_{\it C}$ is used
only to define $\sqsubseteq_{\it CT}$.
\end{definition}
We proceed to define preorders based on simulation of one process by another.

\begin{definition}{simulation}
Assume an LTS.
For a binary relation $R$, let $pR^{-1}q$ iff $qRp$.
\begin{itemise}
\item A binary relation $R$ on processes is a {\em (1-nested) simulation} if
  whenever $pRq$ and $p\goto{a}p'$, then there is a transition
  $q\goto{a}q'$ such that $p'Rq'$.
\item A simulation $R$ is a {\em ready simulation} if
  whenever $pRq$ and $p\gonotto{a}$, then $q\gonotto{a}$.
\item A simulation $R$ is an $n$-nested simulation (for $n\geq 2$)
  if $R^{-1}$ is contained in an $(n-1)$-nested simulation.
\item A {\em bisimulation} is a
  simulation $R$ such that also $R^{-1}$ is a simulation.
\end{itemise}
We write $p \sqsubseteq_{n{\it S}} q$ (for $n\geq 1$), $p \sqsubseteq_{\it RS} q$
or $p\sqsubseteq_{\it B} q$ if there is an $n$-nested
simulation, ready simulation or bisimulation $R$ with $pRq$, respectively.
\end{definition}
Note that $\sqsubseteq_{\it B}$ is symmetric, so that it constitutes
an equivalence relation. In the literature this notion of {\em
bisimulation equivalence} is often denoted by $\bis{}$.

In  \cite{vG90a}, {\sc van Glabbeek} observed that if one
restricts to the domain of finitely branching, concrete, sequential
processes, then most semantic preorders found in the literature ``that
can be defined uniformly in terms of action relations'' coincide with
one of the preorders defined above. He motivated these preorders
by means of testing scenarios, phrased in terms of
`button pushing experiments' on generative and reactive machines.
This gave rise to {\em modal\/} characterizations of
the preorders, characterizations in terms of the {\em observations} that an
experimenter could make during a session with a process.

\begin{definition}{potential-observations}
Assume an action set $A$.
The set $\IO$ of {\em potential observations} or {\em modal formulas}
is defined inductively by:
\begin{list}{}{\labelwidth\leftmargini\advance\labelwidth-\labelsep
	       \topsep 2pt \itemsep 1pt \parsep 1pt}
\item [$\top \in \IO.$] The trivial observation, obtained by
      terminating the session.
\item [$a\varphi \in \IO$] if $\varphi \in \IO$ and $a \in A$.
      The observation of an action $a$, followed by the observation $\varphi$.
\item [$\widetilde{a} \in \IO$] for $a \in A$.
      The observation that the process cannot perform the action $a$.
\item [$\bigwedge_{i\in I}\varphi_i \in \IO$] if $\varphi_i \in \IO$
      for all $i\in I$. The process admits each of the observations
      $\varphi_i$.
\item [$\neg \varphi \in \IO$] if $\varphi \in \IO$. (It can be
      observed that) $\varphi$ cannot be observed.
\end{list}
\end{definition}

\begin{definition}{observations}Let $(\IP,\rightarrow)$ be a
LTS, labelled over $A$. The {\em satisfaction relation} $\models\;
\subseteq \IP \times \IO$ telling which observations are possible for
which process is inductively defined by the clauses below.\vspace{-3pt}
$${\renewcommand{\arraystretch}{1.1}\arraycolsep 2pt \begin{array}{@{}llll@{}}
p \models \top	 	\\
p \models a\varphi	& \mbox{if} & \exists q: p \goto{a} q \wedge q \models \varphi\\
p \models \widetilde{a}	& \mbox{if} & p \gonotto{a}\\
p \models \bigwedge_{i\in I}\varphi_i& \mbox{if}&p \models \varphi_i
\mbox{ for all } i\in I\\
p \models \neg\varphi	& \mbox{if} & p \not\models \varphi\\
\end{array}}$$
\end{definition}
We will use the binary conjunction $\varphi_1 \wedge \varphi_2$ as an
abbreviation of $\bigwedge_{i \in \{1,2\}} \varphi_i$, whereas $\top$
is identified with the empty conjunction. We identify formulas that are
logically equivalent using the laws for conjunction $\top \wedge \varphi
\cong \varphi$ and $\bigwedge_{i \in I}(\bigwedge_{j \in J_i} a_{ij})
\cong \bigwedge_{k \in K}a_k$ where $K=\{ij \mid i \in I \wedge j
\in J_i\}$. This is justified because $\varphi \cong \psi$ implies $p
\models \varphi \Leftrightarrow p \models \psi$.

\begin{definition}{sublanguages}
Below, several sublanguages of the set $\IO$ of observations are defined.
\begin{list}{}{\renewcommand{\makelabel}[1]{#1\hfill}
       \leftmargin 30pt \labelwidth\leftmargin\advance\labelwidth-\labelsep
	       \topsep 2pt \itemsep 1pt \parsep 1pt}
\item [$\IO_{\it T}$] $\varphi ::= \top \mid a\varphi'~(\varphi' \in \IO_{\it T})$
		\hfill {\em (partial) trace} observations
\item [$\IO_{\it CT}$] $\varphi ::= \top \mid a\varphi'~(\varphi' \in \IO_{\it CT}) 
	        \mid \bigwedge_{a \in A} \widetilde{a}$
		\hfill {\em completed trace} observations
\item [$\IO_{\it F}$] $\varphi ::= \top \mid a\varphi'~(\varphi' \in \IO_{\it F})
		\mid \bigwedge_{i\in I} \widetilde{a_i}$ 
		\hfill {\em failure} observations
\item [$\IO_{\it R}$] $\varphi ::= \top \mid a\varphi'~(\varphi' \in \IO_{\it R}) 
		\mid \bigwedge_{i\in I} \widetilde{a_i}
		\wedge \bigwedge_{j\in J} b_j \top$
		\hfill {\em readiness} observations
\item [$\IO_{\it FT}$] $\varphi ::= \top \mid a\varphi'~(\varphi' \in \IO_{\it FT})
		\mid \bigwedge_{i\in I} \widetilde{a_i} 
		\wedge\varphi'~(\varphi' \in \IO_{\it FT})$
		\hfill {\em failure trace} observations
\item [$\IO_{\it RT}$] $\varphi ::= \top \mid a\varphi'~(\varphi' \in \IO_{\it RT})
		\mid \bigwedge_{i\in I} \widetilde{a_i}
		\wedge \bigwedge_{j\in J} b_j \top
		\wedge\varphi'~(\varphi' \in \IO_{\it RT})$
		\hfill {\em ready trace} observations
\item [$\IO_{1{\it S}}$] $\varphi ::= \top \mid a\varphi'~(\varphi' \in \IO_{1{\it S}})
                \mid \bigwedge_{i\in I}\varphi_i~(\varphi_i \in \IO_{1{\it S}})$
		\hfill {\em (1-nested) simulation} observations
\item [$\IO_{\it RS}$] $\varphi ::= \top \mid a\varphi'~(\varphi' \in \IO_{\it RS})
		\mid \widetilde{a}
		\mid \bigwedge_{i\in I}\varphi_i~(\varphi_i \in \IO_{\it RS})$
                \hfill {\em ready simulation} observations
\item [$\IO_{n{\it S}}$] $\varphi ::= \top \mid a\varphi'~(\varphi' \in \IO_{n{\it S}})
		\mid \bigwedge_{i\in I}\varphi_i~(\varphi_i \in \IO_{n{\it S}})
		\mid \neg \varphi' ~(\varphi' \in \IO_{(n-1){\it S}})$\newline\mbox{}
                \hfill {\em n-nested simulation} observations (for $n\geq 2$)
\item [$\IO_{\it B}$] $\varphi ::= \top \mid a\varphi'~(\varphi' \in \IO_{\it B})
		\mid \bigwedge_{i\in I}\varphi_i~(\varphi_i \in \IO_{\it B})
		\mid \neg \varphi'~(\varphi' \in \IO_{\it B})$
                \hfill {\em bisimulation} observations
\end{list}
For each of these notions $N$, the set of {\em $N$-observations of
$p$} is $\fO_N(p) := \{\varphi \in \IO_N \mid p \models \varphi\}$.
\end{definition}

\begin{theorem}{modal}
For $N\in\{{\it T},{\it CT},{\it F},{\it R},{\it FT},{\it RT},{\it RS},
n{\it S}~(n\geq 1),{\it B}\}$, $p \sqsubseteq_N q$ iff
$\fO_N (p) \subseteq \fO_N (q)$. 
\end{theorem}
In fact a slight variation of this result will be needed in this paper.

\begin{definition}{conjunctive sublanguages}
For $N$ as above, let $\IO_N^\wedge$ consist of all formulas
$\bigwedge_{i \in I} \varphi_i$ with $\varphi_i \in \IO_N$.
Let $\fO_N^\wedge(p) := \{\varphi \in \IO_N^\wedge \mid p \models \varphi\}$.
\end{definition}
Then clearly $\fO_N (p) \subseteq \fO_N (q) \Leftrightarrow
\fO_N^\wedge (p) \subseteq \fO_N^\wedge (q)$.
In case $N \in \{{\it RS},n{\it S}~(n\geq 1),{\it B}\}$ there is, up to logical
equivalence, no difference between $\IO_N$ and $\IO_N^\wedge$, or
between $\fO_N(p)$ and $\fO_N^\wedge(p)$.

\begin{corollary}{modal}
For $N\in\{{\it T},{\it CT},{\it F},{\it R},{\it FT},{\it RT},{\it RS},
n{\it S}~(n\geq 1),{\it B}\}$, $p \sqsubseteq_N q$ iff
$\fO_N^\wedge (p) \subseteq \fO_N^\wedge (q)$.
\end{corollary}
The following hierarchy follows immediately from the definitions of
the respective preorders in terms of observations; see \cite{vG90a}.
In any LTS,

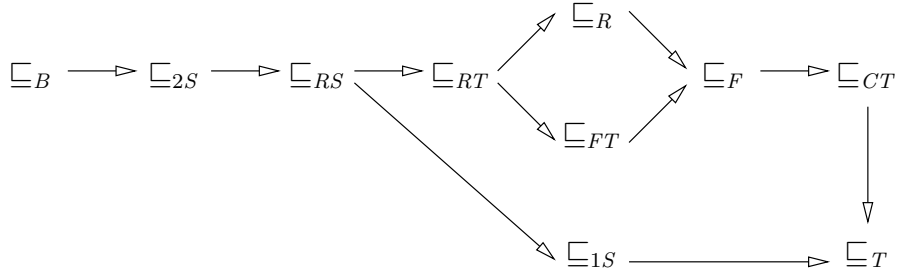
\vspace{3mm}
\begin{picture}(0,0)%
\includegraphics{hierarchy.pstex}%
\end{picture}%
\setlength{\unitlength}{3947sp}%
\begingroup\makeatletter\ifx\SetFigFont\undefined
% extract first six characters in \fmtname
\def\x#1#2#3#4#5#6#7\relax{\def\x{#1#2#3#4#5#6}}%
\expandafter\x\fmtname xxxxxx\relax \def\y{splain}%
\ifx\x\y   % LaTeX or SliTeX?
\gdef\SetFigFont#1#2#3{%
  \ifnum #1<17\tiny\else \ifnum #1<20\small\else
  \ifnum #1<24\normalsize\else \ifnum #1<29\large\else
  \ifnum #1<34\Large\else \ifnum #1<41\LARGE\else
     \huge\fi\fi\fi\fi\fi\fi
  \csname #3\endcsname}%
\else
\gdef\SetFigFont#1#2#3{\begingroup
  \count@#1\relax \ifnum 25<\count@\count@25\fi
  \def\x{\endgroup\@setsize\SetFigFont{#2pt}}%
  \expandafter\x
    \csname \romannumeral\the\count@ pt\expandafter\endcsname
    \csname @\romannumeral\the\count@ pt\endcsname
  \csname #3\endcsname}%
\fi
\fi\endgroup
\begin{picture}(6919,1695)(93,-2425)
\put(901,-1261){\makebox(0,0)[b]{\smash{\SetFigFont{12}{14.4}{rm}$\sqsubseteq_{\it B}$}}}
\put(1801,-1261){\makebox(0,0)[b]{\smash{\SetFigFont{12}{14.4}{rm}$\sqsubseteq_{2{\it S}}$}}}
\put(2701,-1261){\makebox(0,0)[b]{\smash{\SetFigFont{12}{14.4}{rm}$\sqsubseteq_{\it RS}$}}}
\put(4426,-886){\makebox(0,0)[b]{\smash{\SetFigFont{12}{14.4}{rm}$\sqsubseteq_{\it R}$}}}
\put(5251,-1261){\makebox(0,0)[b]{\smash{\SetFigFont{12}{14.4}{rm}$\sqsubseteq_{\it F}$}}}
\put(6151,-1261){\makebox(0,0)[b]{\smash{\SetFigFont{12}{14.4}{rm}$\sqsubseteq_{\it CT}$}}}
\put(4426,-1636){\makebox(0,0)[b]{\smash{\SetFigFont{12}{14.4}{rm}$\sqsubseteq_{\it FT}$}}}
\put(3601,-1261){\makebox(0,0)[b]{\smash{\SetFigFont{12}{14.4}{rm}$\sqsubseteq_{\it RT}$}}}
\put(4426,-2386){\makebox(0,0)[b]{\smash{\SetFigFont{12}{14.4}{rm}$\sqsubseteq_{1{\it S}}$}}}
\put(6151,-2386){\makebox(0,0)[b]{\smash{\SetFigFont{12}{14.4}{rm}$\sqsubseteq_{\it T}$}}}
\end{picture}

\vspace{3mm}
\noindent
where a directed edge from one preorder to another means that the
source of the edge is included in (i.e.\ is a finer preorder than) the target.

For every preorder $\sqsubseteq_N$ defined above there exists an
associated equivalence $=_N$ (the {\em kernel} of $\sqsubseteq_N$)
given by $p =_N q$ iff $p \sqsubseteq_N q \wedge q \sqsubseteq_N p$.  As
$\sqsubseteq_{\it B}$ is symmetric, one has $p =_{\it B} q \Leftrightarrow q
\sqsubseteq_{\it B} p$. Obviously $p =_N q$ iff $\fO_N(p) = \fO_N(p)$ iff
$\fO_N^\wedge(p) = \fO_N^\wedge(p)$. The same inclusion hierarchy
given for the preorders above applies to their kernels.

%%%%%%%%%%%%%%%%%%%%%%%%%%%%%%%%%%%%%%%%%%%%%%%%%%%%%%%%%%%%%%%%%%%%%%%%%%%%%%
\section{Structural operational semantics}
\hlabel{sec-sos}
 
In this paper $V$ and $A$ are two sets of {\em variables} and {\em
actions}. $V$ should be infinite and at least at large as $A$
(i.e.\ $|V| \geq |A|$).
Many concepts that will appear later on are parameterized by the
choice of $V$ and $A$, but as in this paper this choice is fixed, a
corresponding index is suppressed. A syntactic object is called
{\em closed} if it does not contain any variables from $V$.

\begin{definition}{signature}
A {\em signature} is a collection $\Sigma$ of {\em function symbols}
$f \not\in V$, with $|\Sigma| \leq |V|$, equipped with a function $ar:
\Sigma \rightarrow \IN$.
The set $\IT(\Sigma)$ of {\em terms} over a signature $\Sigma$ is
defined recursively by:
\begin{itemise}
\item $V \subseteq \IT(\Sigma)$,
\item if $f \in \Sigma$ and $t_1,\ldots,t_{ar(f)} \in
\IT(\Sigma)$ then $f(t_1,\ldots,t_{ar(f)}) \in \IT(\Sigma)$.
\end{itemise}
A term $c()$ is abbreviated as $c$.  For $t\in \IT(\Sigma)$,
$\var(t)$ denotes the set of variables that occur in $t$. \linebreak[3]
$T(\Sigma)$ is the set of closed terms over $\Sigma$, i.e.\
the terms $t \in \IT(\Sigma)$ with $\var(t)=\emptyset$.
A {\em $\Sigma$-substitution} $\sigma$ is a partial function from $V$
to $\IT(\Sigma)$. If $\sigma$ is a substitution and $S$ is any syntactic
object, then $\sigma(S)$ denotes the object obtained from $S$ by
replacing, for $x$ in the domain of $\sigma$, every occurrence of $x$
in $S$ by $\sigma(x)$. In that case $\sigma(S)$ is called a {\em
substitution instance} of $S$. A $\Sigma$-substitution is {\em closed} if
it is a total function from $V$ to $T(\Sigma)$.
\end{definition}

\begin{definition}{TSS}
Let $\Sigma$ be a signature. A {\em positive $\Sigma$-literal} is an
expression $t \goto{a} t'$ and a {\em negative $\Sigma$-literal} an
expression $t \gonotto{a}$ with $t,t'\in \IT(\Sigma)$ and $a \in
A$. For $t,t' \in \IT(\Sigma)$ the literals $t \goto{a} t'$ and $t
\gonotto{a}$ are said to {\em deny} each other.  A {\em transition
rule} over $\Sigma$ is an expression of the form $\frac{H}{\alpha}$
with $H$ a set of $\Sigma$-literals (the {\em premises} of the
rule) and $\alpha$ a $\Sigma$-literal (the {\em conclusion}). The
left- and right-hand side (if any) of $\alpha$ are called the {\em
source} and the {\em target} of the rule, respectively. A rule
$\frac{H}{\alpha}$ with $H=\emptyset$ is also written $\alpha$.
A {\em transition system specification (TSS)}, written $(\Sigma,R)$,
consists of a signature $\Sigma$ and a collection $R$
of transition rules over $\Sigma$.
A TSS is {\em standard} if all its rules have positive conclusions,
and {\em positive} if moreover all premises of its rules are positive.
\end{definition}
The concept of a positive TSS was introduced by {\sc Groote \&
Vaandrager} \cite{GrV92}; negative premises were added by {\sc Groote}
\cite{Gr93}.  The resulting notion constitutes the first formalisation
of {\sc Plotkin}'s {\em structural operational semantics} \cite{Pl81}
that is sufficiently general to cover most of its applications.
TSSs with negative conclusions are introduced here, because they are needed
as intermediate steps in our proofs for standard TSSs.

The following definition tells when a literal is provable from a
TSS\@. It generalises the standard definition (see e.g.\ \cite{GrV92})
by allowing the derivation of transition rules. The
derivation of a literal $\alpha$ corresponds to the derivation
of the transition rule $\frac{H}{\alpha}$ with $H=\emptyset$.
The case $H \neq \emptyset$ corresponds to the derivation of
$\alpha$ under the assumptions $H$.

\begin{definition}{proof}
Let $P=(\Sigma,R)$ be a TSS. An {\em irredundant proof}
of a transition rule $\frac{H}{\alpha}$ from $P$ is a well-founded, upwardly
branching tree of which the nodes are labelled by $\Sigma$-literals,
and some of the leaves are marked ``hypothesis'', such that:
\begin{itemise}
\item the root is labelled by $\alpha$,
\item $H$ is the set of labels of the hypotheses, and
\item if $\beta$ is the label of a node $q$ which is not a hypothesis
and $K$ is the set of labels of the nodes directly above $q$, then
$\frac{K}{\beta}$ is a substitution instance of a transition rule in $R$.
\end{itemise}
A {\em proof} of $\frac{K}{\alpha}$ from $P$ is an
irredundant proof of $\frac{H}{\alpha}$ from $P$ with $H\subseteq K$.
If an (irredundant) proof of $\frac{K}{\alpha}$ from $P$ exists, then
$\frac{K}{\alpha}$ is {\em (irredundantly) provable} from $P$,
notation $P \vdash \frac{K}{\alpha}$ (resp.\ $P \vdash_{\rm irr}
\frac{K}{\alpha}$).
We write $P \vdash \frac{K}{H}$ (resp.\ $P \vdash H$) for sets of
literals $K$ and $H$ if $P \vdash \frac{K}{\alpha}$ (resp.\ $P \vdash
\alpha$) for all $\alpha \in H$.
\end{definition}
The main purpose of a TSS $(\Sigma,R)$ is to specify
a {\em transition relation} over $\Sigma$, this being
a set of closed positive $\Sigma$-literals ({\em
transitions}).  A positive TSS specifies a transition relation in a
straightforward way as the set of all provable transitions. But as
pointed out by {\sc Groote} \cite{Gr93}, it is much less trivial to
associate a transition relation to a TSS with negative premises; in
particular there are TSSs that appear not to specify a transition
relation in a meaningful way at all. In {\sc van Glabbeek} \cite{vG95}
eleven answers to the questions ``{\em Which TSSs are meaningful, and
which transition relations do they specify?\/}'' are reviewed. The
``most general solution without undesirable properties'' is due to
{\sc van Gelder, Ross \& Schlipf} \cite{GRS91} in the setting of logic
programming, and has been adapted to TSSs by {\sc Bol \& Groote}
\cite{BolG96}. In \cite{vG95} it has been reformulated in terms of
completeness with respect to a notion of provability of closed
literals that incorporates a form of {\em negation as failure}.

\begin{definition}{wsp}
Let $P=(\Sigma,R)$ be a standard TSS. A {\em
well-supported proof} of a closed literal ${\alpha}$ from $P$ is a
well-founded, upwardly branching tree of which the nodes are labelled
by closed $\Sigma$-literals, such that:
\begin{itemize}\vspace{-1ex}
\item the root is labelled by $\alpha$, and\vspace{-1ex}
\item if $\beta$ is the label of a node $q$ and $K$ is the set of
labels of the nodes directly above $q$, then
\begin{enumerate}\vspace{-1ex}
\item either $\frac{K}{\beta}$ is a closed substitution instance of a
transition rule in $R$
\vspace{-1ex}
\item or $\beta$ is negative and for every set $N$ of negative
closed literals such that $P \vdash \frac{N}{\gamma}$ for $\gamma$ a closed
literal denying $\beta$, a literal in $K$ denies one in $N$.\vspace{-1ex}
\end{enumerate}
\end{itemize}
$\alpha$ is {\em $ws$-provable} from $P$, notation $P\vdash_{ws}
\alpha$, if a well-supported proof of $\alpha$ from $P$ exists.
\end{definition}
Note that the proof-steps 1 and 2 establish the validity of $\beta$
when $K$ is the set of literals established earlier. Step 2 allows to
infer $t\gonotto{a}$ whenever it is manifestly impossible to infer
$t\goto{a}t'$ for some term $t'$ (because every conceivable proof of
$t\goto{a}t'$ involves a premise that has already been refuted).
This practice is sometimes referred to as {\em negation as failure}
\cite{Cl78}.

\begin{definition}{complete}
A standard TSS $P$ is {\em complete} if for any closed literal
$t\gonotto{a}$ either $P\vdash_{ws} t\goto{a}t'$ for some closed term
$t'$ or $P\vdash_{ws} t\gonotto{a}$.
\end{definition}
Now a standard TSS is meaningful, in the sense that it specifies a
transition relation, iff it is complete. The specified transition
relation is then the set of all $ws$-provable transitions.

In the present paper this solution is extended by considering all
standard TSSs to be meaningful. However, following {\sc van Gelder,
Ross \& Schlipf} \cite{GRS91}, the meaning of an incomplete TSS is now
not given by a two-valued transition relation as defined above, but
by a three-valued transition relation, in which a potential transition
can be {\em true}, {\em false} or {\em unknown}. In fact, a slight
abstraction of this notion will suffice, in which a transition
relation is simply defined as a set of closed, positive and negative,
literals.

\begin{definition}{transition}
Let $\Sigma$ be a signature. A {\em (3-valued) transition relation}
over $\Sigma$ is a set of closed $\Sigma$-literals, not containing
literals that deny each other. A transition relation $\rightarrow$ is {\em
2-valued} if it satisfies $(t \gonotto{a}) \in \;\rightarrow ~ \Leftrightarrow
~\neg\exists t'\in T(\Sigma) : (t \goto{a} t') \in \;\rightarrow$.
The transition relation {\em associated} to a standard TSS $(\Sigma,R)$
is the set of closed $\Sigma$-literals that are $ws$-provable
from that TSS.
\end{definition}
In \cite{vG95} is has been shown that $\vdash_{ws}$ is consistent, in
the sense that no standard TSS admits well-supported proofs of two
literals that deny each other. Thus the transition relation associated
to a standard TSS is indeed a transition relation as defined above.
Note that if a standard TSS is complete, its associated transition
relation is 2-valued. This means that the negative literals in
its associated transition relation are completely determined by the
positive ones; hence the transition relation can be simply given by
its positive part.

In \sect{preorders} several preorders have been defined on
labelled transition systems, and provided with modal
characterisations.  These definitions and results apply immediately to
the 2-valued transition relation $\rightarrow$ associated to a complete
TSS $(\Sigma,R)$, for such a transition relation gives
rise to the LTS $(T(\Sigma),\rightarrow)$.
In the case of $N\in\{T,CT,F,R,FT,RT,1S,RS\}$, the definition of the
{\em $N$-preorder induced by a TSS} even extends to incomplete TSSs,
namely as follows.

\begin{definition}{preorders}
Let $P=(\Sigma,R)$ be a standard TSS. The {\em satisfaction
relation} $\models_P\; \subseteq T(\Sigma) \times \IO_{\it RS}$ is
inductively defined by the clauses below (in which $p,q \in
T(\Sigma)$).\vspace{-3pt}
$${\renewcommand{\arraystretch}{1.1}\arraycolsep 2pt \begin{array}{@{}llll@{}}
p \models_P \top	 	\\
p \models_P a\varphi	& \mbox{if} & \exists q: P \vdash_{ws} p \goto{a} q \wedge q \models_P \varphi\\
p \models_P \widetilde{a}	& \mbox{if} & P \vdash_{ws} p \gonotto{a}\\
p \models_P \bigwedge_{i\in I}\varphi_i& \mbox{if}&p \models_P \varphi_i
\mbox{ for all } i\in I\\
\end{array}}$$
For $N\in\{T,CT,F,R,FT,RT,1S,RS\}$, the set of {\em $N$-observations
of $p \in T(\Sigma)$} is given by $\fO_N^P(p) := \{\varphi \in \IO_N
\mid p \models_P \varphi\}$. Likewise,
$\fO_N^{\wedge P}(p) := \{\varphi \in \IO_N^\wedge \mid p \models_P \varphi\}$.
The {\em $N$-preorder induced by $P$} is
defined by $p \sqsubseteq_N^P q$ if $\fO_N^P(p) \subseteq \fO_N^P(q)$.
When clear from the context, sub- or superscripts $P$ will be omitted.
\end{definition}
Note that in case $P$ is complete, $\sqsubseteq_N^P$ is indeed the
$N$-preorder as defined in \sect{preorders} on the LTS
$(T(\Sigma),\rightarrow)$, where $\rightarrow$ is the 2-valued
transition relation associated to $P$.
In the general case, the 3-valued transition relation associated to
$P$ gives rise to a structure
$(T(\Sigma),\rightarrow,\not\rightarrow)$ with $\rightarrow\;
\subseteq \IP \times A \times \IP$ and $\not\rightarrow\; \subseteq
\IP \times A$. Such a structure could be called a {\em 3-valued LTS}.
Now $p \gonotto{a}$ stands for $(p,a) \in\; \not\rightarrow$, and
\thm{modal} can be interpreted as the {\em definition} of
$\sqsubseteq_N$ (agreeing with \df{preorders} above).
Note that for $N\in\{T,CT,F,R,FT,RT,1S,RS\}$ \cor{modal} extends to
the preorders induced by incomplete TSSs.

\df{preorders} does not capture the modality $\neg$
of \df{potential-observations}; hence the bisimulation equivalence and
$n$-nested simulation preorders (for $n \geq 2$) induced by a TSS are
defined for complete TSSs only. (Extending \df{preorders} in
a simple-minded way would invalidate the hierarchy of
\sect{preorders}.)

\begin{definition}{precongruence}
Let $\Sigma$ be a signature. A preorder $\sqsubseteq$ on $T(\Sigma)$
is a {\em precongruence} if for all $f \in \Sigma$
$$p_i \sqsubseteq q_i \mbox{ for } i=1,\ldots,ar(f) ~~\Rightarrow~~
f(p_1,\ldots,p_{ar(f)}) \sqsubseteq f(q_1,\ldots,q_{ar(f)}).$$
\end{definition}
This is equivalent to the requirement that for all $t \in \IT(\Sigma)$
and closed substitutions $\sigma, \sigma': V \rightarrow T(\Sigma)$
$$\sigma(x) \sqsubseteq \sigma'(x) \mbox{ for } x\in \var(t)
~~\Rightarrow~~ \sigma(t) \sqsubseteq \sigma'(t).$$
In case $\sqsubseteq$ is an equivalence relation as well as a
precongruence, it is called a {\em congruence}. Note that if
$\sqsubseteq_N$ is a precongruence, its kernel is a congruence.
Thus by establishing precongruence results for the preorders
$\sqsubseteq_N$, we also obtain congruence results for the associated
equivalences $=_N$.

%%%%%%%%%%%%%%%%%%%%%%%%%%%%%%%%%%%%%%%%%%%%%%%%%%%%%%%%%%%%%%%%%%%%%%%%%%%%%%
\section{Precongruence formats}
\hlabel{sec-formats}

In this section we define the formats for TSSs that play a
r\^ole in this paper and state the precongruence results that we are
going to establish.

\begin{definition}{ntyft-ntyxt}
An {\em ntytt rule} is a transition rule in which
the right-hand sides of positive premises are variables that are all
distinct, and that do not occur in the source.
An ntytt rule is an {\em ntyxt rule}
if its source is a variable, and an {\em ntyft rule} if its
source contains exactly one function symbol and no multiple
occurrences of variables.
An ntytt rule (resp.\ ntyft rule) is an {\em nxytt rule} (resp.\ {\em
nxyft rule}) if the left-hand sides of its
premises are variables. An ntytt rule (resp.\ ntyft rule) is an
{\em xyntt rule} (resp.\ {\em xynft rule})
if the left-hand sides of its positive premises are variables.
\end{definition} 

\begin{definition}{free}
A transition rule has {\em no lookahead} if the variables occurring
in the right-hand sides of its positive premises do not occur in
the left-hand sides of its premises.
A variable occurring in a transition rule is {\em free} if it
does not occur in the source nor in the right-hand sides of
the positive premises of this rule.
We say that a transition rule is {\em decent} if it has no lookahead and
does not contain free variables.
\end{definition}
Each combination of syntactic restrictions on transition rules
induces a corresponding syntactic format for TSSs of the same name.
For instance, a TSS is in {\em decent ntyft} format if it contains
decent ntyft rules only. We proceed to define further syntactic
formats for TSSs.

\begin{definition}{formats-old}
A TSS is in {\em ntyft/ntyxt format}
if it contains only ntyft and ntyxt rules.
A TSS is in {\em tyft/tyxt format}
if it is positive and in ntyft/ntyxt format.
A TSS is in {\em ready simulation format} if it is in
ntyft/ntyxt format and its transition rules have no lookahead.
\end{definition}

\begin{theorem}{bisimulation}\cite{FvG96}
If a complete standard TSS is in ntyft/ntyxt format, then the bisimulation
equivalence that it induces is a congruence.
\end{theorem}

\begin{proof} In {\sc Bol \& Groote} \cite{BolG96} it is established that
for any complete standard TSS in ntyft/ntyxt format that satisfies a
condition called {\em well-foundedness}, the bisimulation equivalence
that it induces is a congruence. This extends an earlier result of
{\sc Groote \& Vaandrager} \cite{GrV92} for well-founded TSSs
in tyft/tyxt format. In \cite{FvG96} it was shown that for every complete
standard TSS in ntyft/ntyxt format there exists a complete standard
TSS in xynft format that is well-founded, and that induces the same
transition relation. From this the theorem follows immediately.
\end{proof}

\begin{theorem}{nested-simulation}
If a TSS is in tyft/tyxt format, then the $n$-nested simulation
preorders that it induces for $n\geq 1$ are precongruences.
\end{theorem}

\begin{proof} {\sc Groote \& Vaandrager} \cite{GrV92}
stated that for any well-founded TSS in tyft/tyxt format, the
$n$-nested simulation equivalences that it induces are congruences.
They remarked that this result can be established
in exactly the same way as the fact that the bisimulation equivalence
that a well-founded TSS in tyft/tyxt format induces is a congruence.

We remark that in exactly the same way one can show that the
$n$-nested simulation {\em preorders} that a well-founded TSS in
tyft/tyxt format induces are {\em pre}congruences. In \cite{FvG96} it
was shown that for every TSS in tyft/tyxt format there exists a positive
TSS in xynft format that is well-founded, and that induces the same
transition relation. From this the theorem follows immediately.
\end{proof}

\begin{theorem}{ready-simulation}
If a standard TSS is in ready simulation format, then the ready simulation
preorder that it induces is a precongruence.
\end{theorem}

\begin{definition}{propagated}
An occurrence of a variable in an ntytt rule is {\em propagated}
if the occurrence is either in the target, or in the left-hand side of
a positive premise whose right-hand side occurs in the target.
An occurrence of a variable in an ntytt rule is {\em polled} if the
occurrence is in the left-hand side of a premise that does not have
a right-hand side occurring in the target.
\end{definition}
Consider for instance the transition rules of \ex{ex2} in
\sect{counterexamples}. In the second rule both occurrences of $x$
in the premisses are propagated, i.e.\ the variable $x$ is propagated
twice.  In the third rule the variables $x_1$ and $x_2$ are polled
once each.  We can think of a process, represented by a variable in a
transition rule, as being {\em copied} if the variable is propagated
more than once. The process is {\em tested} if the variable is either
propagated or polled.

Our precongruence formats for decorated trace preorders
operate by keeping track of which variables
represent running processes, and which do not. For example, it is
semantically reasonable to copy a process before it starts,
effectively getting information about all the conjuncts in a $\phi \in
\IO^\wedge$.
However, copying a running process would give information about the
branching structure of the process, which is incompatible with any
form of decorated trace semantics.
We introduce a predicate $\Lambda$ as the basis for determining the
$\Lambda$-{\em floating} variables, which represent processes that may
be running.

\begin{definition}{floating}
Let $\Sigma$ be a signature, and $\Lambda$ a unary predicate on $\{(f,i)
\mid 1 \leq i \leq ar(f),~f \in \Sigma \}$. If $\Lambda(f,i)$, then
we say that argument $i$ of $f$ is {\em liquid}; otherwise it is {\em frozen}.
An occurrence of a variable $x$ in a term $t \in \IT(\Sigma)$ is
{\em at a $\Lambda$-liquid position} if either $t=x$, or $t=f(t_1,\ldots,t_{ar(f)})$
and the occurrence of $x$ is $\Lambda$-liquid in $t_i$ for some liquid
argument $i$ of $f$.
A variable in an ntytt rule over $\Sigma$ is {\em $\Lambda$-floating}
if either it occurs as the right-hand side of a positive premise,
or it occurs exactly once in the source, at a $\Lambda$-liquid position.
\end{definition}
Note that an occurrence of a variable $x$ in a term $t \in \IT(\Sigma)$
is $\Lambda$-liquid iff $t$ does not contain a subterm
$f(t_1,\ldots,t_{ar(f)})$ such that the occurrence of $x$ is in $t_i$
for a frozen argument $i$ of $f$.

\begin{definition}{formats}
Let $\Lambda$ be a unary predicate on arguments of function symbols.
A standard ntytt rule is {\em $\Lambda$-ready trace safe} if
\begin{itemise}
\item it has no lookahead, and
\item each $\Lambda$-floating variable has at most one propagated occurrence,
      which must be at a $\Lambda$-liquid position.
\end{itemise}
The rule is {\em $\Lambda$-readiness safe} if
\begin{itemise}
\item it is $\Lambda$-ready trace safe, and
\item no $\Lambda$-floating variable has both propagated and polled occurrences.
\end{itemise}
The rule is {\em $\Lambda$-failure trace safe} if
\begin{itemise}
\item it is $\Lambda$-readiness safe, and
\item each $\Lambda$-floating variable has at most one polled occurrence,
      which must be at a $\Lambda$-liquid position in a positive premise.
\end{itemise}
\end{definition}
The second restriction on ``$\Lambda$-ready trace safe" guarantees that
a running process is never copied, and continued to be
marked as running after it has executed.  The ``$\Lambda$-readiness safe"
restriction ensures that only at the end of its execution a running process
is tested multiple times. The ``$\Lambda$-failure trace safe" restriction
further limits to a positive test on a single action.

\begin{definition}{ready-trace}
A standard TSS is in {\em ready trace format} if it is in ntyft/ntyxt format
and its rules are $\Lambda$-ready trace safe with respect to some $\Lambda$.
A standard TSS is in {\em readiness format} if it is in ntyft/ntyxt format
and its rules are $\Lambda$-readiness safe with respect to some $\Lambda$.
A standard TSS is in {\em failure trace format} if it is in ntyft/ntyxt format
and its rules are $\Lambda$-failure trace safe with respect to some $\Lambda$.
\end{definition}
Note that if a standard TSS $P$ is in ready trace format (resp.\ readiness
format or failure trace format), then there is a smallest predicate
$\Lambda_0$ such that the rules of $P$ are $\Lambda_0$-ready trace
safe (resp.\ $\Lambda_0$-readiness safe or $\Lambda_0$-failure trace
safe).  In the context of the ready trace format, for instance,
$\Lambda_0$ can be defined as the smallest predicate $\Lambda$ such
that for all rules of $P$ each $\Lambda$-floating variable is
propagated at $\Lambda$-liquid positions only. Now $P$ is in ready
trace format iff it has no lookahead and in all of its rules each
$\Lambda_0$-floating variable is propagated at most once.  Therefore, in
the context of a given standard TSS and a given format, positions can be called
{\em liquid} and variables {\em floating} without mentioning a
specific predicate $\Lambda$; in such a case $\Lambda_0$ may be
assumed.

\begin{theorem}{ready-trace}
If a standard TSS is in ready trace format, then the ready trace preorder
that it induces is a precongruence.
\end{theorem}

\begin{theorem}{readiness}
If a standard TSS is in readiness format, then the readiness preorder
that it induces is a precongruence.
\end{theorem}

\begin{theorem}{failure-trace}
If a standard TSS is in failure trace format, then the failure trace and
failure preorders that it induces are precongruences.
\end{theorem}
Sections \hhref{sec-preservation}--\hhref{sec-crux} are devoted to the proofs
of Theorems \hhref{thm-ready-simulation}--\hhref{thm-failure-trace}.
\sect{counterexamples} presents a series of counterexamples
showing that the syntactic restrictions formulated above are essential
for the claimed precongruence results. These counterexamples also help in
motivating the definitions above.

For comparison with the literature we point out that a standard TSS is
in GSOS format \cite{BIM95} iff it is finite and in decent nxyft
format, and each rule has finitely many premises. A standard TSS is
in de Simone's format iff it is positive, in decent nxyft format, and
its rules are $\Lambda$-failure trace safe with $\Lambda$ the
universal predicate (making all arguments of function symbols liquid).

%%%%%%%%%%%%%%%%%%%%%%%%%%%%%%%%%%%%%%%%%%%%%%%%%%%%%%%%%%%%%%%%%%%%%%%%%%%%%%
\newcommand{\lvar}{{\it lvar}}
\newcommand{\rhs}{{\it rhs}}

\section{Preservation of syntactic restrictions}
\hlabel{sec-preservation}

Later on, in the proofs of the precongruence theorems, we will
use transition rules with negative conclusions.
For this reason, the ready trace, readiness and failure trace
formats need to be extended to non-standard TSSs.

\begin{definition}{formats negative}
Let $\Lambda$ be a unary predicate on arguments of function symbols.
An ntytt rule with a negative conclusion is {\em $\Lambda$-ready trace
safe} or {\em $\Lambda$-readiness safe} if it has no lookahead.
The rule is {\em $\Lambda$-failure trace safe} if
\begin{itemise}
\item it is $\Lambda$-readiness safe, and
\item $\Lambda$-floating variables are polled only at $\Lambda$-liquid
positions and only in negative premises.
\end{itemise}
\end{definition}
Now \df{ready-trace} applies to non-standard TSSs as well.
Note that for $\Lambda$-ready trace and $\Lambda$-readiness safety the
requirements are the same as in the standard case, for in a rule with
a negative conclusion no variable is propagated. In the definition of
$\Lambda$-failure trace safety, however, rules with positive and negative
conclusions are treated differently.

In the remainder of this section we show that the syntactic restrictions
of the precongruence formats for TSSs in decent ntyft format
are inherited by the ntytt rules irredundantly provable from such TSSs.
The restriction to TSSs in decent ntyft format is justified because
Propositions \hhref{pr-ntyft} and \hhref{pr-decent} in
\sect{xynft} will imply that in the proofs of Theorems
\hhref{thm-ready-simulation}--\hhref{thm-failure-trace} we may, without
limitation of generality, restrict attention to TSSs in decent ntyft format.
The results of this section will be of use in the forthcoming proofs
of the precongruence theorems.

For $H$ a set of literals let $\lvar(H)$ denote the set of variables
occurring in left-hand sides of literals in $H$, and $\rhs(H)$ the set
of right-hand sides of positive literals in $H$.
Note that an ntytt rule \plat{\frac{H}{t \gonotto{a}}} [resp.\
$\frac{H}{t \goto{a} t'}$] is decent iff $\lvar(H) \subseteq \var(t)$
[and $\var(t') \subseteq  \var(t) \cup \rhs(H)$].

\begin{lemma}{preservation-decency}
Let $P=(\Sigma,R)$ be a TSS in decent ntytt format. Then any ntytt rule
irredundantly provable from $P$ is decent.
\end{lemma}

\begin{proof}
We prove a slightly stronger statement, namely that {\em any} transition
rule \plat{\frac{H}{t \gonotto{a}}} [resp.\ $\frac{H}{t \goto{a} t'}$]
irredundantly provable from $P$ satisfies $\lvar(H) \subseteq \var(t)$
[and $\var(t') \subseteq \var(t) \cup \rhs(H)$].  We apply structural
induction with respect to the irredundant proof $\pi$ of such a rule
from $P$.  Let $P \irr\frac{H}{t \goto{a} t'}$; the case of a rule
with a negative conclusion goes similarly.

{\em Induction basis}:
Suppose $\pi$ has only one node. Then $H=\{t\goto a t'\}$, which
implies that $t'$ is a variable.
Clearly $\frac{t\goto a t'}{t\goto a t'}$ has the required properties.

{\em Induction step}:
Let $r\in R$ be the decent ntytt rule and $\rho$ be the substitution
used at the bottom of $\pi$, where $r$ is of the form
$\frac{\{v_k\goto{c_k}y_k\mid k\in K\}\cup
\{w_\ell\gonotto{d_\ell}\mid \ell\in L\}}{u\goto a u'}$. Then
$\rho(u)=t$ and $\rho(u')=t'$.  Moreover, transition rules
$\frac{H_k}{\rho(v_k)\goto{c_k}\rho(y_k)}$ for $k\in K$ and
$\frac{H_\ell}{\rho(w_\ell)\gonotto{d_\ell}}$ for $\ell\in L$ are
irredundantly provable from $P$ by means of strict subproofs of $\pi$,
where $H=\bigcup_{k\in K}H_k\cup\bigcup_{\ell\in L}H_\ell$.  By
induction $\lvar(H_k) \subseteq \var(\rho(v_k))$ for $k \in K$.  As
$r$ is decent we have $\var(v_k) \subseteq \var(u)$, so $\lvar(H_k)
\subseteq \var(\rho(v_k)) \subseteq \var(\rho(u)) = \var(t)$ for $k
\in K$.  Similarly, $\lvar(H_\ell) \subseteq \var(t)$ for $\ell \in
L$, thus $\lvar(H) \subseteq \var(t)$.  

As $r$ is decent we have $\var(u')\subseteq\var(u)\cup \{y_k \mid k
\in K\}$, so
$$\var(t') = \var(\rho(u')) \subseteq \var(t) \cup
\bigcup_{k \in K} \var(\rho(y_k)).$$
By induction we have $\var(\rho(y_k)) \subseteq \var(\rho(v_k)) \cup
\rhs(H_k) \subseteq \var(t) \cup \rhs(H_k)$ for $k \in K$. It follows
that $\var(t') \subseteq \var(t) \cup \rhs(H)$.
\end{proof}

\begin{lemma}{preservation-ready-trace}
Let $P=(\Sigma,R)$ be a TSS in decent ntyft format of which the transition rules
are $\Lambda$-ready trace safe.  Then any ntytt rule $\frac{H}{t\goto
a u}$ irredundantly provable from $P$ is $\Lambda$-ready trace safe.
\end{lemma}

\begin{proof}
We apply structural induction with respect to the irredundant proof
$\pi$ of $\frac{H}{t\goto a u}$ from $P$.

{\em Induction basis}:
Suppose $\pi$ has only one node. Then it must be the case that
\plat{H=\{t\goto a u\}}, which implies that $u$ is a variable.
The $\Lambda$-floating variables in $\frac{t\goto a
u}{t\goto a u}$ are $u$ and the variables that occur in
$\Lambda$-liquid positions in $t$. Both kinds are
propagated only once, in the target and in the left-hand side of the
positive premise, respectively.
These propagations are at $\Lambda$-liquid positions.

{\em Induction step}:
Let $r\in R$ be the decent ntyft rule and $\rho$ be the substitution
used at the bottom of $\pi$, where $r$ is of the form
$\frac{\{v_k\goto{c_k}y_k\mid k\in K\}\cup
\{w_\ell\gonotto{d_\ell}\mid \ell\in L\}}
{f(x_1,\ldots,x_{{\it ar}(f)})\goto a v}$. Then
$f(\rho(x_1),\ldots,\rho(x_{ar(f)}))=t$ and $\rho(v)=u$.  Moreover,
transition rules $r_k = \frac{H_k}{\rho(v_k)\goto{c_k}\rho(y_k)}$ for
$k\in K$ and $r'_\ell = \frac{H_\ell}{\rho(w_\ell)\gonotto{d_\ell}}$
for $\ell\in L$ are irredundantly provable from $P$ by means of strict
subproofs of $\pi$, where $H=\bigcup_{k\in K}H_k\cup\bigcup_{\ell\in
L}H_\ell$.  As $r$ is decent we have $\var(v_k) \subseteq
\{x_1,\ldots,x_{ar(f)}\}$, so $\var(\rho(v_k)) \subseteq \bigcup_{i=1}^{ar(f)}
\var(\rho(x_i)) = \var(t)$ for $k \!\in\! K$.  Since $\frac{H}{t \goto a
u}$ is an ntytt rule, $\rhs(H) \cap \var(t) = \emptyset$.  Hence
$\rhs(H_k) \cap \var(\rho(v_k)) = \emptyset$ for $k \!\in\! K$.  It follows that
the rules $r_k$ for $k \in K$ are ntytt rules. The same holds for the
rules $r'_\ell$ for $\ell \in L$.  By \lem{preservation-decency},
the rules $r_k$ and $r'_\ell$ are decent.

Let $K_0$ denote $\{k\in K\mid y_k\in\var(v)\}$.
We make the following observation.
\begin{itemize}
\item[(A)]
If the right-hand side $y$ of a positive premise in $H$ occurs in $u$,
then there is a $k\in K_0$ such that the premise is in $H_k$ and
$y\in\var(\rho(y_k))$.
\end{itemize}
Namely, $y$ does not occur in $\rho(x_i)$ for $i=1,\ldots,{\it
ar}(f)$. Since $y\in\var(\rho(v))$ and, by the decency of $r$,\linebreak[3]
$\var(v)\subseteq\{x_1,\ldots,x_{{\it ar}(f)}\}\cup\{y_k\mid k\in K_0\}$,
$y$ must occur in $\rho(y_k)$ for some $k\in K_0$.
Furthermore, $\var(v_k)\subseteq\{x_1,\ldots,x_{{\it ar}(f)}\}$ implies
$y\not\in\var(\rho(v_k))$. So by the
decency of \plat{\frac{H_k}{\rho(v_k)\goto{c_k}\rho(y_k)}},
the positive premise in $H$ with right-hand side $y$ is in $H_k$.

Let the variable $z$ be $\Lambda$-floating in $\frac{H}{t\goto a u}$.
We need to prove that $z$ is propagated at most once in this rule, and
at a $\Lambda$-liquid position. Since $z$ is $\Lambda$-floating in
\plat{\frac{H}{t\goto a u}}, we can distinguish two cases.
\begin{itemize}
\item[1.]
Let $z$ occur at a $\Lambda$-liquid position in $\rho(x_i)$, for a liquid
argument $i$ of $f$.

By assumption $z$ occurs only once in $t$,
so $z$ occurs only once in $\rho(x_i)$, and $z\not\in\var(\rho(x_j))$
for $j\not=i$. We make a second observation.
\begin{itemize}
\item[(B)]
If $x_i\not\in\var(v_k)$ for some $k\in K$,
then $z\not\in\var(H_k)$ and $z\not\in\var(\rho(y_k))$.
\end{itemize}
Namely, as $\frac{H}{t\goto{a}u}$ is a ntytt rule, $\var(t) \cap
\rhs(H_k) \subseteq \var(t) \cap \rhs(H) = \emptyset$, hence $z
\not\in\rhs(H_k)$. Furthermore,
$\var(v_k)\subseteq\{x_1,\ldots,x_{{\it ar}(f)}\}$,
$x_i\not\in\var(v_k)$ and $z$ does not occur in $\rho(x_j)$
for $j\not=i$, so $z\not\in\var(\rho(v_k))$.
Hence, by the decency of $\frac{H_k}{\rho(v_k)\goto{c_k}\rho(y_k)}$,
$z$ occurs neither in $H_k$ nor in $\rho(y_k)$.
\\
Once more we distinguish two cases.

\item[1.1]
Let $x_i\not\in\var(v_k)$ for all $k\in K_0$.

By (B), $z\not\in\var(\rho(y_k))$ for $k\in K_0$.
Furthermore, $z$ does not occur in $\rho(x_j)$ for $j\not=i$, and $z$
occurs only once in $\rho(x_i)$, at a $\Lambda$-liquid position. Finally,
since $r$ is $\Lambda$-ready trace safe, and $i$ is a liquid argument
of $f$, $x_i$ occurs at most once in $v$, and at a $\Lambda$-liquid
position. As
$\var(v)\subseteq\{x_1,\ldots,x_{{\it ar}(f)}\}\cup\{y_k\mid k\in K_0\}$,
it follows that $z$ occurs
at most once in $\rho(v)=u$, and at a $\Lambda$-liquid position.

By (B), $z\not\in\var(H_k)$ for $k\in K_0$.
Moreover, by (A), right-hand sides of positive premises in
$H_k$ for $k\in K\backslash K_0$ and $H_\ell$ for $\ell\in L$
do not occur in $u$. Hence, there are no
propagated occurrences of $z$ in left-hand sides of positive premises in
$\frac{H}{t\goto a u}$.

\item[1.2]
Let $x_i\in\var(v_{k'})$ for some $k'\in K_0$.

$r$ is $\Lambda$-ready trace safe, and $i$ is a liquid argument
of $f$, so $x_i\not\in\var(v)$ and $x_i\not\in\var(v_k)$ for
$k\in K_0\backslash\{k'\}$. Moreover, $x_i$ occurs only once in $v_{k'}$,
at a $\Lambda$-liquid position. Since $z$ occurs only once in $\rho(x_i)$,
at a $\Lambda$-liquid position, and $z$ does not occur in $\rho(x_j)$
for $j\not=i$,
it follows that $z$ occurs only once in $\rho(v_{k'})$, at a $\Lambda$-liquid
position. Thus, by induction $z$ is propagated at most once
in $\frac{H_{k'}}{\rho(v_{k'})\goto{c_{k'}}\rho(y_{k'})}$, and at a
$\Lambda$-liquid position. In particular, $z$ occurs at most once
in $\rho(y_{k'})$, and at a $\Lambda$-liquid position.
Moreover, as $r$ is $\Lambda$-ready trace safe, $y_{k'}$ occurs at most
once in $v$, and at a $\Lambda$-liquid position.
By (B), $x_i\not\in\var(v_k)$ implies that
$z\not\in\var(\rho(y_k))$ for $k\in K_0\backslash\{k'\}$. Furthermore,
$x_i\not\in\var(v)$ and $z\not\in\var(\rho(x_j))$ for $j\not=i$.
As $\var(v)\subseteq\{x_1,\ldots,x_{{\it ar}(f)}\}\cup\{y_k\mid k\in K_0\}$,
it follows that $z$ occurs at most once in $\rho(v)=u$, and
at a $\Lambda$-liquid position.

If there are no propagated occurrences of $z$ in left-hand sides
of positive premises of $\frac{H}{t\goto a  u}$, then we are done.
So suppose there is a positive premise \plat{s\goto c y} in
$H$ where $z$ occurs in $s$ and $y\in\var(u)$.  We proceed to prove that
the occurrence of $z$ in $s$ is $\Lambda$-liquid and that this is the
only propagated occurrence of $z$ in $\frac{H}{t\goto a u}$.

By (B), $x_i\not\in\var(v_k)$
implies $z\not\in\var(H_k)$ for $k\in K_0\backslash\{k'\}$.
On the other hand, since \plat{y\in\var(u)}, (A) yields that
$s\goto c y\in H_k$ and $y\in\var(\rho(y_k))$ for some $k\in K_0$.
Hence, $\plat{s\goto c y}\in H_{k'}$ and $y\in\!\var(\rho(y_{k'}))$.
By induction, $z$ is propagated at most once
in $\frac{H_{k'}}{\rho(v_{k'})\goto{c_{k'}}\rho(y_{k'})}$,
and at a $\Lambda$-liquid position.
So the occurrence of $z$ in $s$ is $\Lambda$-liquid.
Moreover, it is the only propagated occurrence of $z$ in the left-hand sides
of the positive premises in $H_{k'}$, and $z\not\in\var(\rho(y_{k'}))$.

$x_i\!\not\in\!\var(v)$, $z$ occurs neither in $\rho(x_j)$ for $j\!\neq\! i$
nor in $\rho(y_k)$ for $k\in K_0\backslash\{k'\}$, and $z\!\not\in\!\var(\rho(y_{k'}))$.
As $\var(v)\subseteq\{x_1,\ldots,x_{{\it ar}(f)}\}\cup\{y_k\mid k\in K_0\}$,
it follows that $z$ does not occur in $\rho(v)=u$.

According to (A), right-hand sides of positive premises in $H_k$ for
$k\in K\backslash K_0$ and in $H_\ell$ for $\ell\in L$ do not occur in $u$.
Furthermore, by (B), $x_i\not\in\var(v_k)$
implies $z\not\in\var(H_k)$ for $k\in K_0\backslash\{k'\}$.
Finally, for positive premises $s'\goto d y'$ in $H_{k'}$ with
$z\in\var(s')$ and $y'\not=y$
we have $y'\not\in\var(\rho(y_{k'}))$, and so by (A) $y'\not\in\var(u)$.
Hence, the occurrence of $z$ in the left-hand side of
$s\goto c y$ is the only propagated occurrence of $z$ in
$\frac{H}{t\goto a u}$.

\item[2.]
Let $z$ be the right-hand side of a positive premise in $H$.

Since $\frac{H}{t\goto a u}$ is decent,
$z$ does not occur in left-hand sides of premises in $H$.
It remains to prove that $z$ occurs at most once in $u$, and
at a $\Lambda$-liquid position.

Since $\frac{H}{t\goto a u}$ is an ntytt rule, $z$ does not occur in
$\rho(x_i)$ for $i=1,\ldots,{\it ar}(f)$. As $r$ is decent, $\var(v_k)
\subseteq \{x_1,\ldots,x_{ar(f)}\}$ for $k \in K$. Hence $z$ does not
occur in $\rho(v_k)$ for $k \in K$.

According to (A), right-hand sides of positive premises in $H_k$ for
$k\in K\backslash K_0$ and in $H_\ell$ for $\ell\in L$ do not occur in $u$.
So we may assume that $z$ is the right-hand side of a positive premise
in $H_{k'}$ for some $k'\in K_0$. Then clearly $z$ does not occur in $H_k$ for
$k\in K_0\backslash\{k'\}$. So by the decency of
$\frac{H_k}{\rho(v_k)\goto{c_k} \rho(y_k)}$,
$z\not\in\var(\rho(y_k))$ for $k\in K_0\backslash\{k'\}$.
By induction, $z$ occurs at most once in
$\rho(y_{k'})$, and at a $\Lambda$-liquid position.
Since $r$ is $\Lambda$-ready trace safe, $y_{k'}$
occurs at most once in $v$, and at a $\Lambda$-liquid position. As
$\var(v)\subseteq\{x_1,\ldots,x_{{\it ar}(f)}\}\cup\{y_k\mid k\in K_0\}$,
it follows that $z$ occurs at most once in $\rho(v)=u$, and
at a $\Lambda$-liquid position.
\end{itemize}
We conclude that $z$ is propagated at most once
in $\frac{H}{t\goto a u}$, and at a $\Lambda$-liquid position.
\end{proof}

\begin{lemma}{preservation-readiness}
Let $P=(\Sigma,R)$ be a TSS in decent ntyft format of which the transition rules
are $\Lambda$-readiness safe.  Then any ntytt rule $\frac{H}{t\goto a
u}$ irredundantly provable from $P$ is $\Lambda$-readiness safe.
\end{lemma}

\begin{proof}
Let the variable $z$ be $\Lambda$-floating in $\frac{H}{t\goto a u}$.
We need to prove that $z$ is not both propagated and polled in this rule.
In case $z$ is the right-hand side of a positive premise, it is not
polled, since the rule is decent, by \lem{preservation-decency}.
So assume $z$ occurs exactly once in $t$, at a $\Lambda$-liquid position.
We apply structural induction with respect to the irredundant proof
$\pi$ of $\frac{H}{t\goto a u}$ from $P$.

{\em Induction basis}:
Suppose $\pi$ has only one node. Then it must be the case that
\plat{H=\{t\goto a u\}}, which implies that $u$ is a variable.
There are no polled variables in $\frac{t\goto a u}{t\goto a u}$.

{\em Induction step}:
Let $r\in R$ be the decent ntyft rule and $\rho$ be the substitution
used at the bottom of $\pi$, where $r$ is of the form
$\frac{\{v_k\goto{c_k}y_k\mid k\in K\}\cup
\{w_\ell\gonotto{d_\ell}\mid \ell\in L\}}
{f(x_1,\ldots,x_{{\it ar}(f)})\goto a v}$. Then
$f(\rho(x_1),\ldots,\rho(x_{ar(f)}))=t$ and $\rho(v)=u$.
Moreover, transition rules $\frac{H_k}{\rho(v_k)\goto{c_k}\rho(y_k)}$
for $k\in K$ and $\frac{H_\ell}{\rho(w_\ell)\gonotto{d_\ell}}$ for
$\ell\in L$ are irredundantly provable from $P$ by means of strict subproofs
of $\pi$, where $H=\bigcup_{k\in K}H_k\cup\bigcup_{\ell\in L}H_\ell$.
As in the proof of \lem{preservation-ready-trace}
it can be shown that these rules are ntytt,
and by \lem{preservation-decency} they are decent.

By assumption, $z$ occurs exactly once in $\rho(x_i)$ for some liquid
argument $i$ of $f$, at a $\Lambda$-liquid position, and
$z\not\in\var(\rho(x_j))$ for $j\not=i$. 
Let $K_0$ denote $\{k\in K\mid y_k\in\var(v)\}$.
We recall two observations from the proof of \lem{preservation-ready-trace}.
\begin{itemize}
\item[(A)]
If the right-hand side $y$ of a positive premise in $H$ occurs in $u$, then
there is a $k\in K_0$ such that the premise is in $H_k$ and $y\in\var(\rho(y_k))$.

\item[(B)]
If $x_i\not\in\var(v_k)$ for some $k\in K$,
then $z\not\in\var(H_k)$ and $z\not\in\var(\rho(y_k))$.
Likewise, if $x_i\not\in\var(w_\ell)$ for some $\ell\in L$,
then $z\not\in\var(H_\ell)$.
\end{itemize}
Suppose $z$ is polled in $\frac{H}{t\goto a u}$, so that we
can distinguish the following two cases. We need to prove that
$z$ is not propagated in \plat{\frac{H}{t\goto a u}}.
\begin{enumerate}
\item
Let $z$ be polled in the left-hand side of a premise in $H_{k'}$ for some
$k'\in K\backslash K_0$ or in $H_{\ell'}$ for some $\ell'\in L$.

By (B), $z\in\var(H_{k'})$ or $z\in\var(H_{\ell'})$
implies $x_i\in\var(v_{k'})$ or $x_i\in\var(w_{\ell'})$,
respectively, so $x_i$ is polled in $r$. Since $r$
is $\Lambda$-readiness safe, and $i$ is a liquid argument of $f$,
it follows that $x_i$ is not propagated in $r$. Hence, $x_i$ occurs
neither in $v$ nor in $v_k$ for $k\in K_0$. By (B), $x_i\not\in\var(v_k)$
implies $z\not\in\var(H_k)$ for $k\in K_0$. So, in view of (A),
occurrences of $z$ in left-hand sides of positive premises in $H$
are not propagated in \plat{\frac{H}{t\goto a u}}.
It remains to prove that $z\not\in\var(u)$.

By (B), $x_i\not\in\var(v_k)$ implies that
$z\not\in\var(\rho(y_k))$ for $k\in K_0$. Moreover, $x_i\not\in\var(v)$
and $z$ does not occur in $\rho(x_j)$ for $j\not=i$.
As $\var(v)\subseteq
\{x_1,\ldots,x_{{\it ar}(f)}\}\cup\{y_k\mid k\in K_0\}$, it follows that
$z$ does not occur in $\rho(v)=u$.

\item
Let $z$ be polled in the left-hand side of a premise in $H_{k'}$
for some $k'\in K_0$.

Then $z$ occurs in the left-hand side of a premise in $H_{k'}$
of the form $s\gonotto c$ or $s\goto c y$ with $y\not\in\var(u)$.
In the latter case, $y_{k'}\in\var(v)$ implies that
$y\not\in\var(\rho(y_{k'}))$. So in both cases, $z$ is polled
in $\frac{H_{k'}}{\rho(v_{k'})\goto{c_{k'}} \rho(y_{k'})}$.

By (B), $z\in\var(H_{k'})$ implies $x_i\in\var(v_{k'})$.
Since $r$ is $\Lambda$-readiness
safe, and $i$ is a liquid argument of $f$, $x_i$ occurs neither in
$v$ nor in $v_k$ for $k\in K_0\backslash\{k'\}$. Moreover, $x_i$ occurs only once
in $v_{k'}$, at a $\Lambda$-liquid position. Since $z$ occurs exactly once
in $\rho(x_i)$, at a $\Lambda$-liquid position, and not at all in
$\rho(x_j)$ for $j\not=i$, it follows that $z$ occurs exactly once in
$\rho(v_{k'})$, at a $\Lambda$-liquid position. Since $z$ is polled
in $\frac{H_{k'}}{\rho(v_{k'})\goto{c_{k'}} \rho(y_{k'})}$, by induction
it is not propagated in this rule. So if $(s\goto{c}y)\in H_{k'}$ and
$z \in \var(s)$ then $y$ does not occur in $\rho(y_{k'})$,
and hence, by (A), not in $u$. Thus,
occurrences of $z$ in left-hand sides of positive premises in $H_{k'}$
are not propagated in $\frac{H}{t\goto a u}$.
Furthermore, by (B), $x_i\not\in\var(v_k)$ implies
$x_i\not\in\var(H_k)$ for $k\in K_0\backslash\{k'\}$.
So, in view of (A), occurrences of $z$ in left-hand sides of positive premises
in $H$ are not propagated in $\frac{H}{t\goto a u}$.
It remains to prove that $z\not\in\var(u)$.

By (B), $x_i\not\in\var(v_k)$ implies that $z\not\in\var(\rho(y_k))$
for $k\in K_0\backslash\{k'\}$. Moreover, since $z$ is
not propagated in $\frac{H_{k'}}{\rho(v_{k'})\goto{c_{k'}} \rho(y_{k'})}$,
$z\not\in\var(\rho(y_{k'}))$. Finally, $x_i\not\in\var(v)$
and $z$ does not occur in $\rho(x_j)$ for $j\not=i$.
As $\var(v)\subseteq
\{x_1,\ldots,x_{{\it ar}(f)}\}\cup\{y_k\mid k\in K_0\}$, it follows that
$z$ does not occur in $\rho(v)=u$.
\end{enumerate}
We conclude that $z$ is not propagated in
$\frac{H}{t\goto a u}$.
\end{proof}

\begin{lemma}{preservation-failure-trace}
Let $P=(\Sigma,R)$ be a TSS in decent ntyft format of which the transition rules
are $\Lambda$-failure trace safe.  Then any ntytt rule
$\frac{H}{t\goto a u}$ irredundantly provable from $P$ is
$\Lambda$-failure trace safe.
\end{lemma}

\begin{proof}
Let the variable $z$ be $\Lambda$-floating in $\frac{H}{t\goto a u}$.
We need to prove that $z$ is polled at most once in this rule, at a
$\Lambda$-liquid position in a positive premise.
In case $z$ is the right-hand side of a positive premise, it is not
polled, since the rule is decent, by \lem{preservation-decency}.
So assume $z$ occurs exactly once in $t$, at a $\Lambda$-liquid position.
We apply structural induction with respect to the irredundant proof
$\pi$ of $\frac{H}{t\goto a u}$ from $P$.

{\em Induction basis}:
Suppose $\pi$ has only one node. Then it must be the case that
\plat{H=\{t\goto a u\}}, where $u$ is a variable.
There are no polled variables in $\frac{t\goto a u}{t\goto a u}$.

{\em Induction step}:
Let $r\in R$ be the decent ntyft rule and $\rho$ be the substitution
used at the bottom of $\pi$, where $r$ is of the form
$\frac{\{v_k\goto{c_k}y_k\mid k\in K\}\cup
\{w_\ell\gonotto{d_\ell}\mid \ell\in L\}}
{f(x_1,\ldots,x_{{\it ar}(f)})\goto a v}$. Then
$f(\rho(x_1),\ldots,\rho(x_{ar(f)}))=t$ and $\rho(v)=u$.
Moreover, transition rules $\frac{H_k}{\rho(v_k)\goto{c_k}\rho(y_k)}$
for $k\in K$ and $\frac{H_\ell}{\rho(w_\ell)\gonotto{d_\ell}}$ for
$\ell\in L$ are irredundantly provable from $P$ by means of strict
subproofs of $\pi$,
where $H=\bigcup_{k\in K}H_k\cup\bigcup_{\ell\in L}H_\ell$.
As in the proof of \lem{preservation-ready-trace}
it can be shown that these rules are ntytt,
and by \lem{preservation-decency} they are decent.

By assumption, $z$ occurs exactly once in $\rho(x_i)$ for some liquid
argument $i$ of $f$, at a $\Lambda$-liquid position, and $z\not\in\var(\rho(x_j))$
for $j\not=i$. We need to prove that $z$ is polled at most once in
$\frac{H}{t\goto a u}$, at a $\Lambda$-liquid position in a positive premise.
We recall an observation from the proofs of
Lemmas \hhref{lem-preservation-ready-trace} and \hhref{lem-preservation-readiness}.
\begin{itemize}
\item[(B)]
If $x_i\not\in\var(v_k)$ for some $k\in K$,
then $z\not\in\var(H_k)$.
Likewise, if $x_i\not\in\var(w_\ell)$ for some $\ell\in L$,
then $z\not\in\var(H_\ell)$.
\end{itemize}
$r$ is $\Lambda$-failure trace safe, and $i$ is a liquid argument
of $f$. So $x_i$ is polled and propagated
at most once in $r$ in total, at a $\Lambda$-liquid
position and not in a negative premise. In particular,
$x_i\not\in\var(w_\ell)$ for $\ell\in L$, so in view of (B),
$z\not\in\var(H_\ell)$ for $\ell\in L$.
Suppose $z\in\var(H_{k'})$ for some $k'\in K$. By (B), $x_i\in\var(v_{k'})$. 
Then $x_i\not\in\var(v_k)$ for $k\in K\backslash\{k'\}$,
so in view of (B), $z\not\in\var(H_k)$ for $k\in K\backslash\{k'\}$.
Furthermore, $x_i$ occurs only once in $v_{k'}$, at a
$\Lambda$-liquid position. Since $z$ does not
occur in $\rho(x_j)$ for $j\not=i$ and exactly once in
$\rho(x_i)$, at a $\Lambda$-liquid position, it follows that
$z$ occurs exactly once in $\rho(v_{k'})$, at a $\Lambda$-liquid position.
By induction together with Lemmas \hhref{lem-preservation-ready-trace} and
\hhref{lem-preservation-readiness}, $z$ occurs at most once in
$H_{k'}$, at a $\Lambda$-liquid position in a positive premise.
We conclude that $z$ occurs at most once in
$H$, at a $\Lambda$-liquid position in a positive premise.
\end{proof}

\begin{lemma}{preservation-negative-failure-trace}
Let $P=(\Sigma,R)$ be a TSS in decent ntyft format of which the
transition rules are $\Lambda$-failure trace safe.  Then any ntytt rule
$\frac{H}{t\gonotto a}$ irredundantly provable from $P$ is
$\Lambda$-failure trace safe.
\end{lemma}

\begin{proof}
With structural induction w.r.t.\ the irredundant proof
$\pi$ of \plat{\frac{H}{t\gonotto a}} from $P$ we establish:
\begin{itemise}\item[]\it
Let $\frac{H}{t\gonotto a}$ be an ntytt rule that is irredundantly
provable from $P$. If all occurrences of the variable $z$ in $t$
are at $\Lambda$-liquid positions, then $z$ is polled only at
$\Lambda$-liquid positions and only in negative premises of this rule.
\end{itemise}
Together with \lem{preservation-decency} this immediately yields the
desired result.

{\em Induction basis}:
Suppose $\pi$ has only one node. Then it must be the case that
\plat{H=\{t\gonotto a\}}.
There are no positive premises in $\frac{t\gonotto a}{t\gonotto a}$.
Moreover, $z$ is polled only at $\Lambda$-liquid positions.

{\em Induction step}:
Let $r\in R$ be the decent ntyft rule and $\rho$ be the substitution
used at the bottom of $\pi$, where $r$ is of the form
$\frac{\{v_k\goto{c_k}y_k\mid k\in K\}\cup
\{w_\ell\gonotto{d_\ell}\mid \ell\in L\}}
{f(x_1,\ldots,x_{{\it ar}(f)})\gonotto a}$. Then
$f(\rho(x_1),\ldots,\rho(x_{ar(f)}))=t$.
Moreover, transition rules $\frac{H_k}{\rho(v_k)\goto{c_k}\rho(y_k)}$
for $k\in K$ and $\frac{H_\ell}{\rho(w_\ell)\gonotto{d_\ell}}$ for
$\ell\in L$ are irredundantly provable from $P$ by means of strict
subproofs of $\pi$,
where $H=\bigcup_{k\in K}H_k\cup\bigcup_{\ell\in L}H_\ell$.
As in the proof of \lem{preservation-ready-trace}
it can be shown that these rules are ntytt,
and by \lem{preservation-decency} they are decent.

$r$ is $\Lambda$-failure trace safe, so variables $x_i$ for
liquid arguments $i$ of $f$ do not occur in the $v_k$ for $k\in K$,
and only at $\Lambda$-liquid positions in the $w_\ell$ for $\ell\in L$.
By assumption $z$ only occurs in $\rho(x_i)$ for liquid
arguments $i$ of $f$, at $\Lambda$-liquid positions, so
$z$ does not occur in the $\rho(v_k)$ for $k\in K$, and only
at $\Lambda$-liquid positions in the $\rho(w_\ell)$ for $\ell\in L$.
The decency of $\frac{H_k}{\rho(v_k)\goto{c_k}\rho(y_k)}$ implies that
$z$ does not occur in the left-hand sides of the premises of $H_k$
for $k\in K$. Moreover, by induction $z$ does not occur in the
positive premises and only at $\Lambda$-liquid positions in the
negative premises in $H_\ell$ for $\ell\in L$.
Hence, $z$ is polled only at $\Lambda$-liquid positions and
only in negative premises in $\frac{H}{t\gonotto a}$.
\end{proof}

%%%%%%%%%%%%%%%%%%%%%%%%%%%%%%%%%%%%%%%%%%%%%%%%%%%%%%%%%%%%%%%%%%%%%%%%%%%%%%
\section{Reducing well-supported proofs to standard proofs}
\hlabel{sec-reducing}

Theorems \hhref{thm-ready-simulation}--\hhref{thm-failure-trace} deal with
preorders induced by standard TSSs through the notion of well-founded
provability. In this section it is shown that without loss of generality
we may use the classical notion of provability (of \df{proof}) instead.
To this end we show that for any given standard TSS $P=(\Sigma,R)$ in ready simulation
format (i.e.\ in ntyft/ntyxt format without lookahead),
there exists a TSS $P^+=(\Sigma,R^+)$ in decent ntyft
format such that $P \vdash_{ws} \alpha \Leftrightarrow P^+ \vdash \alpha$.
Moreover, the relevant formats are preserved under the translation of
$R$ into $R^+$. However, in general $P^+$ will not be a standard TSS.
It is for this reason that rules with a negative conclusion have been
introduced in \df{TSS}, and that the precongruence formats were extended
to non-standard TSSs in \df{formats negative}.

The conversion from $R$ to $R^+$ will be performed in three steps.
In \sect{xynft} we show that for any standard TSS $P=(\Sigma,R)$ in
ready simulation format there exists a standard TSS $P'=(\Sigma,R')$
in decent xynft format with the same class of $ws$-provable literals.
Moreover, if $P$ is in ready trace, readiness, resp.\ failure trace
format, then so is $P'$.
In \sect{supported} we show that for standard TSSs in
decent xynft format the notion of well-supported provability coincides
with a simpler notion of {\em supported provability}.
Finally, in \sect{standard} we show that for any standard TSS $P'=(\Sigma,R')$
in decent xynft format there exists a TSS $P^+=(\Sigma,R^+)$ in decent
ntyft format such that supported provability from $P'$ coincides with
classical provability from $P^+$. Also this translation preserves the
ready trace, readiness and failure trace formats.
Together, this yields the desired result.
 
\subsection{Reducing ntyft/ntyxt rules without lookahead to decent xynft rules}
\hlabel{sec-xynft}

We show that for every standard TSS $P$ in ready simulation format
there exists a standard TSS $P'$ in decent xynft format, such that
\begin{list}{$\bullet$}{\leftmargin 25pt
                        \labelwidth\leftmargini\advance\labelwidth-\labelsep
                        \topsep 4pt \itemsep 2pt \parsep 2pt}
\item [(i)] $P \vdash_{ws} \alpha \Leftrightarrow P' \vdash_{ws} \alpha$ for
any closed literal $\alpha$,
\item [(ii)] if $P$ is in ready trace format, then so is $P'$,
\item [(iii)] if $P$ is in readiness format, then so is $P'$,
\item [(iv)] and if $P$ is in failure trace format, then so is $P'$.
\end{list}
The following proposition helps in establishing the first requirement
above.

\begin{proposition}{transition equivalence}\cite{vG95}
Let $P=(\Sigma,R)$ and $P'=(\Sigma,R')$ be standard TSSs with the
property that $P \vdash \frac{N}{t\goto{a}t'} \Leftrightarrow P'
\vdash \frac{N}{t\goto{a}t'}$ for any closed transition rule
$\frac{N}{t\goto{a}t'}$ with only negative premises. Then $P
\vdash_{ws} \alpha \Leftrightarrow P' \vdash_{ws}
\alpha$ for any closed literal $\alpha$. 
\hfill $\Box$
\end{proposition}

\begin{proof}
By symmetry it suffices to establish ``$\Rightarrow$''. This goes with
structural induction on proofs. Let $\pi$ be a well-supported proof of
$\alpha$ from $P$. In case $\alpha$ is negative, it must be the case
that for every set $N$ of negative closed literals such that $P \vdash
\frac{N}{\gamma}$ for $\gamma$ a closed literal denying $\alpha$, a
strict subproof of $\pi$ proves a literal $\delta$ that denies one in $N$.
Hence for every set $N$ of negative closed literals such that $P' \vdash
\frac{N}{\gamma}$ for $\gamma$ a closed literal denying $\alpha$, a
strict subproof of $\pi$ proves a literal $\delta$ that denies one in
$N$. By induction $P'\vdash_{ws} \delta$ for all those literals $\delta$.
It follows that $P' \vdash_{ws} \alpha$.

In case $\alpha$ is positive, just take the bottom portion of $\pi$
obtained by deleting all nodes above nodes that are labelled with a
negative literal. That portion is a proof of a closed transition rule
$\frac{N}{\alpha}$ from $P$ in which $N$ is a set of closed negative literals.
By assumption, $P' \vdash \frac{N}{\alpha}$. Moreover, $P \vdash_{ws}
\gamma$ for all $\gamma \in N$, by strict subproofs of $\pi$.
Thus by induction $P' \vdash_{ws} \gamma$ for all $\gamma \in N$. By
pasting the proofs of $\gamma$ from $P'$ for $\gamma \in N$ on top of
the proof of $\frac{N}{\alpha}$ from $P'$, a well-supported proof of
$\alpha$ from $P'$ is obtained.
\end{proof}
As a first step in the reduction process we show
that we can refrain from ntyxt rules.

\begin{proposition}{ntyft}
For each standard TSS $P=(\Sigma,R)$ in ready simulation format there exists
a standard TSS $P'=(\Sigma,R')$ in ntyft format without lookahead such that the
requirements (i)-(iv) above are met.
\end{proposition}

\begin{proof}
Replace each ntyxt rule $r$ in $R$ by a collection of ntyft rules
$\{r_f \mid f\in\Sigma\}$, where each $r_f$ is obtained by
substituting $f(x_1,\ldots,x_n)$ for the variable $x$ that constitutes
the source of $r$, with $x_1,\ldots,x_n$
variables that do not yet occur in $r$. Let $R'$ denote the collection
of ntyft rules that is thus obtained.  Note that if a closed
transition rule is provable from a TSS, it has a closed proof.
Moreover, each closed proof from $P$ of a closed transition rule is a
proof from $P'$ of the same transition rule, and vice versa. Hence, by
\pr{transition equivalence}, $P \vdash_{ws} \alpha \Leftrightarrow P'
\vdash_{ws} \alpha$ for any closed literal $\alpha$. As the rules in $R$ have no
lookahead, neither have the rules in $R'$. In order to check requirements
(ii)--(iv) note that the variable $x_i$ is propagated (resp.\ polled)
in $r_f$ exactly when and where $x$ is propagated (resp.\ polled) in
$r$. Moreover, if $x_i$ is $\Lambda$-floating in $r_f$, an occurrence
of $x_i$ in $r_f$ is $\Lambda$-liquid iff the corresponding occurrence
of $x$ in $r$ is.
\end{proof}
Next we show that we can restrict attention to {\em decent} ntyft rules,
i.e.\ we can assume that none of the rules has free variables.

\begin{proposition}{decent}
For each standard TSS $P=(\Sigma,R)$ in ntyft format without lookahead there
exists a standard TSS $P'=(\Sigma,R')$ in decent ntyft format such that the
requirements (i)-(iv) above are met.
\end{proposition}

\begin{proof}
Replace every rule with free variables by a set of new rules. The new
rules are obtained by applying every possible substitution of closed
terms for the free variables in the old rule. Now every closed proof
from $P$ of a closed transition rule is a proof from $P'$ of the same
transition rule, and vice versa. Hence, by \pr{transition equivalence},
$P \vdash_{ws} \alpha \Leftrightarrow P'\vdash_{ws} \alpha$ for any
closed literal $\alpha$. By construction, the rules in $R'$ have no
free variables. As the rules in $R$ have no lookahead, neither have the rules
in $R'$; hence $P'$ is in decent ntyft format.
The requirements (ii)--(iv) hold trivially.
\end{proof}
Finally we have to show that decent ntyft rules can be reduced to
decent xynft rules.

\begin{lemma}{cardinality}
If $A\neq\emptyset$, then the set of all literals over a signature
$\Sigma$ is equally large as the set $V$ of variables.
\end{lemma}

\begin{proof}
Recall from \sect{sos} that $|\Sigma|\leq |V|$ and
$|A|\leq |V|$. Let the {\em size} of a term be the largest number of
nested function symbols in it. In case $\Sigma=\emptyset$ we have
$\IT(\Sigma) = V$, so $|\IT(\Sigma)| = |V|$. Now suppose $\Sigma \neq
\emptyset$. With induction to $n \in \IN$ it follows that there are
$|V|$ terms of size $n$:

{\it Induction basis}: Terms of size 0 are variables. The cardinality
of the set of those terms is $|V|$.

{\it Induction step}: Suppose the set of all terms of size $n$ has
cardinality $|V|$. Then the number of terms of size $n+1$ with leading
function symbol $f$ is $|V|^{ar(f)} = |V|$, using that $|V|$ is infinite.
Thus the number of terms of size $n+1$ is $|\Sigma| \times |V| = |V|$,
using that $|\Sigma| \leq |V|$.

Hence, even if $\Sigma \neq \emptyset$ we have $|\IT(\Sigma)| =
\Sigma_{i=1}^\infty |V| = |V|$.
So the set of literals over $\Sigma$ has cardinality $|V| \times |A| \times |V|
+ |V| \times |A| = |V|$, using that $0 \neq |A| \leq |V|$.
\end{proof}
Next we need a lemma that is very similar to the forthcoming
\pr{ruloid} in \sect{nxytt}. There, some intuitive
explanation can be found as well.

\begin{lemma}{n-ruloid}
Let $P=(\Sigma,R)$ be a standard TSS in decent ntyft format.
If $P\vdash\frac{N}{\sigma(t)\goto a t'}$ with $N$ a set of
negative literals,
then there are a decent xyntt rule
$\frac{H}{t\goto a u}$
and a substitution $\sigma'$ with
$P\vdash_{\rm irr}\frac{H}{t\goto a u}$, \plat{P\vdash\frac{N}{\sigma'(H)}},
$\sigma'(t)=\sigma(t)$ and $\sigma'(u)=t'$.
\end{lemma}

\pf
First, suppose $t$ is a variable. By default, the decent xyntt
rule $\frac{t\goto a y}{t\goto a y}$ is irredundantly provable from $P$. Let
$\sigma'$ be a substitution with $\sigma'(t)=\sigma(t)$ and $\sigma'(y)=t'$.
Clearly, $P\vdash\frac{N}{\sigma'(t\goto a y)}$.

Next, suppose $t=f(t_1,\ldots,t_{{\it ar}(f)})$.
We apply structural induction to a proof $\pi$ of the transition rule
\plat{\frac{N}{\sigma(t)\goto a t'}} from $P$.
Let $r\in R$ be the decent ntyft rule and $\rho$ be the substitution
used at the bottom of $\pi$, where $r$ is of the form
$\frac{\{v_k\goto{c_k}y_k\mid k\in K\}\cup
\{w_\ell\gonotto{d_\ell}\mid \ell\in L\}}
{f(x_1,\ldots,x_{{\it ar}(f)})\goto a v}$.
Then $\rho(x_i)=\sigma(t_i)$ for $i=1,\ldots,{\it ar}(f)$, $\rho(v)=t'$,
$(\rho(w_\ell)\gonotto{d_\ell}) \in N$ for $\ell \in L$,
and $\frac{N}{\rho(v_k)\goto{c_k}\rho(y_k)}$ for $k\in K$
are provable from $P$
by means of strict subproofs of $\pi$.
Since $r$ is decent, $\var(v_k)$ for $k\in K$ and
$\var(w_\ell)$ for $\ell\in L$ are included in $\{x_1,\ldots,x_{{\it ar}(f)}\}$.
Let $\rho_0$ be a substitution with $\rho_0(x_i)=t_i$ for $i=1,\ldots,{\it ar}(f)$.
As $\rho(x_i)=\sigma(t_i)=\sigma(\rho_0(x_i))$ for $i=1,\ldots,{\it
ar}(f)$, we have $\rho(v_k)=\sigma(\rho_0(v_k))$ for $k\in K$ and
$\rho(w_\ell)=\sigma(\rho_0(w_\ell))$ for $\ell\in L$.
So $\frac{N}{\sigma(\rho_0(v_k))\goto{c_k}\rho(y_k)}$ for $k\in K$ are
provable from $P$ by means of strict subproofs of $\pi$. According to
the induction hypothesis, for $k\in K$ there are a decent xyntt rule
$\frac{H_k}{\rho_0(v_k)\goto{c_k}u_k}$ and a substitution $\sigma_k'$
with $P\vdash_{\rm irr}\frac{H_k}{\rho_0(v_k)\goto{c_k}u_k}$,
$P\vdash\frac{N}{\sigma_k'(H_k)}$,
$\sigma_k'(\rho_0(v_k))=\sigma(\rho_0(v_k))$ and $\sigma_k'(u_k)=\rho(y_k)$.
By \lem{cardinality}, we can choose the sets of variables in the
right-hand sides of the positive premises in the $H_k$ (for $k\in K$)
pairwise disjoint, and disjoint from $\var(t)$. This
allows us to define a substitution $\sigma'$ with:
\begin{itemise}
\item
$\sigma'(z)=\sigma(z)$ for $z\in\var(t)$;
\item
$\sigma'(z)=\sigma_k'(z)$ for right-hand sides $z$ of positive premises
in $H_k$ for $k\in K$.
\end{itemise}
Let
\[
H ~~=~~
\bigcup_{k\in K}H_k~ \cup~ \{\rho_0(w_\ell)\gonotto{d_\ell}\mid \ell\in L\}~.
\]
Moreover, let $\rho_1$ be a substitution with $\rho_1(x_i)=t_i$ for
$i=1,\ldots,{\it ar}(f)$ and $\rho_1(y_k)=u_k$ for $k\in K$.
We verify that the rule
$\frac{H}{t\goto a\rho_1(v)}$ together
with the substitution $\sigma'$ satisfy the desired properties.

As $\var(v_k)\subseteq\{x_1,\ldots,x_{{\it ar}(f)}\}$, it follows that
$\var(\rho_0(v_k))\subseteq\var(t)$. Since $\sigma'$ and $\sigma$ agree
on $\var(t)$, $\sigma'(\rho_0(v_k))=\sigma(\rho_0(v_k))=\sigma_k'(\rho_0(v_k))$
for $k\in K$. Thus, by the decency of
\plat{\frac{H_k}{\rho_0(v_k)\goto{c_k}u_k}},
$\sigma'$ and $\sigma'_k$ agree on all variables
occurring in this rule for $k\in K$.

\begin{enumerate}

\item
$\rho_1$ and $\rho_0$ agree on $\var(v_k)\subseteq
\{x_1,\ldots,x_{{\it ar}(f)}\}$, so $\rho_1(v_k)=\rho_0(v_k)$ for $k\in K$.
Likewise, $\rho_1(w_\ell)\linebreak=\rho_0(w_\ell)$ for $\ell\in L$.
Since $P\vdash_{\rm irr} r$, we have $P\vdash_{\rm irr}\rho_1(r)=
\frac{\{\rho_0(v_k)\goto{c_k}u_k\mid k\in K\}\cup
\{\rho_0(w_\ell)\gonotto{d_\ell}\mid \ell\in L\}}{t\goto a\rho_1(v)}$.
Furthermore, $P\vdash_{\rm irr} \frac{H_k}{\rho_0(v_k)\goto{c_k}u_k}$ for $k\in K$.
As $H=\bigcup_{k\in K}H_k\cup\{\rho_0(w_\ell)\gonotto{d_\ell}\mid \ell\in L\}$,
it follows that \plat{P\vdash_{\rm irr}\frac{H}{t\goto a\rho_1(v)}}.

\item
The right-hand sides of the positive premises in any $H_k$
are distinct variables.  By construction, these sets of
variables (one for every $k\in K$) are pairwise
disjoint, and disjoint from $\var(t)$. Hence
$\frac{H}{t\goto a\rho_1(v)}$ is an ntytt rule.
Since the positive premises in $H$ originate from $H_k$ (for $k\in
K$), their left-hand sides are variables. This makes the rule an xyntt
rule. The rule is decent by \lem{preservation-decency}.

\item
$\sigma'$ agrees with $\sigma_k'$ on variables in $H_k$ for $k\in K$,
and with $\sigma$ on variables in $\rho_0(w_\ell)$ for $\ell \in L$.
Since $P\vdash\frac{N}{\sigma_k'(H_k)}$ for
$k\!\in\! K$, and $(\sigma(\rho_0(w_\ell))\gonotto{d_\ell}) \in N$ for
$\ell \!\in\! L$ (using that $\sigma(\rho_0(w_\ell))=\rho(w_\ell)$), we
conclude that $P\vdash\frac{N}{\sigma'(H)}$.

\item
Since $\sigma'$ and $\sigma$ agree on $\var(t)$, $\sigma'(t)=\sigma(t)$.

\item

$\sigma'(\rho_1(x_i))=\sigma'(t_i)=\sigma(t_i)=\rho(x_i)$
for $i=1,\ldots,{\it ar}(f)$.
Moreover, since $\sigma'$ and $\sigma_k'$ agree on $\var(u_k)$,
$\sigma'(\rho_1(y_k))=\sigma'(u_k)=\sigma_k'(u_k)=\rho(y_k)$ for $k\in
K$. As $\var(v)\subseteq
\{x_1,\ldots,x_{{\it ar}(f)}\}\cup\{y_k\mid k\in K\}$, it follows that
$\sigma'(\rho_1(v))=\rho(v)=t'$.
\hfill $\Box$
\end{enumerate}

\begin{lemma}{prov}
If all the rules in a TSS $Q$ are provable from a TSS $P$, then
all the rules that are provable from $Q$ are also provable from $P$.
\end{lemma}

\begin{proof}
Straightforward and left to the reader.
\end{proof}

\begin{proposition}{xynft}
For each standard TSS $P=(\Sigma,R)$ in decent ntyft format there exists a
standard TSS $P'=(\Sigma,R')$ in decent xynft format such that the requirements
(i)--(iv) above are met.
\end{proposition}

\begin{proof}
Let $R'$ consist of all decent xynft rules irredundantly provable from
$P$. In order to establish (i) we show that $P' \vdash
\frac{N}{t\goto{a}t'} \Leftrightarrow P \vdash \frac{N}{t\goto{a}t'}$
for all literals $t\goto{a}t'$ with $t$ closed and all sets of
negative literals $N$.  By \lem{prov} we have $P' \vdash
\frac{N}{t\goto{a}t'} \Rightarrow P \vdash \frac{N}{t\goto{a}t'}$. For
the other direction we use structural induction on $t$.  So assume $P
\vdash \frac{N}{f(t_1,\ldots,t_{ar(f)})\goto{a}t'}$.  By
\lem{n-ruloid} there are a decent xynft rule $r =
\frac{H}{f(x_1,\ldots,x_{ar(f)})\goto a u}$ and a substitution
$\sigma'$ with $P\vdash_{\rm irr} r$, $P\vdash\frac{N}{\sigma'(\alpha)}$ for
$\alpha \in H$, $\sigma'(x_i)=t_i$ for $i=1,\ldots,ar(f)$ and
$\sigma'(u)=t'$. Thus $r \in R'$.  For the negative literals $\alpha$
in $H$ the proof of $\frac{N}{\sigma'(\alpha)}$ is trivial. As
$\frac{H}{f(x_1,\ldots,x_{ar(f)})\goto a u}$ is a decent xynft rule,
the positive literals $\sigma'(\alpha)$ with $\alpha \in H$ have the
form $t_i \goto{b} v$.  Hence by induction $P' \vdash
\frac{N}{\sigma'(\alpha)}$ for $\alpha \in H$.  Moreover,
$\sigma'(r) = \frac{\sigma'(H)}{f(t_1,\ldots,t_{ar(f)})\goto{a}t'}$ is
a substitution instance of a rule in $R'$, so $P' \vdash
\frac{N}{f(t_1,\ldots,t_{ar(f)})\goto{a}t'}$.
This concludes the proof of requirement (i).
Requirements (ii), (iii) and (iv) are immediate corollaries of Lemmas
\hhref{lem-preservation-ready-trace}, \hhref{lem-preservation-readiness}
and \hhref{lem-preservation-failure-trace}, respectively.
\end{proof}
The results of this subsection are combined as follows:

\begin{corollary}{xyntt}
Let $P$ be a standard TSS in ready simulation format.
Then there exists a standard TSS $P'$ in decent xynft format such that
$P' \vdash_{ws} \alpha \Leftrightarrow P \vdash_{ws} \alpha$ for all
closed literals $\alpha$. Moreover if $P$ is in ready trace format
(resp.\ readiness format or failure trace format) then so is $P'$.
\end{corollary}
This also implies that $\sqsubseteq_N^{P'}$ equals $\sqsubseteq_N^P$
for each notion of observability $N$.

\subsection{Reducing well-supported proofs to supported proofs}
\hlabel{sec-supported}

Define the {\em size} of a node in a well-supported proof (or in any
well-founded, upwardly branching tree) to be the supremum of the sizes
of the nodes above it, plus one. Write $P \vdash_{ws}^\kappa \alpha$ if
there is a well-supported proof of the closed literal $\alpha$ from
the TSS $P$ of which the root has size no more than the ordinal
$\kappa$. By straightforward induction, this is equivalent to the
following recursive definition.

\begin{definition}{kappa}
Let $P=(\Sigma,R)$ be a standard TSS\@. $P \vdash_{ws}^{\kappa+1} \alpha$ for
$\alpha$ a closed literal iff
\begin{enumerate}\vspace{-1ex}
\item either there is a closed substitution instance $\frac{H}{\alpha}$
of a rule in $R$ with $P \vdash_{ws}^{\kappa} \beta$ for all
$\beta \in H$,\vspace{-1ex}
\item or $\alpha$ is negative and for every set $N$ of negative
closed literals such that $P \vdash \frac{N}{\gamma}$ for $\gamma$ a closed
literal denying $\alpha$, one has $P \vdash_{ws}^{\kappa} \delta$ for a
closed literal $\delta$ denying a literal in $N$.\vspace{-1ex}
\end{enumerate}
$P \vdash_{ws}^0 \alpha$ never holds, and 
in case $\kappa$ is a limit ordinal, $P \vdash_{ws}^{\kappa} \alpha$
iff $P \vdash_{ws}^{\lambda} \alpha$ for some $\lambda < \kappa$.
\end{definition}
Clearly, $P \vdash_{ws} \alpha$ iff $P \vdash_{ws}^\kappa \alpha$ for
some ordinal $\kappa$. Moreover, if $\lambda < \kappa$ then $P
\vdash_{ws}^\lambda \alpha \Rightarrow P \vdash_{ws}^\kappa \alpha$.
Now we introduce the following concept of {\em supported provability},
that will be shown to coincide with well-supported provability for
standard TSSs in decent xynft format.

\begin{definition}{kappa supported}
Let $P=(\Sigma,R)$ be a standard TSS\@. $P \vdash_{s}^{\kappa+1} \alpha$ for
$\alpha$ a closed literal iff
\begin{enumerate}\vspace{-1ex}
\item either there is a closed substitution instance $\frac{H}{\alpha}$
of a rule in $R$ with $P \vdash_{s}^{\kappa} \beta$ for all
$\beta \in H$,\vspace{-1ex}
\item or $\alpha$ is negative and for every closed substitution
instance $\frac{H}{\gamma}$ of a rule in $R$ with $\gamma$ denying $\alpha$,
one has $P \vdash_{s}^{\kappa} \delta$ for a
closed literal $\delta$ denying a literal in $H$.\vspace{-1ex}
\end{enumerate}
$P \vdash_{s}^0 \alpha$ never holds, and 
in case $\kappa$ is a limit ordinal, $P \vdash_{s}^{\kappa} \alpha$
iff $P \vdash_{s}^{\lambda} \alpha$ for some $\lambda < \kappa$.\\
$\alpha$ is said to be {\em $s$-provable} from $P$, notation $P
\vdash_{s} \alpha$, iff $P \vdash_{s}^\kappa \alpha$ for some ordinal $\kappa$.
\end{definition}
Due to the absence in this paper of literals of the form $t \gonotto{a}t'$, 
the notion of supported provability defined above is in general less
powerful than the notion of supported provability from \cite{vG95}
(cf.\ Counterexample $R$ in that paper). However, on TSSs in decent
xynft format both notions coincide, as will follow from
\pr{well-supported proof}.

\begin{lemma}{ntytt}
Let $\frac{H}{t\goto{a}t'}$ be a closed substitution instance of an
ntytt rule $r$ without lookahead, and let $H'$ be a set of closed literals
that differs from $H$ only in the right-hand sides of its positive
members. Then there exists a closed term $t''$ such that also
$\frac{H'}{t\goto{a} t''}$ is a substitution instance of $r$.
\end{lemma}

\begin{proof}
Straightforward.
\end{proof}

\begin{lemma}{ws-refutable}
Let $P=(\Sigma,R)$ be a standard TSS in ntytt format without lookahead and $\alpha$ be a
negative literal. Let $\kappa$ be an ordinal. Then the following are equivalent.
\begin{list}{$\bullet$}{\labelwidth\leftmargini\advance\labelwidth-\labelsep
                        \topsep 4pt \itemsep 2pt \parsep 2pt}
\item [(i)] For every set $N$ of negative closed literals such that $P
\vdash \frac{N}{\gamma}$ for $\gamma$ a closed literal denying
$\alpha$, one has $P \vdash_{ws}^{\kappa} \delta$ for a closed literal
$\delta$ denying a literal in $N$ (i.e.\ case 2 of \df{kappa}).
\item [(ii)]
For every closed substitution instance $\frac{H}{\gamma}$ of a rule in
$R$ with $\gamma$ denying $\alpha$, $H$ contains either a positive
literal $\beta$ denying a closed literal $\delta$ with
\plat{P\vdash_{ws}^{\kappa+1} \delta}, or a negative literal $\beta$
denying a closed literal $\delta$ with $P \vdash_{ws}^{\kappa} \delta$.
\end{list}
\end{lemma}

\begin{proof}
Suppose (ii) holds. Let $N$ be a set of negative closed literals
such that \plat{P \vdash \frac{N}{\gamma}} for $\gamma$ a closed literal
denying $\alpha$. Let $\frac{H}{\gamma}$ be the closed substitution
instance of the rule in $R$ used at the bottom of the proof of
\plat{\frac{N}{\gamma}}. Then $H$ must contain either a positive
literal $\beta$ denying a closed literal $\delta$ with
\plat{P\vdash_{ws}^{\kappa+1} \delta}, or a negative literal $\beta$
denying a closed literal $\delta$ with $P \vdash_{ws}^{\kappa}
\delta$. In the latter case, using that $P$ is standard, it must be
that $\beta \in N$ and we are done. In the former case, we have $P \vdash
\frac{N}{\beta}$, so by \df{kappa} (taking $\delta$ and $\beta$ for
$\alpha$ and $\gamma$, respectively), one has $P \vdash^\kappa_{ws}
\delta'$ for a closed literal $\delta'$ denying a literal in $N$,
which had to be proved.

Suppose (ii) does not hold. Let \plat{\frac{H}{\gamma}} be a
closed substitution instance of a rule in $R$ with $\gamma$ denying
$\alpha$, such that $H$ does not contain the specified literal
$\beta$, i.e.\ $P \not\vdash_{ws}^{\kappa+1} u\gonotto{c}$ for every
positive premise $u \goto{c} u'$ in $H$, and $P \not
\vdash_{ws}^{\kappa} u\goto{c}u'$ for every negative premise $u
\gonotto{c}$ in $H$ and every closed term $u'$. It suffices to find a
set of negative closed literals $N$ with $P \vdash \frac{N}{\gamma'}$
for $\gamma'$ a closed literal denying $\alpha$, such that $P
\not\vdash_{ws}^{\kappa} \delta$ for all closed literals $\delta$ denying a
literal in $N$.

By \df{kappa}, for every positive premise $\beta = u \goto{c} u'$ in
$H$ there must be a set $N_\beta$ of negative closed literals with $P
\vdash \frac{N_\beta}{u\goto{c}u''}$ for $u''$ a closed term, such
that $P \not\vdash_{ws}^{\kappa} \delta$ for all closed literals $\delta$
denying a literal in $N_\beta$. Let $H^+$ be the set of positive and
$H^-$ the set of negative literals in $H$, and let $N=
\bigcup_{\beta \in H^+} N_\beta \cup H^-$. Then $P
\not\vdash_{ws}^{\kappa} \delta$ for all closed literals $\delta$ denying a
literal in $N$. By \lem{ntytt} there exists a substitution instance
\plat{\frac{H'}{\gamma'}} of a rule in $P$, where $H'$ is obtained from $H$
by replacing the right-hand sides $u'$ of its positive members by
$u''$, and $\gamma'$ denies $\alpha$. Hence, $P \vdash \frac{N}{\gamma'}$.
\end{proof}

\begin{proposition}{well-supported proof}
Let $P=(\Sigma,R)$ be a standard TSS in decent xynft format.
Then $$P \vdash_{s} {\alpha} \Leftrightarrow P \vdash_{ws} {\alpha}.$$
\end{proposition}

\begin{proof}
For ``$\Rightarrow$'' we prove $P \vdash_{s}^\kappa {\alpha}
\Rightarrow P \vdash_{ws}^\kappa {\alpha}$ with induction to $\kappa$.
The case that $\kappa=0$ or $\kappa$ is a limit ordinal is trivial.
So suppose $P \vdash_{s}^{\kappa+1} \alpha$, and consider the two cases
provided by \df{kappa supported}.

{\sc Case 1}: Let $\frac{H}{\alpha}$ be a closed substitution instance
of a rule in $R$ with $P \vdash_{s}^{\kappa} \beta$ for all
$\beta \in H$. Then, by induction, $P \vdash_{ws}^\kappa \beta$ for all
$\beta \in H$, and hence $P \vdash_{ws}^{\kappa+1} \alpha$.

{\sc Case 2}: Suppose case 2 of \df{kappa supported} applies,
i.e.\ for every closed substitution instance $\frac{H}{\gamma}$ of a
rule in $R$ with $\gamma$ denying $\alpha$, $H$ contains a literal
$\beta$ denying a closed literal $\delta$ with $P \vdash_{s}^{\kappa}
\delta$. As the induction hypothesis yields $P \vdash_{s}^{\kappa}
\delta \Rightarrow P \vdash_{ws}^{\kappa} \delta$, and by definition
$P \vdash_{ws}^{\kappa} \delta \Rightarrow P \vdash_{ws}^{\kappa+1}
\delta$, case (ii) of \lem{ws-refutable} applies, and hence also case (i).
It follows that $P \vdash_{ws}^{\kappa+1} \alpha$.

For ``$\Leftarrow$'' we prove $P \vdash_{ws}^\kappa {\alpha}
\Rightarrow P \vdash_s {\alpha}$ with induction to $\kappa$, and a
nested induction to the {\em size} of the left-hand side of $\alpha$,
i.e.\ to the largest number of nested function symbols in it.
Again, the case that $\kappa=0$ or $\kappa$ is a limit ordinal is trivial.
So suppose $P \vdash_{ws}^{\kappa+1} \alpha$, and consider the two cases
provided by \df{kappa}.

{\sc Case 1}: Let $\frac{H}{\alpha}$ be a closed substitution instance
of a rule in $R$ with $P \vdash_{ws}^{\kappa} \beta$ for all
$\beta \in H$. Then, by induction, $P \vdash_{s} \beta$ for all
$\beta \in H$, and hence $P \vdash_s \alpha$.

{\sc Case 2}: Suppose case 2 of \df{kappa} applies.
Let \plat{\frac{H}{\gamma}} be a closed substitution instance of a
rule $r$ in $R$ with $\gamma$ denying $\alpha$.
By \lem{ws-refutable}, $H$ must contain either a positive literal
$\beta$ denying a literal $\delta$ with $P \vdash_{ws}^{\kappa+1}
\delta$, or a negative literal $\beta$ denying a literal $\delta$
with $P \vdash_{ws}^{\kappa} \delta$. In the latter case, the
induction hypothesis gives $P \vdash_s \delta$. In the former case,
using that $r$ is a decent xynft rule, the left-hand side of $\delta$
(and $\beta$) is smaller than the left-hand side of $\alpha$ (and
$\gamma$), so the nested induction hypothesis allows to conclude that
$P \vdash_s \delta$. It follows that $P \vdash_s \alpha$.
\end{proof}
The following counterexample shows that the restriction to decent
xynft format is essential here.

\begin{example}{no free variables}
Let $A=\{a\}$, let $\Sigma$ consist of the constant $c$,
and let $R$ be the decent ntyxt rule
\[
\frac{x\goto a y}{x\goto a y}
\]
No closed rule of the form $\frac{N}{c\goto a u}$, where $N$ contains only
closed negative literals, is provable from $P=(\Sigma,R)$. Hence, $P\vdash_{ws}c\gonotto a$.
However, since $\frac{c\goto a c}{c\goto a c}$ is a closed substitution
instance of the rule in $R$, we have $P\not\vdash_{s}c\gonotto a$.

This shows that \pr{well-supported proof} does not extend to rules in
which the source is a variable, and explains why we needed \pr{ntyft}.
The example also applies when we change $R$ into
\[
\frac{x\goto a y}{c\goto a c}
\]
This shows that \pr{well-supported proof} does not extend to rules
with free variables, and explains why we needed \pr{decent}.
When we change $R$ into
\[
\frac{c\goto a y}{c\goto a y}
\]
the example shows that \pr{well-supported proof} does not extend to
ntyft rules, and explains why we needed \pr{xynft}.
In order to see that \pr{well-supported proof} does not extend to
rules with lookahead, let $\Sigma$ consist of the constant $c$ and the
unary function symbol $f$, and let $R$ be
\[
c\goto a f(c)
\hspace{1cm}
\frac{x\goto a y \goto a z}{f(x)\goto a z}
\]
This time we obtain $P\vdash_{ws}f(c)\gonotto a$, but
$P\not\vdash_{s}f(c)\gonotto a$.
\end{example}

\subsection{Reducing supported proofs to standard proofs}
\hlabel{sec-standard}

We proceed to show that for any given standard TSS $P=(\Sigma,R)$ in decent ntyft
format there exists a TSS $P^+=(\Sigma,R^+)$ in decent ntyft format such that $P
\vdash_{s} \alpha \Leftrightarrow P^+ \vdash \alpha$. 
Again, the translation of $R$ into $R^+$ preserves the ready trace,
readiness and failure trace formats.
The construction below uses the absence of lookahead in an essential way.
\pagebreak[3]

\begin{definition}{pick}
Let $P=(\Sigma,R)$ be a TSS, $t \in \IT(\Sigma)$ and $a \in A$.
\begin{itemise}
\item
$cl(R)$ denotes the collection of closed substitution instances of rules in $R$.
\item
$R \rest (t \goto{a})$ denotes the set of rules in $R$ with conclusion
$t \goto{a} u$ for some $u \in \IT(\Sigma)$.
\item
$pick(R, t\gonotto{a})$ denotes the collection of transition rules
\plat{\frac{H}{t \gonotto{a}}} in which $H$ is obtained by taking one
premise from every rule in $R \rest (t \goto{a})$.
\item
$deny(R)$ denotes the collection of rules obtained from $R$ by changing in each
rule every positive premise $u \goto{c} u'$ into $u \gonotto{c}$, and every
negative premise $u \gonotto{c}$ into $u \goto{c} y$, where the $y$ are
all different variables, not occurring elsewhere in the rule. (That there are
that many different variables follows from \lem{cardinality}; if
needed the variables in the rule that do not occur in its source may
be renamed.)
\item
$prove(R,t \gonotto{a})$ denotes $deny(pick(R, t \gonotto{a}))$.
\item
Let $\overline{R}$ be the union of $cl(R)$ and all collections
$cl(prove(cl(R),\alpha))$ for negative closed literals $\alpha$.
\end{itemise}
\end{definition}
Note that $cl(R) \rest (t \goto{a})$ with $t$ closed is the set of
closed substitution instances \plat{\frac{H}{\gamma}} of rules in $R$
with $\gamma$ denying $t \gonotto{a}$. Thus $cl(prove(cl(R),\alpha))$,
for $\alpha$ a negative literal, contains only closed rules
$\frac{K}{\alpha}$ such that for every closed substitution instance
$\frac{H}{\gamma}$ of a rule in $R$ with $\gamma$ denying $\alpha$, a
literal in $K$ denies a literal in $H$. Moreover, for every closed
rule \plat{\frac{K}{\alpha}} of the latter kind, there is a $K' \subseteq
K$ such that \plat{\frac{K'}{\alpha} \in cl(prove(cl(R),\alpha))}. From
this we obtain:

\begin{lemma}{making support explicit}
Let $P=(\Sigma,R)$ and $P'=(\Sigma,R')$ be TSSs with $P$ standard and
$cl(R') = \overline{R}$.
Then $P' \vdash {\alpha} \Leftrightarrow P \vdash_s {\alpha}$.\hfill $\Box$
\end{lemma}

% \begin{definition}{making support explicit}
% Let $P=(\Sigma,R)$ be a TSS in decent uniform ntyft format. Let $R^+$ be obtained
% from $R$ by adding all ntyft rules $\frac{K}{f(x_1,\ldots,x_{ar(f)})
% \gonotto{a}}$ such that there is a function $\xi$ choosing from every rule
% $\frac{H}{f(x_1,\ldots,x_{ar(f)})\goto{a}t'}$ in $R$ one of its premises,
% and $K$ is obtained from $range(\xi)$ by changing any premise 
% $u \goto{b} u'$ into $u\gonotto{b}$ and any premise $u \gonotto{b}$
% into $u \goto{b} y$ (using a fresh variable $y \in V$).
% \end{definition}

\begin{definition}{uniform}
A TSS $P=(\Sigma,R)$ is in {\em uniform ntyft format}
if it is in ntyft format and for every function symbol $f \in \Sigma$
there is a specific sequence of all different variables $\vec{x}_f =
(x_1, \ldots, x_{ar(f)})$, such that all sources of rules in $R$ have
the form $f(\vec{x}_f)$ for some $f \in \Sigma$.

Let $P$ be in uniform ntyft format. Then $R^+$
denotes the union of $R$ and all collections
$prove(R,f(\vec{x}_f)\gonotto{a})$ for $f \in \Sigma$ and $a \in A$.
\end{definition}

\begin{example}{closure}
Let $A=\{a,b\}$ and let $\Sigma$ consist of the constant $c$ and
the binary function symbol $f$. Let $R$ be:
\[
c\goto a c
\hspace*{1cm}
\frac{x_1\goto{a} y}{f(x_1,x_2)\goto{b} y}
\hspace*{1cm}
\Rule{x_2\goto{a} y ~~~~ x_1 \gonotto{b}}{f(x_1,x_2)\goto{b} y}
\]
Then $R^+$ is $R$ together with:
\[
c\gonotto b
\hspace*{1cm}
f(x_1,x_2)\gonotto a
\hspace*{1cm}
\Rule{x_1\gonotto{a} ~~~~ x_2\gonotto{a}}{f(x_1,x_2)\gonotto{b}}
\hspace*{1cm}
\Rule{x_1\gonotto{a} ~~~~ x_1\goto{b}z}{f(x_1,x_2)\gonotto{b}}
\]
\end{example}

\begin{definition}{equiv}
Write $r \equiv r'$ for transition rules $r$ and $r'$ if they differ
only through a bijective renaming of their variables ({\em
$\alpha$-conversion}).
Write $R \equiv R'$ if for each $r \in R$ there
is an $r' \in R'$ with $r \equiv r'$ and vice versa.

Write $r \approx r'$ for transition rules $r$ and $r'$ if they differ
only in their targets and in the right-hand sides of their premises.
Write $R \approx R'$ if for each $r \in R$ there
is an $r' \in R'$ with $r \approx r'$ and vice versa.
\end{definition}
Note that for any TSS $P=(\Sigma,R)$ in decent ntyft format there is a
TSS $P'=(\Sigma,R')$ in decent uniform ntyft format with $R \equiv
R'$. Moreover, $P \vdash_s \beta
\Leftrightarrow P' \vdash_s \beta$ for any literal $\beta$;
also $P'$ is in ready trace/readiness/failure trace format iff $P$ is.
Finally note that if $R \approx R'$ and all rules in $R$ and $R'$
have negative conclusions then $deny(R) \equiv deny(R')$.

\begin{lemma}{uniform}
Let $P=(\Sigma,R)$ be a TSS in decent uniform ntyft format, and let $\sigma$ be a
closed substitution. Then $\sigma(R \rest ({f(\vec{x}_f)\goto{a}}))
\approx cl(R) \rest (\sigma(f(\vec{x}_f))\goto{a})$.
\end{lemma}

\begin{proof}
``$\subseteq$'': If rule $r\in R$ has conclusion $f(\vec{x}_f)
\goto{a} u$ then $\sigma(r) \in \sigma(R) \subseteq cl(R)$ has
conclusion $\sigma(f(\vec{x}_f))\goto{a} \sigma(u)$.

``$\supseteq$'': Let $r \in cl(R) \rest ({\sigma(f(\vec{x}_f))\goto{a}})$.
Then $r = \rho(r')$ for $r'\in R$ and $\rho$ a closed substitution,
and the conclusion of $r$ is $\sigma(f(\vec{x}_f)) \goto{a} u$ for some
term $u$. As $P$ is in uniform ntyft format this implies that the
conclusion of $r'$ is $f(\vec{x}_f) \goto{a} u'$ for some term $u'$, 
and $\rho(x_i)=\sigma(x_i)$ for $i=1,\ldots, ar(f)$. As $r$ is
decent, the left-hand sides of its premises contain no other variables
than $x_1, \ldots ,x_{ar(f)}$. Therefore $r=\rho(r') \approx
\sigma(r') \in \sigma(R \rest (f(\vec{x}_f)\goto{a}))$.
\end{proof}

\begin{proposition}{supported proof}
Let $P=(\Sigma,R)$ be a standard TSS in decent uniform ntyft format. Then also
$P^+=(\Sigma,R^+)$ is in decent uniform ntyft format, and $P^+ \vdash {\alpha}
\Leftrightarrow P \vdash_s {\alpha}$.
\end{proposition}

\begin{proof}
The first statement is immediate from the construction. For the second,
using \lem{making support explicit}, it suffices to show that $cl(R^+)
\equiv \overline{R}$. First of all, for any $f \in \Sigma$, $a \in A$ and
closed substitution $\sigma$,
\\
$\sigma(pick(R, f(\vec{x}_f)\gonotto{a})) = \\
\{\frac{\sigma(H)}{\sigma(f(\vec{x}_f)) \gonotto{a}} \mid H \mbox{ is
obtained by taking one premise from each rule in } R \rest
{f(\vec{x}_f)\goto{a}}\}
= \\
\{\frac{K}{\sigma(f(\vec{x}_f)) \gonotto{a}} \mid K \mbox{ is
obtained by taking one premise from each rule in }
\sigma(R \rest {f(\vec{x}_f)\goto{a}})\}
\!\stackrel{\rm \lem{uniform}}{\approx} \\[2pt]
\{\frac{K}{\sigma(f(\vec{x}_f)) \gonotto{a}} \mid K \mbox{ is
obtained by taking one premise from each rule in }
cl(R) \rest {\sigma(f(\vec{x}_f))\goto{a}}\}
= \\[2pt]
pick(cl(R), \sigma(f(\vec{x}_f)\gonotto{a})).$

Therefore, $cl(R^+) = 
cl(R) \cup \bigcup_{a\in A} \bigcup_{f\in \Sigma} cl(prove(R,f(\vec{x}_f)\gonotto{a})) =\\
cl(R) \cup \bigcup_{a\in A} \bigcup_{f\in \Sigma} cl(deny(pick(R,f(\vec{x}_f)\gonotto{a}))) = \\
cl(R) \cup \bigcup_{a\in A} \bigcup_{f\in \Sigma} cl(deny(cl(pick(R,f(\vec{x}_f)\gonotto{a})))) = \\
cl(R) \cup \bigcup_{a\in A} \bigcup_{f\in \Sigma}
cl(deny(\bigcup_{\sigma:V \rightarrow T(\Sigma)}\sigma(pick(R,f(\vec{x}_f)\gonotto{a}))))
\equiv \\
cl(R) \cup \bigcup_{a\in A} \bigcup_{f\in \Sigma} cl(deny(
\bigcup_{\sigma:V \rightarrow T(\Sigma)}pick(cl(R),\sigma(f(\vec{x}_f)\gonotto{a})))) = \\
cl(R) \cup \bigcup_{a\in A} \bigcup_{f\in \Sigma} cl(deny(
\bigcup_{t_1,\ldots,t_{ar(f)}\in T(\Sigma)}pick(cl(R),f(t_1,\ldots,t_{ar(f)})\gonotto{a}))) = \\
cl(R) \cup \bigcup_{a\in A} \bigcup_{f\in \Sigma} \bigcup_{t_1,\ldots,t_{ar(f)}\in T(\Sigma)}
cl(deny(pick(cl(R),f(t_1,\ldots,t_{ar(f)})\gonotto{a}))) = \\
cl(R) \cup \bigcup_{a\in A} \bigcup_{f\in \Sigma} \bigcup_{t_1,\ldots,t_{ar(f)}\in T(\Sigma)}
cl(prove(cl(R),f(t_1,\ldots,t_{ar(f)})\gonotto{a})) = \\
cl(R) \cup \bigcup_{\alpha \rm~a~closed~negative~literal}
cl(prove(cl(R),\alpha)) = \overline{R}$.
\end{proof}
The following counterexample shows that absence of free variables
is essential here.

\begin{example}{decency needed}
Let $A=\{a\}$, let $\Sigma$ consist of the constants $a$, $c$ and
$\epsilon$, and let $R$ be the following collection of rules:
$$a \goto{a} \epsilon \hspace{1cm} \Rule{x\goto{a}y}{c\goto{a}y}$$
Note that $P=(\Sigma,R)$ is a standard TSS in uniform ntyft format.
Clearly $P \vdash c \goto{a} \epsilon$ and thus certainly
$P \vdash_{s} c \goto{a} \epsilon$ and thus
$P \not\vdash_{s} c \gonotto{a}$. Nevertheless, $R^+$ would be $R$
together with the rules $\epsilon \gonotto{a}$ and
$\Rule{x\gonotto{a}}{c\gonotto{a}}$, so we would have
$(\Sigma,R^+) \vdash c \gonotto{a}$.
\end{example}
The following counterexample shows that
absence of lookahead is essential too.

\begin{example}{no lookahead}
Let $A=\{a\}$, let $\Sigma$ consist of the constants $a$ and
$\epsilon$ and the unary function symbol $f$, and let $R$ be the
following collection of rules:
$$a\goto{a}\epsilon \hspace{1cm} \Rule{x\goto{a}y \gonotto{a}}{f(x)\goto{a}y}$$
Again $P=(\Sigma,R)$ is a standard TSS in uniform ntyft format.
As there are no rules with conclusion $\epsilon \goto a t$ for some
term $t$, one obtains $P \vdash_s \epsilon \gonotto a$, and hence
$P \vdash_s f(a) \goto a \epsilon$. Nevertheless, $R^+$ would be $R$
together with the rules $\epsilon \gonotto{a}$,
$\Rule{x\gonotto{a}}{f(x)\gonotto{a}}$ and
$\Rule{y\goto{a}z}{f(x)\gonotto{a}}$, so we would have
$(\Sigma,R^+) \vdash f(c) \gonotto{a}$.
\end{example}
It remains to show the relevant formats are preserved under the
translation of $R$ into $R^+$. Recall that these formats were extended
to non-standard TSSs in \df{formats negative}.

\begin{proposition}{supported preserved}
Let $P=(\Sigma,R)$ be a TSS in decent uniform ntyft format. If $P$ is in ready
trace format (resp.\ readiness format or failure trace format), then so
is $P^+=(\Sigma,R^+)$.
\end{proposition}

\begin{proof}
By \df{formats negative} the statements for the ready trace and
readiness formats are trivial.

Let $P$ be in failure trace format, let $f \in \Sigma$ and $a \in A$,
and let $z$ be $\Lambda$-floating in a rule $r$ in
$prove(R,f(\vec{x}_f) \gonotto{a})$. In case $z$ is the right-hand
side of a positive premise, it is not polled in $r$. So assume $z$
occurs in the source of $r$. As in the rules of
$R\rest(f(\vec{x}_f)\goto{a})$ the variable $z$ is also
$\Lambda$-floating, it is polled only in $\Lambda$-liquid positions in
those rules, and only in positive premises. The same holds for the
rules in $pick(R,f(\vec{x}_f)\gonotto{a})$. Hence in the rules of
$deny(pick(R,f(\vec{x}_f)\gonotto{a}))$ it is polled only in
$\Lambda$-liquid positions in negative premises.
\end{proof}
The results of this section can now be combined as follows,
extending the definition of $R^+$ to TSSs $(\Sigma,R)$ in ready simulation
format.

\begin{definition}{plus}
Let $P=(\Sigma,R)$ be a TSS in ready simulation format. Then $P^+=(\Sigma,R^+)$
is defined as the TSS obtained by subsequently
\begin{itemise}
\item convert $P$ to ntyft format without lookahead following the
construction in the proof of \href{pr-ntyft}{Prop.~\ref{pr-ntyft}},
\item convert the result to decent ntyft format following the
construction in the proof of \pr{decent},
\item convert the result to decent xynft format following the
construction in the proof of \pr{xynft},
\item convert the result to decent uniform xynft format by bijectively
renaming the variables in each of its rules,
\item and applying the $^+$ construction of \df{uniform}.
\end{itemise}
\end{definition}
Note that $R^+$ is fully determined by $R$ up to $\alpha$-conversion.

\begin{corollary}{plus}
Let $P=(\Sigma,R)$ be a standard TSS in ready simulation format.
Then $P^+=(\Sigma,R^+)$ is in decent ntyft format,
$P^+ \vdash \alpha \Leftrightarrow P \vdash_{ws} \alpha$ for all
closed literals $\alpha$, and if $P$ is in ready trace format
(resp.\ readiness format or failure trace format) then so is $P^+$.
\end{corollary}

%%%%%%%%%%%%%%%%%%%%%%%%%%%%%%%%%%%%%%%%%%%%%%%%%%%%%%%%%%%%%%%%%%%%%%%%%%%%%%
\section{Reducing decent ntyft rules to decent nxytt rules}
\hlabel{sec-nxytt}

The following proposition says, for $P$ a TSS in decent ntyft
format, that for any function symbol $f$ there are a number of decent
nxyft rules with source $f(x_1,\ldots,x_{ar(f)})$ irredundantly
provable from $P$ that are ``just right'' for $f$. Here ``just
right'' means that for any (closed) literal \plat{f(t_1,\ldots,t_{ar(f)})
\gonotto a} or $f(t_1,\ldots,t_{ar(f)}) \goto a t'$ that is provable
from $P$ there is a (closed) proof using one of those rules as the
last step. Moreover, the same holds not only for function symbols $f$,
but for arbitrary open terms. If for an open term $t$ with
$\var(t)=\{x_1,\ldots,x_n\}$ we would introduce an $n$-ary function
symbol $f_t$ such that $f_t(x_1,\ldots,x_{n})$ is just a shorthand for
$t$, then $P$ also irredundantly proves a set of decent nxyft rules
that are just right for $f_t$.

\pr{ruloid} and its proof below can be read in four ways:
pertaining to negative or positive literals (with the material
pertaining only to positive literals enclosed in square brackets), and
with and without the adjective ``closed''.

\begin{proposition}{ruloid}
Let $P=(\Sigma,R)$ be a TSS in decent ntyft format.
If $P\vdash\sigma(t)\gonotto a$ [resp.\ $P\vdash\sigma(t)\goto a t'$]
for $t$ a term, [$t'$ a (closed) term] and $\sigma$ a (closed) substitution,
then there are a decent nxytt rule $\frac{H}{t\gonotto a}$
[resp.\ $\frac{H}{t\goto a u}$]
and a (closed) substitution $\sigma'$ with $P\irr\frac{H}{t\gonotto a}$
[resp.\ $P\irr\frac{H}{t\goto a u}$], $P\vdash\sigma'(H)$,
$\sigma'(t)=\sigma(t)$ [and $\sigma'(u)=t'$].
\end{proposition}

\pf
First, suppose $t$ is a variable. By default, the decent nxytt
rule $\frac{t\gonotto a}{t\gonotto a}$ [resp.\ $\frac{t\goto a
y}{t\goto a y}$] is irredundantly provable from $P$. Let $\sigma'$
be a (closed) substitution with $\sigma'(t)=\sigma(t)$ [and $\sigma'(y)=t'$].
Clearly, $P\vdash\sigma'(t\gonotto a)$ [resp.\ $P\vdash\sigma'(t\goto a y)$].

Next, suppose $t=f(t_1,\ldots,t_{{\it ar}(f)})$.
We apply structural induction to a (closed) proof $\pi$ of the literal
$\sigma(t)\gonotto a$ [resp.\ $\sigma(t)\goto a t'$] from $P$.
Let $r\in R$ be the decent ntyft rule and $\rho$ be the (closed)
substitution used at the bottom of $\pi$, where $r$ is of the form
$\frac{\{v_k\goto{c_k}y_k\mid k\in K\}\cup
\{w_\ell\gonotto{d_\ell}\mid \ell\in L\}}
{f(x_1,\ldots,x_{{\it ar}(f)})\gonotto a}$
[resp.\ $\frac{\{v_k\goto{c_k}y_k\mid k\in K\}\cup
\{w_\ell\gonotto{d_\ell}\mid \ell\in L\}}
{f(x_1,\ldots,x_{{\it ar}(f)})\goto a v}$].
Then $\rho(x_i)=\sigma(t_i)$ for $i=1,\ldots,{\it ar}(f)$, [$\rho(v)=t'$]
and $\rho(v_k)\goto{c_k}\rho(y_k)$ for $k\in K$ and
$\rho(w_\ell)\gonotto{d_\ell}$ for $\ell\in L$ are provable from $P$
by means of strict subproofs of $\pi$.
Since $r$ is decent, $\var(v_k)$ for $k\in K$ and
$\var(w_\ell)$ for $\ell\in L$ are included in $\{x_1,\ldots,x_{{\it ar}(f)}\}$.
Let $\rho_0$ be a substitution with $\rho_0(x_i)=t_i$ for $i=1,\ldots,{\it ar}(f)$.
As $\rho(x_i)=\sigma(t_i)=\sigma(\rho_0(x_i))$ for $i=1,\ldots,{\it
ar}(f)$, we have $\rho(v_k)=\sigma(\rho_0(v_k))$ for $k\in K$ and
$\rho(w_\ell)=\sigma(\rho_0(w_\ell))$ for $\ell\in L$.
So $\sigma(\rho_0(v_k))\goto{c_k}\rho(y_k)$ for $k\in K$ and
$\sigma(\rho_0(w_\ell))\gonotto{d_\ell}$ for $\ell\in L$ are provable from $P$
by means of strict subproofs of $\pi$. According to the induction hypothesis,
for $k\in K$ there are a decent nxytt rule
$\frac{H_k}{\rho_0(v_k)\goto{c_k}u_k}$ and a (closed) substitution $\sigma_k'$
with $P\irr\frac{H_k}{\rho_0(v_k)\goto{c_k}u_k}$,
$P\vdash\sigma_k'(H_k)$, $\sigma_k'(\rho_0(v_k))=\sigma(\rho_0(v_k))$
and $\sigma_k'(u_k)=\rho(y_k)$. Likewise, for $\ell\in L$ there are a
decent nxytt rule $\frac{H_\ell}{\rho_0(w_\ell)\gonotto{d_\ell}}$
and a (closed) substitution $\sigma_\ell'$
with $P\irr\frac{H_\ell}{\rho_0(w_\ell)\gonotto{d_\ell}}$,
$P\vdash\sigma_\ell'(H_\ell)$ and
$\sigma_\ell'(\rho_0(w_\ell))=\sigma(\rho_0(w_\ell))$.
By \lem{cardinality}, we can choose the sets of variables in the
right-hand sides of the positive premises in the $H_k$ (for $k\in K$)
and $H_\ell$ (for $\ell\in L$) pairwise disjoint, and disjoint from
$\var(t)$. This allows us to define a (closed) substitution $\sigma'$ with:
\begin{itemise}
\item
$\sigma'(z)=\sigma(z)$ for $z\in\var(t)$;
\item
$\sigma'(z)=\sigma_k'(z)$ for right-hand sides $z$ of positive premises
in $H_k$ for $k\in K$;
\item
$\sigma'(z)=\sigma_\ell'(z)$ for right-hand sides $z$ of positive premises
in $H_\ell$ for $\ell\in L$.
\end{itemise}
Let
\[
H ~~=~~
\bigcup_{k\in K}H_k~\cup~\bigcup_{\ell\in L}H_\ell ~.
\]
Moreover, let $\rho_1$ be a substitution with $\rho_1(x_i)=t_i$ for
$i=1,\ldots,{\it ar}(f)$ and $\rho_1(y_k)=u_k$ for $k\in K$.\vspace{1pt}
We verify that the rule \plat{\frac{H}{t\gonotto a}}
[resp.\ \plat{\frac{H}{t\goto a\rho_1(v)}}] together
with the substitution $\sigma'$ satisfy the desired properties.

As $\var(v_k)\subseteq\{x_1,\ldots,x_{{\it ar}(f)}\}$, it follows that
$\var(\rho_0(v_k))\subseteq\var(t)$. Since $\sigma'$ and $\sigma$ agree
on $\var(t)$, $\sigma'(\rho_0(v_k))=\sigma(\rho_0(v_k))=\sigma_k'(\rho_0(v_k))$
for $k\in K$. Thus, by the decency of $\frac{H_k}{\rho_0(v_k)\goto{c_k}u_k}$,
$\sigma'$ and $\sigma'_k$ agree on all variables
occurring in this rule for $k\in K$. Likewise,
$\sigma'$ and $\sigma'_\ell$ agree on all variables occurring in
$\frac{H_\ell}{\rho_0(w_\ell)\gonotto{d_\ell}}$ for $\ell \in L$.

\begin{enumerate}

\item
$\rho_1$ and $\rho_0$ agree on $\var(v_k)\subseteq
\{x_1,\ldots,x_{{\it ar}(f)}\}$, so $\rho_1(v_k)=\rho_0(v_k)$ for $k\!\in\! K$.
Likewise, $\rho_1(w_\ell)\linebreak[3]=\rho_0(w_\ell)$ for $\ell\!\in\! L$.
\vspace{-2pt}
Since $P\irr r$, we have $P\irr\rho_1(r)=
\frac{\{\rho_0(v_k)\goto{c_k}u_k\mid k\in K\}\cup
\{\rho_0(w_\ell)\gonotto{d_\ell}\mid \ell\in L\}}{t\gonotto a}$
[resp.\ $\frac{\{\rho_0(v_k)\goto{c_k}u_k\mid k\in K\}\cup
\{\rho_0(w_\ell)\gonotto{d_\ell}\mid \ell\in L\}}{t\goto a\rho_1(v)}$].
Furthermore, $P\irr \frac{H_k}{\rho_0(v_k)\goto{c_k}u_k}$ for $k\in K$ and
$P\irr\frac{H_\ell}{\rho_0(w_\ell)\gonotto{d_\ell}}$ for $\ell\in L$.
As \plat{\displaystyle H=\bigcup_{k\in K}H_k\cup\bigcup_{\ell\in
L}H_\ell}, it follows that
$P\irr\frac{H}{t\gonotto a}$ [resp.\ $P\irr\frac{H}{t\goto a\rho_1(v)}$].

\item
The right-hand sides of the positive premises in any $H_k$
or $H_\ell$ are distinct variables.  By construction, these sets of
variables (one for every $k\in K$ and $\ell\in L$) are pairwise
disjoint, and disjoint from $\var(t)$. Hence $\frac{H}{t\gonotto a}$
[resp.\ $\frac{H}{t\goto a\rho_1(v)}$] is an ntytt rule.
Since the premises in $H$ originate from $H_k$ (for $k\in K$) and
$H_\ell$ (for $\ell\in L$), their left-hand sides are variables.
This makes the rule an nxytt rule. The rule is decent by
\lem{preservation-decency}.

\item
$\sigma'$ agrees with $\sigma_k'$ on variables
in $H_k$ for $k\in K$, and $\sigma'$ agrees with $\sigma_\ell'$
on variables in $H_\ell$ for $\ell\in L$.
Since $P\vdash\sigma_k'(H_k)$ for $k\in K$ and $P\vdash\sigma_\ell'(H_\ell)$
for $\ell\in L$, we conclude that $P\vdash\sigma'(H)$.

\item
Since $\sigma'$ and $\sigma$ agree on $\var(t)$, $\sigma'(t)=\sigma(t)$.

\item

[$\sigma'(\rho_1(x_i))=\sigma'(t_i)=\sigma(t_i)=\rho(x_i)$
for $i=1,\ldots,{\it ar}(f)$.
Moreover, since $\sigma'$ and $\sigma_k'$ agree on $\var(u_k)$,
$\sigma'(\rho_1(y_k))=\sigma'(u_k)=\sigma_k'(u_k)=\rho(y_k)$ for $k\in
K$. As $\var(v)\subseteq
\{x_1,\ldots,x_{{\it ar}(f)}\}\cup\{y_k\mid k\in K\}$, it follows that
$\sigma'(\rho_1(v))=\rho(v)=t'$.]
\hfill $\Box$
\end{enumerate}
Note the similarity between \lem{n-ruloid} and \pr{ruloid} and their
proofs. The essential difference is that \lem{n-ruloid} employs a
definition of ``just right'' that deals with transition rules with
negative premises instead of mere literals.  The price to be paid for
that is that the constructed rules may contain arbitrary negative
premises, instead of premises of the form $x \gonotto{a}$. Furthermore
\lem{n-ruloid} deals only with standard TSSs (and does not mention the
closed case).

The following corollary will not be needed in the remainder of this
paper; however it may be interesting in its own right.

\begin{corollary}{nxyft}
Let $P=(\Sigma,R)$ be a TSS in decent ntyft format.
Then there exists a TSS $P'=(\Sigma,R')$ in decent nxyft format such that
$P' \vdash \alpha \Leftrightarrow P \vdash \alpha$ for all
closed literals $\alpha$. Moreover if $P$ is in ready trace format
(resp.\ readiness format or failure trace format) then so is $P'$.
\end{corollary}

\begin{proof}
Let $R'$ consist of all decent nxyft rules irredundantly provable from
$P$. That $P' \vdash \alpha \Rightarrow P \vdash \alpha$ follows
immediately from \lem{prov}. For the other direction we use
structural induction on the source of $\alpha$.  So assume $P \vdash
\alpha$ with $\alpha$ being ${f(t_1,\ldots,t_{ar(f)})\gonotto{a}}$ or
${f(t_1,\ldots,t_{ar(f)})\goto{a}t'}$.  By \pr{ruloid} there are a
decent nxyft rule $r = \frac{H}{f(x_1,\ldots,x_{ar(f)})\gonotto a}$
[resp.\ $\frac{H}{f(x_1,\ldots,x_{ar(f)})\goto a u}$] and a
substitution $\sigma'$ with $P \vdash_{\rm irr} r$, $P
\vdash{\sigma'(H)}$, $\sigma'(x_i)=t_i$ for $i=1,\ldots,ar(f)$ and
$\sigma'(u)=t'$. Thus $r \in R'$. As $r$ is a decent nxyft rule, the
literals $\sigma'(\beta)$ with $\beta \in H$ have the form $t_i
\goto{b} v$ or $t_i \gonotto{b}$.  Hence by induction
$P' \vdash {\sigma'(H)}$.  Moreover, $\sigma'(r) =
\frac{\sigma'(H)}{\alpha}$ is a substitution instance of a rule in
$R'$, so $P' \vdash \alpha$.  This concludes the proof of the first
statement.  The second is an immediate corollary of Lemmas
\hhref{lem-preservation-ready-trace},
\hhref{lem-preservation-readiness}, \hhref{lem-preservation-failure-trace}
and \hhref{lem-preservation-negative-failure-trace}.
\end{proof}

%%%%%%%%%%%%%%%%%%%%%%%%%%%%%%%%%%%%%%%%%%%%%%%%%%%%%%%%%%%%%%%%%%%%%%%%%%%%%%
\section{Crux}
\hlabel{sec-crux}

This section witnesses the proofs of Theorems 4-7; see Corollary 5.
The bulk of this section consists of lemmas building up to this corollary.

\begin{definition}{ruloid-crux}
Let $P=(\Sigma,R)$ be a standard TSS in ready simulation format.  A {\em $P$-ruloid}
is a decent nxytt rule, irredundantly provable from $P^+=(\Sigma,R^+)$.
\end{definition}

\begin{lemma}{ruloid}
Let $P$ be a standard TSS in ready simulation format. Then 
$P\vdash_{ws}\sigma(t)\gonotto a$ [resp.\ $P\vdash_{ws}\sigma(t)\goto a t'$]
for $t$ a term, [$t'$ a closed term] and $\sigma$ a closed substitution,
iff there are a $P$-ruloid $\frac{H}{t\gonotto a}$
[resp.\ $\frac{H}{t\goto a u}$]
and a closed substitution $\sigma'$ with $P\vdash_{ws}\sigma'(H)$,
$\sigma'(t)=\sigma(t)$ [and $\sigma'(u)=t'$].
\end{lemma}

\begin{proof}
By \cor{plus}, $P^+$ is in decent ntyft format and $P^+ \vdash \alpha
\Leftrightarrow P \vdash_{ws} \alpha$ for all closed literals $\alpha$.
Using this, ``if'' follows immediately from \lem{prov}, and ``only if''
from \pr{ruloid}.
\end{proof}
The following definition assigns to each term and each observation in
$\IO_{\it RS}$ a collection of mappings from variables to $\IO_{\it RS}$.
This construct plays a crucial r\^ole in the technical developments to
follow. Intuitively, $t^{-1}_P(\varphi)$ consists of observational
reformulations of the $P$-ruloids with source $t$, that validate the
observation $\varphi$ for the term $\sigma(t)$, for any closed
substitution $\sigma$, in terms of the observations that can be made
for the terms $\sigma(x)$ for $x \in \var(t)$.

\begin{definition}{inverse}
Let $P=(\Sigma,R)$ be a standard TSS in ready simulation format.
Then $\cdot^{-1}_P:\IT(\Sigma)\rightarrow(\IO_{\it RS}\rightarrow
\pow(V\rightarrow\IO_{\it RS}))$ is defined by:
\begin{itemize}
\item $t^{-1}_P (\top) = \{\psi\}$ with $\psi(x)=\top$ for $x \in V$.
\item $\psi \in t^{-1}_P(\widetilde{a})$ iff there is a $P$-ruloid
	$\frac{H}{t\gonotto{a}}$ and
	$\psi: V \rightarrow \IO$ is given by $$\begin{array}{ll}
        \displaystyle \psi(x) = \bigwedge_{(x \gonotto{b})
	\in H}\!\!\!\!\!\widetilde{b} \,\,\,\, \wedge
	\bigwedge_{(x\goto{c}y)\in H}\!\!\!\!\! c\top
        & \mbox{~for~} x \in \var(t)\\
	\psi(x) = \top & \mbox{~for~} x \not\in \var(t).\end{array}$$
\item $\psi \in t^{-1}_P(a\varphi)$ iff there are a $P$-ruloid
	$\frac{H}{t\goto{a} u}$ and a $\chi \in u^{-1}_P (\varphi)$ and
	$\psi: V \rightarrow \IO$ is given by $$\begin{array}{ll}
        \displaystyle \psi(x) = \chi(x) \wedge \bigwedge_{(x \gonotto{b})
	\in H}\!\!\!\!\!\widetilde{b} \,\,\,\, \wedge
	\bigwedge_{(x\goto{c}y)\in H}\!\!\!\!\! c\chi(y)
        & \mbox{~for~} x \in \var(t)\\
	\psi(x) = \top & \mbox{~for~} x \not\in \var(t).\end{array}$$
\item $t^{-1}_P (\bigwedge_{i\in I}\varphi_i)=\{\bigwedge_{i\in I}
	\psi_i \mid \psi_i \in t^{-1}_P (\varphi_i) \mbox{ for } i\in I\}$.
% \item $t^{-1}_P (\bigvee_{i\in I}\varphi_i) = \bigcup_{i \in I}
%	t^{-1}_P (\varphi_i )$. 
% \item $\psi \in t^{-1}_P (\neg \varphi )$ iff there is a function $h:
%	t^{-1}_P (\varphi ) \rightarrow V(t)$, and $\psi: V \rightarrow \IO$
%	is given by $$\psi(x) = \neg \bigvee h^{-1}_P(x) \mbox{ for } x\in V.$$
\end{itemize}
When clear from the context, the subscript $P$ will be omitted.
\end{definition}
It is not hard to see that if $\psi\in t^{-1}(\varphi)$ then
$\psi(x)\cong\top$ for all $x\not\in\var(t)$.

\begin{lemma}{inverse} Let $P=(\Sigma,R)$ be a standard TSS in ready simulation
format.\newline Let $\varphi \in \IO$.
For any term $t \in \IT(\Sigma)$ and closed substitution $\sigma: V
\rightarrow T(\Sigma)$ one has
$$\sigma(t) \models_P \varphi ~~\Leftrightarrow~~ \exists \psi \in t^{-1}_P
(\varphi)~\forall x\in \var(t): \sigma(x) \models_P \psi(x).$$
\end{lemma}
\pf With induction on the structure of $\varphi$.
\begin{itemize}
	\item $\varphi = \top$. In this case $\sigma(t) \models \top$
and for the only parameterized observation $\psi \in t^{-1} (\top)$ one
has $\sigma(x) \models \psi(x)=\top$ for all $x \in \var(t)$.
	\item Suppose $\sigma(t)\models \widetilde{a}$.
Then $P \vdash_{ws} \sigma(t) \gonotto{a}$, by \df{preorders}.
Thus, by \lem{ruloid} there must be a $P$-ruloid $\frac{H}{t \gonotto{a}}$
and a closed substitution $\sigma'$ with $P \vdash_{ws} \sigma'(H)$ and
$\sigma'(t)=\sigma(t)$, i.e.\ $\sigma'(x) = \sigma(x)$ for $x \in \var(t)$.
Define $\psi$ as indicated in \df{inverse}.
By definition, $\psi \in t^{-1} (\widetilde{a})$. 
Let $x \in \var(t)$. % We have to show that $\sigma(x) \models \psi(x)$.
For $(x\goto{c}y)\in H$ one has $P \vdash_{ws} \sigma'(x) \goto{c}
\sigma'(y)$, so $\sigma'(x) \models c\top$.
Moreover, for $(x\gonotto{b})\in H$ one has $P \vdash_{ws} \sigma'(x)
\gonotto{b}$, so $\sigma'(x) \models \widetilde{b}$.
Hence $\sigma(x) = \sigma'(x) \models \psi(x)$.

\hspace{15pt}
Now suppose that there is a $\psi \in t^{-1} (\widetilde{a})$ such that
$\sigma(x) \models \psi(x)$ for all $x \in \var(t)$. This means that there
is a $P$-ruloid $\displaystyle \frac{\{x\goto{a_i}y_i\mid i\in
I_x,~x\in\var(t)\} \cup \{x\gonotto{b_j}\mid j\in J_x,~x\in\var(t)\}}
{t \gonotto{a}}$ such that
$\displaystyle \sigma(x) \models \bigwedge_{j \in J_x}
\widetilde{b_j} \wedge \bigwedge_{i\in I_x} a_i \top$ for all $x
\in \var(t)$. Thus, for $x\in \var(t)$ and $i\in I_x$,
$P \vdash_{ws} \sigma(x) \goto{a_i} t_i$ for some $t_i \in T(\Sigma)$,
and for $x\in \var(t)$ and $j\in J_x$,
\plat{P \vdash_{ws} \sigma(x) \gonotto{b_j}}.
Let $\sigma'$ be a closed substitution with
\begin{tabular}{ll@{}}
$\sigma'(x) = \sigma(x)$ & for $x \in \var(t)$\\
$\sigma'(y_i) = t_i$ & for $i \in I_x$ and $x\in \var(t)$.
\end{tabular}\linebreak[3]
Here we use that the variables $x$ and $y_i$ are all different.
Now $P \vdash_{ws} \sigma'(x) \goto{a_i} \sigma'(y_i)$ for
$x\in \var(t)$ and $i\in I_x$, and $P \vdash_{ws} \sigma'(x) \gonotto{b_j}$ 
for $x\in \var(t)$ and $j\in J_x$. So by \lem{ruloid}
$P \vdash_{ws} \sigma'(t) \gonotto{a}$, which implies $\sigma(t) =
\sigma'(t) \models \widetilde{a}$.
	\item Suppose $\sigma(t)\models a\varphi$.
Then by \df{preorders} there is
a $t'\in T(\Sigma)$ with $P \vdash_{ws} \sigma(t) \goto{a} t'$ and
$t' \models \varphi$.
Thus, by \lem{ruloid} there must be a $P$-ruloid $\frac{H}{t \goto{a} u}$
and a closed substitution $\sigma'$ with $P \vdash_{ws} \sigma'(H)$,
$\sigma'(t)=\sigma(t)$, i.e. $\sigma'(x) = \sigma(x)$ for $x \in
\var(t)$, and $\sigma'(u)=t'$.
Since $\sigma'(u) \models \varphi$, the induction hypothesis can be
applied, and there must be a $\chi \in u^{-1} (\varphi)$ such that
$\sigma'(z) \models \chi(z)$ for all $z \in \var(u)$. Furthermore
$\sigma'(z) \models \chi(z)\cong\top$ for all $z \not\in \var(u)$.
Now define $\psi$ as indicated in \df{inverse}.
By definition, $\psi \in t^{-1} (a\varphi)$. 
Let $x \in \var(t)$. % We have to show that $\sigma(x) \models \psi(x)$.
For $(x\goto{c}y)\in H$ one has $P \vdash_{ws} \sigma'(x) \goto{c} \sigma'(y)
\models \chi(y)$, so $\sigma'(x) \models c\chi(y)$.
Moreover, for $(x\gonotto{b})\in H$ one has $P \vdash_{ws} \sigma'(x)
\gonotto{b}$, so $\sigma'(x) \models \widetilde{b}$.
It follows that $\sigma(x) = \sigma'(x) \models \psi(x)$.

\hspace{15pt}
Now suppose that there is a $\psi \in t^{-1} (a\varphi)$ such that $\sigma(x)
\models \psi(x)$ for all $x \in \var(t)$. This means that there is a $P$-ruloid
$\displaystyle \frac{\{x\goto{a_i}y_i\mid i\in I_x,~x\in\var(t)\} \cup
\{x\gonotto{b_j}\mid j\in J_x,~x\in\var(t)\}}{t \goto{a} u}$
and a parameterized observation $\chi \in u^{-1} (\varphi)$ such that
$\displaystyle \sigma(x) \models \chi(x) \wedge \bigwedge_{j \in J_x}
\widetilde{b_j} \wedge \bigwedge_{i\in I_x} a_i \chi(y_i)$ for all $x
\in \var(t)$. It follows that, for $x\in \var(t)$ and $i\in I_x$,
$P \vdash_{ws} \sigma(x) \goto{a_i} t_i$ for some $t_i \in T(\Sigma)$
with $t_i \models \chi(y_i)$. Let $\sigma'$ be a closed substitution with
\begin{tabular}{@{~}l@{~~}l@{}}
$\sigma'(x) = \sigma(x)$ & for $x \in \var(t)$\\
$\sigma'(y_i) = t_i$ & for $i \!\in\! I_x$ and $x\in \var(t)$.
\end{tabular}
Here we use that the variables $x$ and $y_i$ are all different.
Now $\sigma'(z) \models \chi(z)$ for $z\in \var(u)$, using that $u$
contains only variables that occur in the premises of the ruloid.
Thus the induction hypothesis can be applied, and $\sigma'(u) \models
\varphi$. Moreover, $P \vdash_{ws} \sigma'(x) \goto{a_i} \sigma'(y_i)$ for
$x\in \var(t)$ and $i\in I_x$, and $P \vdash_{ws} \sigma'(x) \gonotto{b_j}$ 
for $x\in \var(t)$ and $j\in J_x$. So by \lem{ruloid} $P \vdash_{ws}
\sigma'(t) \goto{a} \sigma'(u)$, which implies $\sigma(t) = \sigma'(t)
\models a\varphi$.
	\item $\sigma(t) \models \bigwedge_{i\in I}\varphi_i
~\Leftrightarrow~ \forall i\!\in\! I: \sigma(t) \models \varphi_i
~\Leftrightarrow~ \forall i\!\in\! I ~\exists \psi_i \!\in\! t^{-1} 
(\varphi_i) ~\forall x\!\in\!\var(t): \sigma(x) \models \psi_i(x) ~\Leftrightarrow$
$\Leftrightarrow~ \exists \psi \!\in\! t^{-1}(\bigwedge_{i\in I}\varphi_i)
~\forall x\!\in\! \var(t): \sigma(x) \models \psi(x).$
\hfill$\Box$
%	 \item $\sigma(t) \models \bigvee_{i\in I}\varphi_i
% ~~\Leftrightarrow~~ \exists i\in I: \sigma(t) \models \varphi_i
% ~~\Leftrightarrow~~ \exists i\in I ~\exists \psi_i \in t^{-1} 
% (\varphi_i) ~\forall x\in V: \sigma(x) \models \psi_i(x) ~~\Leftrightarrow$
% $\Leftrightarrow~~ \exists \psi \in t^{-1}(\bigvee_{i\in I}\varphi_i)
% ~\forall x\in V: \sigma(x) \models \psi(x).$
% 	\item $\sigma(t) \models \neg\varphi
% ~~\Leftrightarrow~~ \sigma(t) \not\models \varphi
% ~~\Leftrightarrow~~ \forall \psi \!\in\! t^{-1}
% (\varphi) ~\exists x\!\in\! V(t): \sigma(x) \not\models \psi(x)
% ~~\Leftrightarrow~~ \exists h: t^{-1}(\varphi)\rightarrow V(t)$
% $\forall x\!\in\! V(t): \sigma(x) \models \neg\bigvee h^{-1}(x)
% ~~\Leftrightarrow~~ \exists \psi \!\in\! t^{-1}(\neg\varphi)
% ~\forall x\!\in\! V: \sigma(x) \models \psi(x).$\hfill $\Box$
\end{itemize}

\begin{example}{inverse}
Let $A=\{a,b\}$ and let $\Sigma$ consist of the constant $c$ and
the unary function symbol $f$. Let $R$ be:
\[
c\goto a c
\hspace*{1cm}
\frac{x\goto{a} y}{f(x)\goto{b}y}
\hspace*{1cm}
\Rule{x\goto{b} y}{f(x)\goto{a}f(y)}
\]
Suppose $\psi\in f(f(x))^{-1}(ba\top)$. The only $P$-ruloid with
a conclusion of the form $f(f(x))\goto{b}\_$ is
$\displaystyle\frac{x\goto{b}y}{f(f(x))\goto{b}f(y)}$.\vspace{-1ex}
So $\psi(x)=\chi(x)\wedge b\chi(y)$ with $\chi\in f(y)^{-1}(a\top)$.
The only $P$-ruloid with a conclusion $f(y)\goto{a}\_$ is
$\displaystyle\frac{y\goto{b}z}{f(y)\goto{a}f(z)}$.
So $\chi(y)=\chi'(y)\wedge b\chi'(z)$ with $\chi'\in f(z)^{-1}(\top)$.
Since $\chi'(y)=\chi'(z)=\top$ we have $\chi(y)\cong b\top$. Moreover
$x\not\in\var(f(y))$ implies $\chi(x)\cong\top$. Hence
$\psi(x)\cong bb\top$.

By \lem{inverse} a closed term $f(f(u))$ can execute a $b$ followed
by an $a$ iff the closed term $u$ can execute two consecutive $b$'s.
\end{example}
In order to arrive at the desired precongruence results we need to
know that if a standard TSS is in the desired format $N$, and
$\varphi$ is a potential $N$-observation of a closed term $\sigma(t)$,
then the observations of $\sigma(x)$ for $x \in \var(t)$ that determine
whether or not $\varphi$ is an $N$-observation of $\sigma(t)$ are also
$N$-observations. This is established in the following two lemmas.
In fact, our formats have been found by investigating what was needed
to make these lemmas hold. 

The work is divided over two lemmas. \lem{formula preservation} deals
with those variables of $t$ that are floating in the $P$-ruloids with
source $t$. As the proof inductively refers to terms that can be
thought of as successors of $t$ after performing a number of actions,
and as these terms may contain variables $y$ representing successors of
the arguments of $t$ after performing several actions, the observations
employed should be the ones from \df{sublanguages}. \lem{formula
preservation frozen} extends this to arbitrary variables. As
non-floating variables represent processes in their initial state only,
that lemma may use the richer language of observations employed in
\df{conjunctive sublanguages}, which is much less cumbersome. This
enables the absence of restrictions on non-floating variables in the
definitions of the formats.

\begin{lemma}{formula preservation}
Let $P=(\Sigma,R)$ be a standard TSS in ready simulation format, and let
$\Lambda$ be an unary predicate on arguments of function symbols in $\Sigma$.
Let $t \in \IT(\Sigma)$, $\varphi \in \IO_{\it RS}$, $\psi \in
t^{-1}_P(\varphi)$ and $x \in \var(t)$, such that $x$ occurs only once
in $t$, and at a $\Lambda$-liquid position.
\begin{itemise}
\item
If the rules in $R^+$ are $\Lambda$-ready trace safe and
$\varphi \in \IO_{\it RT}$ then $\psi(x) \in \IO_{\it RT}$.
\item
If the rules in $R^+$ are $\Lambda$-readiness safe and
$\varphi \in \IO_{\it R}$ then $\psi(x) \in \IO_{\it R}$.
\item
If the rules in $R^+$ are $\Lambda$-failure trace safe and
$\varphi \in \IO_{\it FT}$ then $\psi(x) \in \IO_{\it FT}$.
\item
If the rules in $R^+$ are $\Lambda$-failure trace safe and
$\varphi \in \IO_{\it F}$ then $\psi(x) \in \IO_{\it F}$.
\end{itemise}
\end{lemma}

\pf
Let $P=(\Sigma,R)$ be a standard TSS in ready simulation format, $t \in
\IT(\Sigma)$ and $x \in \var(t)$. 
Note that if $\psi \in t^{-1}(\widetilde{a})$ then $\psi(x)$ has the
form $\bigwedge_{j \in J} \widetilde{b_j} \wedge \bigwedge_{k \in K}
c_k \top$. Thus if $\psi \in t^{-1}( \bigwedge_{i \in I}\widetilde{a_i})$
then $\psi(x)$ has the form $\bigwedge_{i \in I}(\bigwedge_{j \in J_i}
\widetilde{b_{ij}} \wedge \bigwedge_{k \in K_i} c_{ik} \top)$. The
latter formula is equivalent to one of the form $\bigwedge_{j \in J'}
\widetilde{b'_j} \wedge \bigwedge_{k \in K'} c'_k \top$.

Furthermore note that if $\psi \in t^{-1}(a\top)$ then $\psi(x)$ has
the form $\top \wedge \bigwedge_{j \in J} \widetilde{b_j} \wedge
\bigwedge_{k \in K} c_j \top \cong
\bigwedge_{j \in J} \widetilde{b_j} \wedge \bigwedge_{k \in K} c_k \top$.
Here it is used that if $\chi \in u^{-1}(\top)$ for some $u \in
\IT(\Sigma)$ then $\chi(x) = \top$. Thus if $\psi \in t^{-1}(
\bigwedge_{i \in I} a_i \top )$ then $\psi(x)$ is again equivalent to
a formula of the form $\bigwedge_{j \in J'}
\widetilde{b'_j} \wedge \bigwedge_{k \in K'} c'_k \top$.

Combining these observations it follows that if $\psi \in t^{-1}(
\bigwedge_{i \in I}\widetilde{a_i} \wedge \bigwedge_{j \in J} b_j \top )$
then also $\psi(x)$ is equivalent to a formula of the form above.
\begin{itemise}
\item Let the rules in $R^+$ be $\Lambda$-ready trace safe
and $\varphi \in \IO_{\it RT}$. We apply structural induction on $\varphi$.
Take $t \in \IT(t)$, $\psi \in t^{-1}(\varphi)$ and $x \in \var(t)$,
such that $x$ occurs only once in $t$, and at a $\Lambda$-liquid position.
\begin{itemise2}
\item In case $\varphi = \top$ we have $\psi(x)=\top \in \IO_{\it RT}$.
\item Let $\varphi = \bigwedge_{i \in I} \widetilde{a_i} \wedge
\bigwedge_{j \in J} b_j\top \wedge \varphi'$ with $\varphi' \in \IO_{\it RT}$.
Then $\psi(x) \cong \bigwedge_{k \in K} \widetilde{c_k} \wedge
\bigwedge_{\ell \in L} d_\ell \top \wedge \psi'(x)$, where $\psi' \in
t^{-1}(\varphi')$. By induction $\psi'(x) \in \IO_{\it RT}$, and
hence also $\psi(x) \in \IO_{\it RT}$.
\item Let $\varphi = a\varphi'$ with $\varphi' \in \IO_{\it RT}$.
Then there are a $P$-ruloid $\frac{H}{t\goto{a} u}$ and $\chi \in
u^{-1} (\varphi')$ such that
	$$\psi(x) = \chi(x) \wedge \bigwedge_{(x \gonotto{b})
	\in H}\!\!\!\!\!\widetilde{b} \,\,\,\, \wedge
	\bigwedge_{(x\goto{c}y)\in H}\!\!\!\!\! c\chi(y).$$
By \lem{preservation-ready-trace}, $x$ is propagated at most once in
$\frac{H}{t \goto{a} u}$, and only at a $\Lambda$-liquid position.
This implies that of the set of variables $W= \{x\} \cup \{y \mid \exists
c: (x \goto{c} y) \in H\}$ at most one member, say $z$, occurs in $\var(u)$.
Since $z$ is $\Lambda$-floating in $\frac{H}{t\goto{a} u}$ and
occurs at most once in $t$,
\lem{preservation-ready-trace} moreover guarantees that $z$ occurs
only once in $u$, and at a $\Lambda$-liquid position. By induction we
obtain $\chi(z) \in \IO_{\it RT}$. For all other variables $w \in W$
we have $w\not\in\var(u)$ and so $\chi(w)\cong\top$.
Thus $\psi(x)$ is of the form $\bigwedge_{j
\in J} \widetilde{b_j} \wedge \bigwedge_{k \in K} c_k \top$ or
$\chi(z) \wedge \bigwedge_{j \in J} \widetilde{b_j} \wedge
\bigwedge_{k \in K} c_k \top$ or $\bigwedge_{j \in J} \widetilde{b_j}
\wedge \bigwedge_{k \in K} c_k \top \wedge c \chi(z)$ with $\chi(z)
\in \IO_{\it RT}$. In all three cases $\psi(x) \in \IO_{\it RT}$.
\end{itemise2}
\item Let the rules in $R^+$ be $\Lambda$-readiness safe
and $\varphi \in \IO_{\it R}$. We apply structural induction on $\varphi$.
Take $t \in \IT(t)$, $\psi \in t^{-1}(\varphi)$ and $x \in \var(t)$,
such that $x$ occurs only once in $t$, and at a $\Lambda$-liquid position.
\begin{itemise2}
\item In case $\varphi = \top$ we have $\psi(x)=\top \in \IO_{\it R}$.
\item Let $\varphi = \bigwedge_{i \in I} \widetilde{a_i} \wedge
\bigwedge_{j \in J} b_j\top$.
Then $\psi(x) \cong \bigwedge_{k \in K} \widetilde{c_k} \wedge
\bigwedge_{\ell \in L} d_\ell \top \in \IO_{\it R}$.
\item Let $\varphi = a\varphi'$ with $\varphi' \in \IO_{\it R}$.
Then there are a $P$-ruloid $\frac{H}{t\goto{a} u}$ and $\chi \in
u^{-1} (\varphi')$ such that
	$$\psi(x) = \chi(x) \wedge \bigwedge_{(x \gonotto{b})
	\in H}\!\!\!\!\!\widetilde{b} \,\,\,\, \wedge
	\bigwedge_{(x\goto{c}y)\in H}\!\!\!\!\! c\chi(y).$$
By \lem{preservation-ready-trace}, $x$ is propagated at most once in
$\frac{H}{t \goto{a} u}$, and only at a $\Lambda$-liquid position.
Moreover, by \lem{preservation-readiness}, $x$ is not both propagated
and polled in $\frac{H}{t \goto{a} u}$. We consider three cases.
\begin{itemise3}
\item
Suppose $x \in \var(u)$. Then $x$ is propagated, so it occurs only
once in $u$, and at a $\Lambda$-liquid position. By induction $\chi(x)
\in \IO_{\it R}$. Furthermore, $H$ has no premises of the form
\plat{x\gonotto{b}} or $x\goto{c}y$.  Hence $\psi(x) \cong \chi(x)\in \IO_{\it R}$.
\item
Suppose $x$ is propagated, but does not occur in $u$. Then $\chi(x)\cong\top$.
Furthermore, $H$ contains no premises of the form \plat{x
\gonotto{b}} and exactly one of the form $x \goto{c} y$, where $y$
occurs in $u$. So $\psi(x) \cong c \chi(y)$.
As $y$ is $\Lambda$-floating in $\frac{H}{t\goto{a} u}$ and does
not occur in $t$, \lem{preservation-ready-trace} guarantees that $y$ occurs
only once in $u$, and at a $\Lambda$-liquid position. By induction
$\chi(y) \in \IO_{\it R}$, so $\psi(x) \in \IO_{\it R}$.
\item
Suppose $x$ is not propagated. Then none of the variables $y$ with
$(x\goto{c}y)\in H$ for some $c \in A$ occurs in $u$. Hence for all
those variables we have $\chi(y)\cong\top$. Moreover $x\not\in\var(u)$
so $\chi(x)\cong\top$.
Thus $\psi(x)$ is of the form $\bigwedge_{j\in J} \widetilde{b_j}
\wedge \bigwedge_{k \in K} c_k \top \in \IO_{\it R}$.
\end{itemise3}
\end{itemise2}
\item Let the rules in $R^+$ be $\Lambda$-failure trace safe
and $\varphi \in \IO_{\it FT}$. We apply structural induction on $\varphi$.
Take $t \in \IT(t)$, $\psi \in t^{-1}(\varphi)$ and $x \in \var(t)$,
such that $x$ occurs only once in $t$, and at a $\Lambda$-liquid position.
\begin{itemise2}
\item In case $\varphi = \top$ we have $\psi(x)=\top \in \IO_{\it FT}$.
\item Let $\varphi = \bigwedge_{i \in I} \widetilde{a_i} \wedge
\varphi'$ with $\varphi' \in \IO_{\it FT}$.
Then $\psi(x) = \bigwedge_{i \in I} \psi_i(x) \wedge \psi'(x)$ where
$\psi_i(x) \in t^{-1}(\widetilde{a_i})$ for $i \in I$ and $\psi'(x)
\in t^{-1}(\varphi')$. For $i \in I$
there is a $P$-ruloid $\frac{H}{t\gonotto{a_i}}$ such that
	$$\psi_i(x) = \bigwedge_{(x \gonotto{b})
	\in H}\!\!\!\!\!\widetilde{b} \,\,\,\, \wedge
	\bigwedge_{(x\goto{c}y)\in H}\!\!\!\!\! c\top.$$
By \lem{preservation-negative-failure-trace}, $H$ has no premises of
the form $x \goto{c} y$. Therefore $\psi_i(x) \cong \bigwedge_{j \in J}
\widetilde{b_j}$. Furthermore, by induction $\psi'(x) \in \IO_{\it FT}$,
so $\psi(x) = \bigwedge_{i \in I} \psi_i(x) \wedge \psi'(x)\in\IO_{\it FT}$.
\item Let $\varphi = a\varphi'$ with $\varphi' \in \IO_{\it FT}$.
Then there are a $P$-ruloid $\frac{H}{t\goto{a} u}$ and $\chi \in
u^{-1} (\varphi')$ such that
	$$\psi(x) = \chi(x) \wedge \bigwedge_{(x \gonotto{b})
	\in H}\!\!\!\!\!\widetilde{b} \,\,\,\, \wedge
	\bigwedge_{(x\goto{c}y)\in H}\!\!\!\!\! c\chi(y).$$
Using Lemmas \hhref{lem-preservation-ready-trace} and
\hhref{lem-preservation-readiness}, the same case distinction as in the
readiness case applies, and in the first two cases we find $\psi(x) \in
\IO_{\it FT}$. Consider the third case, in which $x$ is not propagated.
Then $x\not\in\var(u)$ so $\chi(x)\cong\top$.
By \lem{preservation-failure-trace}, $x$ is polled at most once
in $\frac{H}{t\goto{a}u}$, and in a positive premise.  Hence $H$
contains no literals of the form \plat{x \gonotto b}, and no more than
one literal \plat{x \goto{c} y}. If there is such a literal, $y$ does
not occur in $u$, so $\chi(y)\cong\top$. Hence $\psi(x) \cong \top$ or
$\psi(x) \cong c\top$. In both cases $\psi(x) \in \IO_{\it FT}$.
\end{itemise2}
\item The proof of the last statement (with $\varphi \in \IO_{\it F}$) is a trivial
simplification of the previous one. This proof is left to the reader.
\hfill $\Box$
\end{itemise}

\begin{lemma}{formula preservation frozen}
Let $P=(\Sigma,R)$ be a standard TSS in ready simulation format,
$t \in \IT(\Sigma)$, $\varphi \in \IO_{\it RS}$, $\psi \in
t^{-1}_P(\varphi)$ and $x \in \var(t)$.
Then $\psi(x) \in \IO_{\it RS}^\wedge$ and the following hold:
\begin{itemise}
\item
If $P$ is in ready trace format and
$\varphi \in \IO_{\it RT}^\wedge$ then $\psi(x) \in \IO_{\it RT}^\wedge$.
\item
If $P$ is in readiness format and
$\varphi \in \IO_{\it R}^\wedge$ then $\psi(x) \in \IO_{\it R}^\wedge$.
\item
If $P$ is in failure trace format and
$\varphi \in \IO_{\it FT}^\wedge$ then $\psi(x) \in \IO_{\it FT}^\wedge$.
\item
If $P$ is in failure trace format and
$\varphi \in \IO_{\it F}^\wedge$ then $\psi(x) \in \IO_{\it F}^\wedge$.
\end{itemise}
\end{lemma}

\pf That $\psi(x) \in \IO_{\it RS} = \IO_{\it RS}^\wedge$ is immediate
from the definitions.
\begin{itemise}
\item Let $P$ be in ready trace format
and $\varphi \in \IO_{\it RT}$. We apply structural induction on $\varphi$.
Take $t \in \IT(t)$, $\psi \in t^{-1}(\varphi)$ and $x \in \var(t)$.
\begin{itemise2}
\item In case $\varphi = \top$ we have $\psi(x)=\top \in \IO_{\it
RT} \subseteq \IO_{\it RT}^\wedge$.
\item Let $\varphi = \bigwedge_{i \in I} \widetilde{a_i} \wedge
\bigwedge_{j \in J} b_j\top \wedge \varphi'$ with $\varphi' \in
\IO_{\it RT}$.
Then $\psi(x) \cong \bigwedge_{k \in K} \widetilde{c_k} \wedge
\bigwedge_{\ell \in L} d_\ell \top \wedge \psi'(x)$, where $\psi' \in
t^{-1}(\varphi')$. Clearly $\widetilde{c_k} \in \IO_{\it RT}$ for $k \in
K$ and $d_\ell\top \in \IO_{\it RT}$ for $\ell \in L$. By induction
$\psi'(x) \in \IO_{\it RT}^\wedge$, and hence also $\psi(x) \in
\IO_{\it RT}^\wedge$. 
\item Let $\varphi = a\varphi'$ with $\varphi' \in \IO_{\it RT}$.
Then there are a $P$-ruloid $\frac{H}{t\goto{a} u}$ and $\chi \in
u^{-1} (\varphi')$ such that
	$$\psi(x) = \chi(x) \wedge \bigwedge_{(x \gonotto{b})
	\in H}\!\!\!\!\!\widetilde{b} \,\,\,\, \wedge
	\bigwedge_{(x\goto{c}y)\in H}\!\!\!\!\! c\chi(y).$$
As $P$ is in ready trace format, by \cor{plus} so is $P^+$,
so the rules in $R^+$
are $\Lambda$-ready trace safe for some unary predicate on arguments
of function symbols $\Lambda$. The variables $y$ with $(x \goto{a} y)
\in H$ are $\Lambda$-floating in $\frac{H}{t\goto{a} u}$ and do not occur in $t$.
Thus, by \lem{preservation-ready-trace}, each of them occurs at most
once in $u$, and at a $\Lambda$-liquid position. In case $y$ does not
occur in $u$ we have $\chi(y) \cong \top \in \IO_{\it RT}$; otherwise
\lem{formula preservation} yields $\chi(y) \in \IO_{\it RT}$.
Hence $c\chi(y) \in \IO_{\it RT}$ for each $(x \goto{c} y) \in H$.
Clearly $\bigwedge_{(x \gonotto{b})\in H}\widetilde{b} \in \IO_{\it RT}$,
and by induction $\chi(x) \in \IO_{\it RT}^\wedge$. Therefore $\psi(x)
\in \IO_{\it RT}^\wedge$.
\item Let $\varphi = \bigwedge_{i\in I}\varphi_i$ with $\varphi_i \in
\IO_{\it RT}$ for $i \in I$. Then $\psi(x) = \bigwedge_{i \in I}
\psi_i(x)$ with $\psi_i(x) \in t^{-1}(\varphi_i)$ for $i \in I$.
By induction (using the three cases treated above) $\psi_i(x) \in
\IO_{\it RT}^\wedge$ for $i \in I$. Therefore $\psi(x) \in \IO_{\it
RT}^\wedge$.
\end{itemise2}
\item The proofs of the remaining three statements proceed in exactly
the same way.
% , except for the cases $\varphi = \bigwedge_{i \in I}
% $\widetilde{a_i} \wedge \varphi'$ with $\varphi' \in \IO_{\it
% FT}^\wedge$, and $\varphi = \bigwedge_{i \in I} \widetilde{a_i}$ when
% dealing with $\IO_{\it F}^\wedge$, which proceed as in the proof of
% \lem{formula preservation}.
\hfill $\Box$
\pagebreak[3]
\end{itemise}

\begin{trivlist}
\item
Now we are ready to prove Theorems
\hhref{thm-ready-simulation}--\hhref{thm-failure-trace}.
In the light of \df{precongruence} these theorems can be reformulated
as in the following corollary, where $N$ ranges over $\{$ready
simulation, ready trace, readiness, failure trace, failure$\}$ and
{\em failure format} means failure trace format.
\end{trivlist}

\begin{corollary}{congruence}{\bf (Precongruence)}
Let $P=(\Sigma,R)$ be a standard TSS in $N$ format, $t \in \IT(\Sigma)$ and $\sigma,
\sigma'$ closed substitutions.
If $\sigma(x) \sqsubseteq_N^P \sigma'(x)$ for $x \in \var(t)$, then
$\sigma(t) \sqsubseteq_N^P \sigma'(t)$.
\end{corollary}

\begin{proof}
By \cor{modal} we have to show that 
$\fO_N^\wedge(\sigma(x)) \subseteq \fO_N^\wedge(\sigma'(x))$ for $x
\!\in\! \var(t)$ implies $\fO_N^\wedge(\sigma(t)) \subseteq
\fO_N^\wedge(\sigma'(t))$.
Suppose $\fO_N^\wedge(\sigma(x)) \subseteq \fO_N^\wedge(\sigma'(x))$
for $x \!\in\! \var(t)$. Let $\varphi \in \fO_N^\wedge(\sigma(t))$, i.e.\
$\varphi \!\in\! \IO_N^\wedge$ and $\sigma(t) \!\models \!\varphi$.
By \lem{inverse} $\exists \psi \in t^{-1}
(\varphi) ~\forall x\!\in\! \var(t): \sigma(x) \models \psi(x)$.
By \lem{formula preservation frozen} $\forall x \!\in\! \var(t): \psi(x)
\in \IO_N^\wedge$. Thus $\forall x \!\in\! \var(t): \psi(x) \in
\fO_N^\wedge (\sigma(x)) \subseteq \fO_N^\wedge (\sigma'(x))$.
Hence $\forall x \!\in\! \var(t): \sigma'(x) \models \psi(x)$.
So by \lem{inverse} $\sigma'(t) \models \varphi$.
It follows that $\varphi \in \fO_N^\wedge (\sigma'(t))$, which had to
be proved.
\end{proof}

%%%%%%%%%%%%%%%%%%%%%%%%%%%%%%%%%%%%%%%%%%%%%%%%%%%%%%%%%%%%%%%%%%%%%%%%%%%%%%
\section{Counterexamples}
\hlabel{sec-counterexamples}

This section presents a string of counterexamples of complete standard
TSSs in ntyft/ntyxt format, to show that the syntactic restrictions of
our precongruence formats are essential. In \cite{GrV92} a series of
counterexamples can be found showing that the syntactic restrictions
of the ntyft/ntyxt format are essential as well.

\subsection{Basic process algebra}

The examples in this section assume basic process algebra \cite{BK84}.
We assume a collection ${\it Act}$ of constants, called {\em atomic actions},
representing indivisible behaviour, and two special constants:
the {\em deadlock} $\delta$ does not display any behaviour, while the
{\em empty process} $\varepsilon$ \cite{Vra97} terminates successfully.
Basic process algebra moreover includes function symbols $\_+\_$ and
$\_\cdot\_$ of arity two, called {\em alternative composition} and
{\em sequential composition}, respectively. Intuitively, $t_1+t_2$
executes either $t_1$ or $t_2$, while $t_1\cdot t_2$ first executes $t_1$
and upon successful termination executes $t_2$. These intuitions are made precise
by means of the standard transition rules for BPA$_{\delta\varepsilon}$ presented
in \tab{bpa}, where the label $v$ ranges over the set ${\it Act}$ of
atomic actions together with a special label $\surd$, representing
successful termination.
Note that this TSS is positive and in ready simulation format.
It is not hard to check that the TSS is in failure trace format (and
so by default in ready trace and readiness format),
if we take at least the first argument
of sequential composition to be liquid. In particular, in the first
rule for sequential composition the floating variable $y$ occurs in a liquid
argument of the target, and in the second rule for sequential composition the
floating variable $x_1$ is polled only once and not propagated.

\begin{table}[ht]
\centering
$
\begin{array}{|cc|}
\hline
&\\
v\goto v \varepsilon\hspace{5mm}(v\not=\surd)\hspace{5mm}&
\varepsilon\goto\surd\delta\\
&\\
\Rule{x_1\goto v y}{x_1+x_2\goto v y}\hspace{5mm}&
\Rule{x_2\goto v y}{x_1+x_2\goto v y}\\
&\\
\Rule{x_1\goto v y}{x_1\cdot x_2\goto v y\cdot x_2}\hspace{5mm}(v\not=\surd)\hspace{5mm}&
\Rule{x_1\goto \surd y_1~~~~x_2\goto v y_2}{x_1\cdot x_2\goto v y_2}\\
&\\
\hline
\end{array}
$
\caption{Transition rules for BPA$_{\delta\varepsilon}$}
\hlabel{tab-bpa}
\end{table}

Terms $t_1\cdot t_2$ are abbreviated to $t_1 t_2$.  Brackets are used
for disambiguation only, assuming associativity of $+$ and $\cdot$,
and letting $\cdot$ bind stronger than $+$.  In the remainder of this
section we assume that ${\it Act}=\{a,b,c,d\}$. Moreover, we assume unary
function symbols $f$ and $h$ and a binary function symbol $g$.

\subsection{Lookahead}

The following counterexample shows that the ready simulation
format (and its more restrictive analogues) cannot allow lookahead.

\begin{example}{lookahead}
We extend BPA$_{\delta\varepsilon}$ with the following rule,
containing lookahead:
\[
\frac{x\goto b y_1\hspace{5mm}y_1\goto c y_2}{f(x)\goto a \delta}
\]
It is easy to see that $bd\sqsubseteq_{\it RS}bc+bd$
(so {\em a fortiori} $bd\sqsubseteq_N bc+bd$ for
$N\in\{{\it RT},{\it R},{\it FT},{\it F}\}$).
The empty trace is a completed trace of $f(bd)$
but not of $f(bc+bd)$ (as $f(bc+bd)\goto a \delta$).
Hence, $f(bd)\not\sqsubseteq_{\it CT}f(bc+bd)$
(and {\em a fortiori} $f(bd)\not\sqsubseteq_N f(bc+bd)$
for $N\in\{{\it RS},{\it RT},{\it R},{\it FT},{\it F}\}$).
\end{example}

\subsection{Multiple propagations}

The following counterexample shows that the
ready trace format (and its more restrictive analogues) cannot allow
a liquid argument of the source
to be propagated more than once in the left-hand sides of
the positive premises. 

\begin{example}{ex2}
Let the arguments of $f$ and $g$ be liquid.
We extend BPA$_{\delta\varepsilon}$ with the rules:
\[
\frac{x\goto a y}{f(x)\goto a f(y)}
\hspace{1cm}
\frac{x\goto b y_1\hspace{5mm}x\goto b y_2}{f(x)\goto b g(y_1,y_2)}
\hspace{1cm}
\frac{x_1\goto c y_1\hspace{5mm}x_2\goto d y_2}{g(x_1,x_2)\goto d \delta}
\]
In the second rule, the liquid argument $x$ of
the source is propagated in the left-hand sides of both
premises.

It is easy to see that $a(bc+bd)\sqsubseteq_{\it RT}abc+abd$
(so {\em a fortiori} $a(bc+bd)\sqsubseteq_N abc+abd$ for
$N\in\{{\it R},{\it FT},{\it F}\}$).
Note that $abd$ is a trace of $f(a(bc+bd))$
(as $f(a(bc+bd))\goto a f(bc+bd)\goto b
g(c,d)\goto d \delta$), but not of $f(abc+abd)$.
Hence, $f(a(bc+bd))\not\sqsubseteq_{T}f(abc+abd)$ (and
{\em a fortiori} $f(a(bc+bd))\not\sqsubseteq_N f(abc+abd)$
for $N\in\{{\it RT},{\it R},{\it FT},{\it F}\}$).
\end{example}
A similar example can be given to show that the ready trace
format cannot allow a liquid argument of the source or a right-hand
side of a positive premise to occur more than once in the target.
Likewise, an example can be given to show that the ready trace
format cannot allow a liquid argument of the source to be propagated
in the left-hand side of a positive premise and at the same time to
occur in the target.

\subsection{Propagation at a non-liquid position}

If in the example above the argument of $f$ were defined
to be frozen, then in the first rule the right-hand side
$y$ of the premise would occur in a non-liquid position in the target.
This shows that the ready trace format cannot allow
right-hand sides of positive premises to occur at
non-liquid positions in the target. Variants of \ex{ex2}
show that the ready trace format cannot allow liquid arguments
of the source to be propagated at non-liquid positions either.
\begin{example}{non-liquid propagation}
Replace the second rule in \ex{ex2} by the two rules
\[
\frac{h(x)\goto b y}{f(x)\goto b y}
\hspace{1cm}
\frac{x\goto b y_1\hspace{5mm}x\goto b y_2}{h(x)\goto b g(y_1,y_2)}
\]
Taking the arguments of $f$ and $g$ liquid, but that of $h$ frozen,
the resulting TSS sins against the ready trace format only in that in
the first rule above the floating variable $x$ is propagated at a
non-liquid position, namely as argument of $h$. Clearly the same mishap
as in \ex{ex2} ensues.\pagebreak[3]
The same argument applies when replacing the second rule in \ex{ex2}
by the two rules
\[
f(x)\goto a h(x)
\hspace{1cm}
\frac{x\goto b y_1\hspace{5mm}x\goto b y_2}{h(x)\goto b g(y_1,y_2)}
\]
\end{example}
The examples above show that if a floating variable $x$ is propagated
in a term $f(x)$, then the argument of $f$ should be classified as
liquid.  Furthermore, if $x$ is propagated in a term $f(h(x))$, then
{\em both} the argument of $f$ and that of $h$ should be classified as
liquid.  Namely, if only the argument of $f$ would be liquid, a rule with
conclusion \plat{h(x) \goto{b} g(x,x)} could have fatal consequences, and
if only the argument of $h$ would be liquid, a rule with conclusion
\plat{f(x) \goto{b} g(x,x)} could be fatal. (It is left to the reader
to fill in the details.) This justifies the definition of
$\Lambda$-liquid in \sect{formats}.

\subsection{Propagation in combination with polling}

The following counterexample shows that the readiness
format cannot allow that a liquid
argument of the source is both propagated and polled.
(The TSS in this example is in a flawed
congruence format for failure equivalence from \cite{vG93a}.)

\begin{example}{ex3}
Let the arguments of $f$ and $h$ be liquid.
We extend BPA$_{\delta\varepsilon}$ with the rules:
\[
\frac{x\goto a y}{f(x)\goto a f(y)}
\hspace{1cm}
\frac{x\goto b y}{f(x)\goto b h(x)}
\hspace{1cm}
\frac{x\goto c y}{h(x)\goto c h(y)}
\hspace{1cm}
\frac{x\goto d y}{h(x)\goto d \delta}
\]
In the second rule, the liquid argument $x$
of the source is both propagated and polled.

It is easy to see that $a(b+cd)+ac$ and
$a(b+c)+acd$ are readiness and
failure equivalent (but not ready trace or failure trace equivalent).
Note that $abcd$ is a trace of $f(a(b+cd)+ac)$
(as \plat{f(a(b+cd)+ac)\goto a f(b+cd)
\goto b h(b+cd)\goto c h(d)\goto d \delta}),
but not of $f(a(b+c)+acd)$. Hence,
$f(a(b+cd)+ac)$ and $f(a(b+c)+acd)$
are not even trace equivalent.

It is easy to see that $a(b+c)+ab+ac$ and $ab+ac$ are failure trace and
failure equivalent (but not ready trace or readiness equivalent).
Note that $abc$ is a trace of $f(a(b+c)+ab+ac)$
(as $f(a(b+c)+ab+ac)\goto a f(b+c) \goto b h(b+c)\goto c h(\varepsilon)$),
but not of $f(ab+ac)$. Hence, $f(a(b+c)+ab+ac)$ and $f(ab+ac)$
are not even trace equivalent.
\end{example}

\subsection{Multiple pollings}

The following counterexample shows that the failure trace
format cannot allow that a liquid argument of the source
is polled more than once.

\begin{example}{ex4}
Let the argument of $f$ be liquid.
We extend BPA$_{\delta\varepsilon}$ with the rules:
\[
\frac{x\goto a y}{f(x)\goto a f(y)}
\hspace{1cm}
\frac{x\goto b y_1\hspace{5mm}x\goto c  y_2}{f(x)\goto d \delta}
\]
In the second rule, the liquid argument $x$
of the source is polled in the two positive premises.

We recall that $a(b+c)+ab+ac$
and $ab+ac$ are failure trace and failure equivalent.
Note that $ad$ is a trace of $f(a(b+c)+ab+ac)$
(as \plat{f(a(b+c)+ab+ac)\goto a f(b+c)
\goto d \delta}), but not of $f(ab+ac)$. Hence,
$f(a(b+c)+ab+ac)$ and $f(ab+ac)$
are not even trace equivalent.
\end{example}

\subsection{Polling at a non-liquid position}

The following variant of \ex{ex4} shows that the failure trace
format cannot allow that a liquid argument of the source is polled
at non-liquid positions.

\begin{example}{ex6}
Let the argument of $f$ be liquid and the argument of $h$ be frozen.
We extend BPA$_{\delta\varepsilon}$ with the rules:
\[
\frac{x\goto a y}{f(x)\goto a f(y)}
\hspace{1cm}
\frac{h(x)\goto b y}{f(x)\goto b \delta}
\hspace{1cm}
\frac{x\goto b y_1\hspace{5mm}x\goto c  y_2}{h(x)\goto d \delta}
\]
In the second rule, the liquid argument $x$
of the source is polled at a non-liquid position.
Clearly the same mishap as in \ex{ex4} ensues.
\end{example}

\subsection{Polling in a negative premise}

The following counterexample shows that the failure trace
format cannot allow that a liquid argument of the source
is polled in a negative premise.

\begin{example}{ex5}
Let the argument of $f$ be liquid.
We extend BPA$_{\delta\varepsilon}$ with the rules:
\[
\frac{x\goto a y}{f(x)\goto a f(y)}
\hspace{1cm}
\frac{x\gonotto b}{f(x)\goto d \delta}
\hspace{1cm}
\frac{x\gonotto c}{f(x)\goto d \delta}
\]
In the second and third rule, the liquid argument $x$
of the source is polled in the negative premise.

We recall that $a(b+c)+ab+ac$
and $ab+ac$ are failure trace and failure equivalent.
Note that $a$ is a completed trace of $f(a(b+c)+ab+ac)$
(as \plat{f(a(b+c)+ab+ac)\goto a f(b+c) \gonotto{d}}), but not of $f(ab+ac)$.
Hence, $f(a(b+c)+ab+ac)$ and $f(ab+ac)$
are not even completed trace equivalent.
\end{example}

%%%%%%%%%%%%%%%%%%%%%%%%%%%%%%%%%%%%%%%%%%%%%%%%%%%%%%%%%%%%%%%%%%%%%%%%%%%%%%
\section{Applications}
\hlabel{sec-applications}

This section contains some applications of our precongruence
formats to TSSs from the literature.

\subsection{Priority}
\hlabel{sec-priority}

{\em Priority} \cite{BBK86} is a unary function symbol that assumes an ordering
on transition labels. The term $\Theta(t)$ executes the transitions of $t$,
with the restriction that a transition $t'\goto a t_1$ only gives rise
to a transition $\Theta(t')\goto a \Theta(t_1)$ if there does not exist
a transition \plat{t'\goto b t_2} with $a<b$. This intuition is
captured by the rule for the priority operator in \tab{priority},
which is added to the rules for BPA$_{\delta\varepsilon}$ in \tab{bpa}.
The resulting standard TSS is in ready simulation format.

\begin{table}[htb]
\centering
$
\begin{array} {|c|}
\hline
\\
\Rule{x\goto v y~~~~x\gonotto w\mbox{ for } v<w}
{\Theta(x)\goto v \Theta(y)}\\
\\
\hline
\end{array}
$
\caption{Transition rule for priority}
\hlabel{tab-priority}
\end{table}

As the floating variable $y$ in the rule for priority is propagated in
$\Theta(y)$, the argument of $\Theta$ has to be liquid. The
floating variable $x$ is propagated only once (and trivially
at a liquid position). Hence the TSS for BPA$_{\delta\varepsilon}$
with priority is in ready trace format.

\begin{corollary}{priority}
The ready simulation and ready trace preorders are precongruences
with respect to BPA$_{\delta\varepsilon}$ with priority.
\end{corollary}
The TSS for BPA$_{\delta\varepsilon}$ with priority is not in readiness
format. Namely, in the case of a non-trivial ordering on transition labels,
the floating variable $x$ in the rule for priority is both propagated
and polled. We show that the readiness, failure trace and failure preorders
are not precongruences with respect to BPA$_{\delta\varepsilon}$ with priority.
Recall that $a(b+c)+acd$ and
$a(b+cd)+ac$ are readiness equivalent.
\vspace{-2pt}
If the ordering consists of $c<b$, then $acd$ is a trace of
$\Theta(a(b+c)+acd)$ (as it can execute $\goto a \Theta(cd)\goto c
\Theta(d)\goto d\Theta(\varepsilon)$), but not of
$\Theta(a(b+cd)+ac)$. Hence, $\Theta(a(b+c)+acd)$ and
$\Theta(a(b+cd)+ac)$ are not even trace equivalent.

Clearly $ab+a(c+d)+a(b+c+d)$ and $ab+a(c+d)$ are failure trace
equivalent. If the ordering consists of $c<b$, then
$\emptyset\,a\,\{c\}\,d\,\emptyset$ is a failure trace of
$\Theta(ab+a(c+d)+a(b+c+d))$ (as it can execute
\plat{\goto a \Theta(b+c+d)\goto d\Theta(\varepsilon)}), but not of
$\Theta(ab+a(c+d))$. Hence, $\Theta(ab+a(c+d)+a(b+c+d))$ and
$\Theta(ab+a(c+d))$ are not failure trace equivalent.

\subsection{Initial priority}

{\em Initial priority} is a unary function symbol that assumes an ordering
on transition labels. The term $\theta(t)$ executes the transitions of $t$,
with the restriction that an initial transition $t\goto a t_1$ only gives
rise to an initial transition $\theta(t)\goto a t_1$ if there does not exist
an initial transition \plat{t\goto b t_2} with $a<b$. This intuition is
captured by the rule for the initial priority operator in \tab{initial},
which is added to the rules for BPA$_{\delta\varepsilon}$ in \tab{bpa}.
The resulting standard TSS is in ready simulation format.

\begin{table}[htb]
\centering
$
\begin{array} {|c|}
\hline
\\
\Rule{x\goto v y~~~~x\gonotto w\mbox{ for } v<w}
{\theta(x)\goto v y}\\
\\
\hline
\end{array}
$
\caption{Transition rule for initial priority}
\hlabel{tab-initial}
\end{table}

We take the argument of initial priority to be frozen. 
Now the TSS for BPA$_{\delta\varepsilon}$ with initial
priority is in failure trace format. Note that in the
rule for initial priority the variable $x$, which
is polled and propagated, is non-floating.

\begin{corollary}{initial-priority}
The ready simulation, ready trace, readiness, failure trace and
failure preorders are precongruences with respect to
BPA$_{\delta\varepsilon}$ with initial priority.
\end{corollary}
In the case of a non-trivial ordering on transition labels,
the rule for initial priority is outside
de Simone's format, due to the presence of negative premises.

\subsection{Binary Kleene star}

The {\em binary Kleene star} $t_1{}^\ast t_2$ \cite{Kle56} repeatedly executes
$t_1$ until it executes $t_2$. This operational behaviour is captured by the
rules in \tab{BKS}, which are added to the rules for
BPA$_{\delta\varepsilon}$ in \tab{bpa}. The resulting positive
TSS is in failure trace format if we take the first argument of
sequential composition to be liquid and the first argument of the
binary Kleene star to be frozen.
Note that in the first rule for the binary Kleene star the floating variable
$y$ is propagated only once, at a liquid position, and not polled.
Moreover in this rule the variable $x_1$, which
is propagated twice, is non-floating.

\begin{table}[htb]
\centering
$
\begin{array} {|cc|}
\hline
&\\
\Rule{x_1\goto v y}{x_1{}^\ast x_2\goto v y\cdot(x_1{}^\ast x_2)}
~(v\not=\surd)&\Rule{x_2\goto v y}{x_1{}^\ast x_2\goto v y}\\
&\\
\hline
\end{array}
$
\caption{Transition rules for the binary Kleene star}
\hlabel{tab-BKS}
\end{table}

\begin{corollary}{iteration}
The $n$-nested simulation (for $n\geq 1$), ready simulation,
ready trace, readiness, failure trace and failure preorders are
precongruences w.r.t.\ BPA$_{\delta\varepsilon}$ with the binary Kleene star.
\end{corollary}
It was noted in \cite{AFI98} that the second rule for the binary Kleene star
does not fit the congruence format for ready trace equivalence from
\cite{vG93a}. Namely, in this rule the variables $x_1$ and $y$ in the target
are connected by the premise. Neither does this rule fit
de Simone's format, due to the fact that
$x_1$ occurs in the left-hand side of the premise and in the target.

\subsection{Sequencing}

{\em Sequencing} $t_1;t_2$ executes $t_1$ until it can do no further
transitions, after which it starts executing $t_2$.
Basic process algebra with sequencing contains the atomic actions in ${\it Act}$,
alternative composition and sequencing, while the deadlock and
the empty process are fused to a single constant {\bf 0}.
The transition rules for BPA$_{\bf 0}^;$ are presented in
\tab{sequencing}, where $v$ and $w$ range over ${\it Act}$.
This standard TSS is in ready simulation format.
The rules for sequencing were taken from \cite{Bl94}.

\begin{table}[ht]
\centering
$
\begin{array}{|cc|}
\hline
&\\
v \goto v {\bf 0} &\\
&\\
\Rule{x_1\goto v y}{x_1+x_2\goto v y}\hspace{3mm}&
\Rule{x_2\goto v y}{x_1+x_2\goto v y}\\
&\\
\Rule{x_1\goto v y}{x_1;x_2\goto v y;x_2}\hspace{3mm}&
\Rule{x_1\gonotto v\mbox{ for }v\in {\it Act}~~~~x_2\goto w y}{x_1;x_2\goto w y}\\
&\\
\hline
\end{array}
$
\caption{Transitions rules for BPA$_{\bf 0}^;$}
\hlabel{tab-sequencing}
\end{table}

As the floating variable $y$ in the first rule for sequencing
is propagated in $y;x_2$, the first argument of sequencing
has to be liquid. In the first rule for sequencing the
floating variable $x_1$ is propagated only once (and trivially
at a liquid position), and not polled. In the second rule for sequencing
the floating variable $x_1$ is not propagated, while the floating variable
$y$ is propagated only once (and trivially at a liquid position), and
not polled. Hence the TSS for BPA$_{\bf 0}^;$ is in readiness format.

\begin{corollary}{sequencing}
The ready simulation, ready trace and readiness preorders are
precongruences with respect to BPA$_{\bf 0}^;$.
\end{corollary}
The TSS for BPA$_{\bf 0}^;$ is not in failure trace
format. Namely, in the second rule for sequencing the floating
variable $x_1$ is polled in negative premises.
We show that the failure preorder is not a precongruence with
respect to BPA$_{\bf 0}^;$.
Let sequencing bind stronger than alternative composition.
Clearly $a;(b+c)+a;c+a$ and $a;(b+c)+a$ are failure equivalent.
Note that $(a,\{b\})$ is a failure pair of $(a;(b+c)+a;c+a);b$
(as it can execute $\goto a {\bf 0};c;b\gonotto b$), but not of $(a;(b+c)+a);b$.
Hence, $(a;(b+c)+a;c+a);b$ and $(a;(b+c)+a);b$ are not failure equivalent.

The failure trace preorder is a precongruence with respect to
BPA$_{\bf 0}^;$. Namely $X_0 a_1 X_1 \cdots a_n X_n$ is a failure
trace of $p;q$ iff either $X_0 a_1 X_1 \cdots a_{m-1} X_{m-1} a_m {\it Act}$
is a failure trace of $p$ for some $m\leq n$ and $X_m a_{m+1} X_{m+1} \cdots a_n
X_n$ is a failure trace of $q$, or $X_0 a_1 X_1 \cdots a_n X_n b
\emptyset$ is a failure trace of $p$ for some $b \in {\it Act}$.
Thus sequencing is an example of an operator from the literature
that preserves failure traces but that lies outside the scope of the
failure trace format.

\subsection{Action refinement}

The binary {\em action refinement} operator $t_1[a\rightarrow t_2]$ for
atomic actions $a$, based on a similar operator in \cite{vGW89},
replaces each $a$ transition in $t_1$
by $t_2$. Its transition rules are presented in
\tab{refinement}, which are added to the rules for
BPA$_{\bf 0}^;$ in \tab{sequencing}.
If we take the first argument of sequencing and of action refinement
to be liquid and the second argument of action refinement to be frozen,
then this standard TSS is in readiness format.
Note that in the second rule for action refinement the floating
variable $y_1$ is propagated at a liquid position, while
the variable $x_2$, which is propagated twice, is non-floating.

\begin{table}[htb]
\centering
$
\begin{array} {|cc|}
\hline
&\\
\Rule{x_1\goto w y}{x_1[v\rightarrow x_2]\goto w y[v\rightarrow x_2]}
\hspace{5mm}(v\not=w)\hspace{7mm} &
\Rule{x_1\goto v y_1~~~~x_2\goto w y_2}{x_1[v\rightarrow x_2]
\goto w y_2;(y_1[v\rightarrow x_2])} \\
&\\
\hline
\end{array}
$
\caption{Transition rules for action refinement}
\hlabel{tab-refinement}
\end{table}

\begin{corollary}{action refinement}
The ready simulation, ready trace and readiness preorders are precongruences
with respect to BPA$_{\bf 0}^;$ with action refinement.
\end{corollary}

%%%%%%%%%%%%%%%%%%%%%%%%%%%%%%%%%%%%%%%%%%%%%%%%%%%%%%%%%%%%%%%%%%%%%%%%%%%%%%
\section{Partial traces}
\hlabel{sec-partial}

The proof technique developed in this paper can be generally applied to
generate a precongruence format for a preorder from the observational
definition of this preorder. As an example we sketch how this technique
yields a precongruence format for the partial trace preorder. The details
are left to the reader. We say that a TSS is in {\em partial trace
format} if it is positive and in failure trace format.

\begin{theorem}{partial-trace preorder}
If a TSS is in partial trace format,
then the partial trace preorder that it induces is a precongruence.
\end{theorem}
The outline of the proof of \thm{partial-trace preorder} is as follows.
Assume a predicate $\Lambda$ on arguments of function symbols over a signature
$\Sigma$. We say that an ntytt rule is {\em $\Lambda$-partial trace safe}
if it is $\Lambda$-failure trace safe and either its conclusion
and premises are all positive, or they are all negative. In line with Lemmas
\hhref{lem-preservation-ready-trace}-\hhref{lem-preservation-negative-failure-trace}
we obtain: if $P$ is a TSS in decent ntyft format of which the
transition rules are $\Lambda$-partial trace safe, then any ntytt rule
irredundantly provable from $P$ is $\Lambda$-partial trace safe.
As in \cor{plus} we obtain: if $P=(\Sigma,R)$ is a standard TSS in partial
trace format, then $P^+=(\Sigma,R^+)$ is in partial trace format and
$P^+\vdash\alpha\Leftrightarrow P\vdash_{ws}\alpha$.
As in \lem{formula preservation} we obtain: if the rules
in $R^+$ are $\Lambda$-partial trace safe and
$\varphi \in \IO_{\it T}$ then $\psi(x) \in \IO_{\it T}$.
(Note that, as observations $\widetilde{a}$ do not occur in $\IO_{\it T}$,
the case $\psi \in t_P^{-1}(\widetilde{a})$ need not be considered.)
As in \lem{formula preservation frozen} we obtain:
if $P$ is in partial trace format and
$\varphi \in \IO_{\it T}^\wedge$ then $\psi(x) \in \IO_{\it T}^\wedge$.
Finally, as in \cor{congruence} we obtain: if $P$ is a standard TSS
in partial trace format and $\sigma(x) \sqsubseteq_{\it T}^P \sigma'(x)$ for
$x \in \var(t)$ then $\sigma(t) \sqsubseteq_{\it T}^P \sigma'(t)$.
This immediately implies \thm{partial-trace preorder}.

In the precongruence format for the partial trace preorder one could in
principle allow lookahead.
We leave it as an open question whether the partial trace format
extended with lookahead is
indeed a precongruence format for the partial trace preorder.

The following counterexample shows that in the case of the partial trace
{\em preorder}, the restriction to positive TSSs in the partial trace
format is essential.

\begin{example}{ex7}
Let the argument of $f$ be frozen.
We extend BPA$_{\delta\varepsilon}$ with the rule:
\[
\frac{x\gonotto b}{f(x)\goto c \delta}
\]
Clearly $a\sqsubseteq_{\it T} a+b$.
Note that $c$ is a trace of $f(a)$
(as \plat{f(a)\goto c \delta}), but not of $f(a+b)$.
Hence, $f(a)\not\sqsubseteq_{\it T} f(a+b)$.
\end{example}

According to the following theorem, in the case of partial trace {\em equivalence}
one can allow TSSs with negative premises.

\begin{theorem}{partial-trace equivalence}
If a standard TSS is in failure trace format,
then the partial trace equivalence that it induces is a congruence.
\end{theorem}
The outline of the proof of \thm{partial-trace equivalence} is as follows.
Let $\IO_{\it T}^{\wedge\sim}$ consist of all formulas
\[
\bigwedge_{i\in I}\varphi_i\,\wedge\,\bigwedge_{j\in J}\widetilde{a_j}
\]
with $\varphi_i\in\IO_{\it T}$ and $a_j\in A$. Let
$O_{\it T}^{\wedge\sim}(p):=\{\varphi\in\IO_{\it T}^{\wedge\sim}
\mid p\models\varphi\}$. If $p\sqsubseteq_{\it T} q$ and
$q\models\bigwedge_{j\in J}\widetilde{a_j}$
then $p\models\bigwedge_{j\in J}\widetilde{a_j}$.
Thus $p=_{\it T} q$ iff $O_{\it T}^{\wedge\sim}(p)=
O_{\it T}^{\wedge\sim}(q)$. As in \lem{formula preservation}
we obtain: if the rules in $R^+$ are $\Lambda$-failure
trace safe and $\varphi\in\IO_{\it T}$ then $\psi(x)\in\IO_{\it T}$.
As in \lem{formula preservation frozen} we obtain:
if $P$ is in failure trace format and $\varphi\in\IO_{\it T}^{\wedge\sim}$
then $\psi(x)\in\IO_{\it T}^{\wedge\sim}$.
Finally, as in \cor{congruence} we obtain:
if $P$ is a standard TSS in failure trace format
and $\sigma(x) =_{\it T}^P \sigma'(x)$ for $x \in \var(t)$
then $\sigma(t) =_{\it T}^P \sigma'(t)$.
This immediately implies \thm{partial-trace equivalence}.

%%%%%%%%%%%%%%%%%%%%%%%%%%%%%%%%%%%%%%%%%%%%%%%%%%%%%%%%%%%%%%%%%%%%%%%%%%%%%%
\section{Conservative extension}
\hlabel{sec-conservativity}

Traditionally, papers on congruence formats also introduce syntactic restrictions
on TSSs to ensure that one TSS is a {\em conservative extension} of another
\cite{BolG96,Gr93,GrV92}.
Here, we extend results on conservative extension from those three
papers to incomplete TSSs, building on propositions obtained in earlier sections.
However, since these propositions were proved for incomplete TSSs in ready
simulation format only, our conservative extension result is restricted to that format.

\begin{definition}{conservative}
Let $P_1=(\Sigma_1,R_1)$ and $P_2=(\Sigma_2,R_2)$ be standard TSSs
with $\Sigma_1 \subseteq \Sigma_2$. $P_2$ is said to be a {\em
conservative extension} of $P_1$ if $P_2 \vdash_{ws} \alpha
\Leftrightarrow P_1 \vdash_{ws} \alpha$ for any closed $\Sigma_2$-literal
$\alpha$ whose left-hand side is a term over $\Sigma_1$.
\end{definition}
This definition is equivalent to the ones in \cite{BolG96,Gr93,GrV92},
except that there only the case is considered where $R_1 \subseteq R_2$.

\begin{lemma}{conservative}
Let $P_1=(\Sigma_1,R_1)$ and $P_2=(\Sigma_2,R_2)$ be TSSs with
$\Sigma_1 \subseteq \Sigma_2$ and $R_1 \subseteq R_2$, such that each
rule in $R_1$ is decent and each rule in $R_2-R_1$ has a function
symbol $f \in \Sigma_2-\Sigma_1$ in its source. Then $P_2 \vdash_{\rm irr}
\frac{H}{\alpha} \Leftrightarrow P_1 \vdash_{\rm irr}
\frac{H}{\alpha}$ for any transition rule $\frac{H}{\alpha}$ over
$\Sigma_2$ whose source is a term over $\Sigma_1$.
Thus in particular $P_2 \vdash \alpha \Leftrightarrow P_1 \vdash \alpha$
for any $\Sigma_2$-literal $\alpha$ whose left-hand side is a term over
$\Sigma_1$.
\end{lemma}

\begin{proof}
``$\Leftarrow$'': Trivially, any proof of $\frac{H}{\alpha}$ from $P_1$
is also a proof of $\frac{H}{\alpha}$ from $P_2$.

``$\Rightarrow$'': with induction on the structure of irredundant
proofs. Let $\pi$ be an irredundant proof of $\frac{H}{\alpha}$ from
$P_2$.  We show that $\pi$ is also an irredundant proof of
$\frac{H}{\alpha}$ from $P_1$.  Let $K$ be the set of labels of the
nodes directly above the root node $q$ of $\pi$ (which is labelled
with $\alpha$). Then $\frac{K}{\alpha}$ is a substitution instance of
a transition rule $\rho$ in $R_2$.  As the left-hand side of $\alpha$
is a term over $\Sigma_1$, the rule $\rho$ must be in $R_1$.  Thus
$\rho$ is decent, which implies that all variables occurring in the
left-hand sides of its premises also occur in its source. Therefore,
any function symbol occurring in the left-hand side of a literal in
$K$ also occurs in the left-hand side of $\alpha$ and hence belongs to
$\Sigma_1$.  For all $\beta \in K$ there is an irredundant proof
$\pi_\beta$, which is a proper part of $\pi$, of a rule
\plat{\frac{H_\beta}{\beta}} from $P_2$. We have $\bigcup_{\beta \in
K} H_\beta = H$. Moreover, the left-hand side of ${\beta}$ is a term
over $\Sigma_1$. So, by induction, $\pi_\beta$ is an irredundant proof
of ${\beta}$ from $P_1$ for all $\beta \in K$. Thus $\beta$ is a
$\Sigma_1$-literal. As $\rho$ is decent, all variables occurring in
its target also occur in its source or in one of its premises.
Therefore, any function symbol occurring in the right-hand side of
$\alpha$ also occurs in its left-hand side or in $K$, and hence
belongs to $\Sigma_1$.  Hence $\pi$ is an irredundant proof of
$\alpha$ from $P_1$.
\end{proof}

\begin{theorem}{conservative}
Let $P_1=(\Sigma_1,R_1)$ and $P_2=(\Sigma_2,R_2)$ be standard TSSs in
decent ntyft format with $\Sigma_1 \subseteq \Sigma_2$ and $R_1
\subseteq R_2$, such that each rule in $R_2-R$ has a function symbol
$f \in \Sigma_2-\Sigma_1$ in its source. Then $P_2$ is a conservative
extension of $P_1$.
\end{theorem}

\begin{proof}
Let $P_1=(\Sigma_1,R_1)$ and $P_2=(\Sigma_2,R_2)$ be as stipulated. Let
$P'_1=(\Sigma_1,R'_1)$ and $P'_2=(\Sigma,R'_2)$ be the TSSs in decent
xynft format obtained from $P_1$ and $P_2$ as indicated in the proof of
\pr{xynft}: $R'_1$ (resp.\ $R'_2$) consists of all decent xynft rules
irredundantly provable from $P_1$ (resp.\ $P_2$).  Now clearly $R'_1
\subseteq R'_2$, and by \lem{conservative} each rule in $R'_2-R'_1$
has a function symbol $f \in \Sigma_2-\Sigma_1$ in its source.  By
means of a coordinated bijective renaming of the variables in $R'_1$
and $R'_2$ we obtain TSSs in decent uniform xynft format $P''_1$ and
$P''_2$ that still satisfy the stipulations of \thm{conservative}.  On
these the construction of \df{uniform} may be applied to obtain
$P_1^+=(\Sigma,R_1^+)$ and $P_2^+=(\Sigma_2,R_2^+)$. By Definitions
\hhref{df-pick} and \hhref{df-uniform} we have $R_1^+
\subseteq R_2^+$ and each rule in $R_2^+-R_1^+$ has a function symbol
$f \in \Sigma_2-\Sigma_1$ in its source. Thus, by Propositions
\hhref{pr-xynft}, \hhref{pr-well-supported proof} and
\hhref{pr-supported proof} and \lem{conservative} we have  
$P_2 \vdash_{ws} {\alpha} \Leftrightarrow
P'_2 \vdash_{ws} {\alpha} \Leftrightarrow
P'_2 \vdash_{s} {\alpha} \Leftrightarrow P''_2 \vdash_{s} {\alpha}
\Leftrightarrow P_2^+ \vdash {\alpha} \Leftrightarrow P_1^+ \vdash {\alpha}
\Leftrightarrow \cdots
\Leftrightarrow P_1 \vdash_{ws} {\alpha}$
for any closed $\Sigma_2$-literal $\alpha$ whose left-hand side is a
term over $\Sigma_1$.
\end{proof}

%%%%%%%%%%%%%%%%%%%%%%%%%%%%%%%%%%%%%%%%%%%%%%%%%%%%%%%%%%%%%%%%%%%%%%%%%%%%%%
\section{Possible extensions}
\hlabel{sec-conclusion}

In this paper we have presented precongruence formats for a range
of preorders, which are {\em a fortiori} also congruence formats
for the induced equivalences. We are not aware of any
further plausible extensions of the precongruence formats
(apart from conceptual extensions such as higher-orderness
and syntactic sugar such as predicates and terms as transition
labels).

It may be possible to formulate more liberal congruence formats
for ready simulation equivalence and the decorated trace equivalences.
Namely, one could in principle allow lookahead for frozen arguments
of the source. To be more precise, one could (re)define that an occurrence
of a variable in an ntytt rule is {\em propagated} if the occurrence
is either in the target or in the left-hand side of a positive
premise of which the right-hand side is polled or propagated.
Furthermore, one could define that a positive premise in an ntytt rule
has {\em lookahead} if its right-hand side occurs in the left-hand side
of some premise in this rule. Assume a predicate $\Lambda$ on
arguments of function symbols. Then a standard ntytt rule is
{\em $\Lambda$-ready simulation equivalence safe} if
each $\Lambda$-floating variable is only propagated
at $\Lambda$-liquid positions, and not in positive premises
with lookahead. A standard ntytt rule is
{\em $\Lambda$-ready trace equivalence safe} if moreover
each $\Lambda$-floating variable is propagated at most once,
at a $\Lambda$-liquid position. A standard ntytt rule is
{\em $\Lambda$-readiness equivalence safe} if moreover
each $\Lambda$-floating variable is not both propagated and polled.
A standard ntytt rule is
{\em $\Lambda$-failure trace equivalence safe} if moreover
each $\Lambda$-floating variable is polled at most once,
at a $\Lambda$-liquid position in a positive premise.
A TSS is in $N$ equivalence format
(for $N\in\{\mbox{ready simulation},
\mbox{ready trace},\mbox{readiness},\mbox{failure trace}\}$) if
it is in ntyft/ntyxt format and its rules are $\Lambda$-$N$ equivalence
safe with respect to some $\Lambda$.

We leave it as an open question whether these four
formats are indeed congruence formats for ready simulation equivalence,
ready trace equivalence, readiness equivalence, failure trace
equivalence and failure equivalence.
The proof technique employed in this paper uses the
absence of lookahead in an essential way. In particular, as
illustrated by \ex{no lookahead}, the construction
of $R^+$ in \sect{reducing} only works in the absence of
lookahead. So the open question cannot be answered by a simple
adaptation of the precongruence proofs in this paper.
\ex{lookahead} shows that the {\em pre}congruence formats cannot allow lookahead.

%%%%%%%%%%%%%%%%%%%%%%%%%%%%%%%%%%%%%%%%%%%%%%%%%%%%%%%%%%%%%%%%%%%%%%%%%%%%%%

\end{document}